\documentclass[twocolumn]{autart}    
\usepackage{graphicx}          
\usepackage{amsmath}
\usepackage{mathtools}
\usepackage{amssymb}
\usepackage{amsfonts}
\usepackage{upgreek}
\usepackage{hyperref}
\usepackage{cite}
\usepackage{savesym}
\savesymbol{AND}
\usepackage{algorithmic}
\usepackage{algorithm}
\usepackage{algorithm}
\usepackage{algorithmic}
\usepackage{graphicx}
\usepackage{mathdots}
\usepackage{bm}
\usepackage{cases}
\usepackage{caption}
\usepackage{subcaption}
\usepackage{color}
\usepackage{soul}
\usepackage{enumitem}
\newtheorem{definition}{Definition}
\newtheorem{lemma}{Lemma}
\newtheorem{remark}{Remark}
\newtheorem{theorem}{Theorem}

\begin{document}
	
	\begin{frontmatter}
		\title{Prescribed Performance Distance-Based Formation Control of Multi-Agent Systems (Extended Version)}
		
		\author[Belgium]{Farhad Mehdifar}\ead{farhad.mehdifar@uclouvain.be},    
		\author[Greece]{Charalampos P. Bechlioulis}\ead{chmpechl@mail.ntua.gr},               
		\author[Iran]{Farzad Hashemzadeh}\ead{hashemzadeh@tabrizu.ac.ir}, 
		\author[Iran]{Mahdi Baradarannia}\ead{mbaradaran@tabrizu.ac.ir} 
		
		\address[Belgium]{INMA, ICTEAM, UCLouvain, Louvain-la-Neuve, Belgium.}  
		\address[Greece]{School of Mechanical Engineering, National Technical University of Athens, Athens, Greece.}             
		\address[Iran]{Faculty of Electrical and Computer Engineering, University of Tabriz, Tabriz, Iran}        


		\begin{abstract}                          
			This paper presents a novel control protocol for robust distance-based formation control with prescribed performance in which agents are subjected to unknown external disturbances. Connectivity maintenance and collision avoidance among neighboring agents are also handled by the appropriate design of certain performance bounds that constrain the inter-agent distance errors. As an extension to the proposed scheme, distance-based formation centroid maneuvering is also studied for disturbance-free agents, in which the formation centroid tracks a desired time-varying velocity. The proposed control laws are decentralized, in the sense that each agent employs local relative information regarding its neighbors to calculate its control signal. Therefore, the control scheme is implementable on the agents' local coordinate frames. Using rigid graph theory, input-to-state stability, and Lyapunov based analysis, the results are established for minimally and infinitesimally rigid formations in 2-D or 3-D space. Furthermore, it is argued that the proposed approach increases formation robustness against shape distortions and can prevent formation convergence to incorrect shapes, which is likely to happen in conventional distance-based formation control methods. Finally, extensive simulation studies clarify and verify the proposed approach.
		\end{abstract}
		
	\end{frontmatter}
	
	\section{Introduction}
	During the past several years, multi-agent systems (MAS) and particularly cooperative control of MAS, which deals with achieving a global group behavior that is beyond each individual capabilities through local interactions, have attracted increasing attention from control scientists and engineers due to its broad applications \cite{darmanin2017review}.  Control problems in MAS are mainly classified into consensus, formation, containment, flocking, coverage, and rendezvous \cite{dorri2018multi,cao2013overview,wang2016multi,anderson2008rigid}. Formation control refers to the design of appropriate control protocols for stabilizing agents’ positions with respect to each other so that they set up and maintain a predefined geometrical shape. Based on a recent survey \cite{oh2015survey}, the existing approaches can be categorized into position-, displacement-, and distance-based formation control schemes, depending on the sensing and controlled variables. Among them, distance-based formation control is considered to be an attractive architecture since it imposes less implementation issues compared to other methods.  In distance-based formation control, agents measure the relative positions with respect to their neighbors and actively control their inter-agent distances in order to reach a desired predefined shape. This approach enables us to design formation control laws in agents’ local coordinate frames, which do not require global position measurements (e.g., using GPS) nor pre-alignment of agents’ local coordinate frames (e.g., using a compass) \cite{meng2016formation,oh2015survey}.  In particular, this formation control approach is advantageous not only due to lower agents' costs (since they use less complex equipment for sensing and local interactions) but also for operation in GPS-denied environments, where unmanned multi-agent systems are used for search and rescue operations, planetary explorations, indoor
	navigation, and so on \cite{ramazani2017rigidity}.
	
	An introduction to distance-based formation control for both undirected and directed inter-agent sensing graphs is found in \cite{anderson2008rigid}. Some early results in distance-based formation control are also given in \cite{krick2009stabilisation,dorfler2010geometric,cai2014rigidity}. In these works, formation acquisition (convergence to a stationary shape) for single integrator agents with undirected minimally and infinitesimally rigid interaction graphs is considered. The aforementioned problem is also analyzed under directed minimally persistent graphs in \cite{yu2009control}. In recent years, various modified controllers for undirected distance based formation acquisition were also developed considering event-triggered control \cite{sun2019cooperative}, quantized distance measurements \cite{sun2018quantization}, exponential convergence \cite{sun2016exponential}, finite-time convergence \cite{sun2015finite}, source-seeking \cite{barogh2017cooperative}, nonlinear dynamics \cite{cai2015adaptive,vaddipalli2018multi}, and formation scaling \cite{yang2019stress}. Sun et al in \cite{sun2018distributed} presented a distance-based formation acquisition with prescribed orientation by using displacement-based formation control for a portion of agents. More recently, optimal distance-based formation control is also addressed in \cite{babazadeh2019distance}. Moreover, a multi layered version of distance-based formation acquisition for 3-dimensional shapes are developed in \cite{ramazani2017rigidity,vaddipalli2018multi}. A method for dealing with the problem of convergence to incorrect equilibrium points (undesired shapes) of distance-based formation acquisition controllers was recently proposed in \cite{anderson2017formation}, introducing an additional control variable (the signed area of a triangle) for triangular shapes. Later, Liu et al \cite{liu2019further} tried to generalize this method for planar formations with $n > 3$ agents. Furthermore, the authors in \cite{mou2015undirected} have analyzed the effects of distance mismatches between neighboring agents in distance-based formation stabilization. Towards the same direction, \cite{de2014controlling} proposed a solution taking into account the distance mismatch effects in the closed loop system. Later, de Marina et al \cite{de2016distributed} exploited the artificially imposed inter-agent distance mismatches to control formations with rotational and translation maneuvers.
	
	In addition to formation acquisition control, there have been attempts to solve distance-based formation control problems with moving shapes (such as formation tracking and formation maneuvering). In \cite{deghat2015combined}, by combining distance-based formation control with consensus protocols, agents move in a target formation shape with a common constant speed. The results are enhanced taking into account the collision avoidance problem and finite time convergence scheme in \cite{wang2018distance} and \cite{sun2015finite}, respectively. Recently, in \cite{yang2018distributed}, a weighted centroid formation tracking with distance-based control laws was introduced, where the formation centroid tracks a predefined path with the agents employing finite time estimators to calculate the centroid of the formation. However, this methodology requires each agent to sense its relative orientation with respect to its neighbors. Moreover, \cite{babazadeh2019distance} developed a distance-based optimal formation tracking using State Dependent Riccati Equation with energy constraints, nonetheless, the design methodology relies on a centralized control approach. A distance-based formation maneuvering controller is proposed in \cite{cai2015formation} provided that all agents have direct access to the desired time-varying swarm velocity, which is not practical and also requires pre-alignment of agent's local coordinate systems. Later, \cite{mehdifar2018finite} and \cite{khaledyan2019flocking} utilized distributed velocity estimators in distance-based formation maneuvering for single integrator and unicycle agent models, respectively. In these works, agents estimate the desired group velocity that is only available to a leader in order to relax the requirement of direct access to the desired time-varying group velocity. Nevertheless, these schemes also require inter-agent relative orientation measurements in order to be applicable in arbitrary oriented local coordinate systems. In \cite{rozenheck2015proportional}, a distance-based centroid formation maneuvering with a leader is investigated, where the centroid of the formation tracks a constant (or at best a very slowly varying) desired reference velocity.
	
	The existence of external disturbances that affect the agents dynamics is a significant issue of practical interest for MAS applications. It is noteworthy to mention that in distance-based formation control problems, none of the aforementioned works have taken into account external disturbances. Recently, \cite{bae2018disturbance} studied the disturbance attenuation problem with the LMI approach in distance-based formation control. However, its results depend on certain LMI feasibility tests, which are not favorable in practice and may increase complexity in the controller design. Moreover, as a practical problem, collision avoidance among agents has been addressed partially in a few of the above mentioned works, such as \cite{wang2018distance,barogh2017cooperative,babazadeh2019distance}. In addition, none of these works has addressed connectivity maintenance among neighboring agents, which is also critical since agents have limited sensing capabilities in practice. Apparently, tackling distance-based formation control under external disturbances along with collision avoidance and connectivity maintenance is a more challenging problem. Finally, another crucial issue concerns the transient response of the MAS. Towards this direction, Prescribed Performance Control (PPC) \cite{bechlioulis2008robust}, proposes a simple and constructive procedure based on which the transient performance of the closed-loop system is predetermined by certain user defined performance bounds. This method has been also applied in MAS control \cite{karayiannidis2012multi,macellari2017multi,bechlioulis2018distributed,guan2014guaranteed}. Recently, PPC has been utilized for formation control problems as well, however, these results are mainly applied to displacement-based formation control methods \cite{bechlioulis2014robust,bechlioulis2016decentralized,wang2018guaranteed,wei2018learning}. A recent paper \cite{verginis2019robust} has also employed PPC for MAS, however, rather than solving a formation control problem, by controlling inter-agent relative orientations and distances it solves a distance- and orientation- based multi-agent coordination problem (that cannot ensure convergence to a specific pre-defined shape) where inter-agent interactions are modeled by undirected tree graphs.
	
	In this paper, we propose a robust distance-based formation acquisition control protocol with guaranteed transient performance, connectivity maintenance and collision avoidance among neighboring agents. The target formation is assumed to be minimally and infinitesimally rigid in two or three dimensional space. User defined performance guarantees on the system's response are achieved by employing time-varying decreasing performance bounds on the inter-agent distance errors. We prove that the proper choice of performance bounds handles the problems of connectivity maintenance and collision avoidance among neighboring agents as well. More specifically, an error transformation technique based on a time-varying mapping is used to convert the original constrained error system into a new equivalent unconstrained one, whose stability only ensures satisfaction of certain time-varying constraints of the inter-agent distance errors. Afterwards, we develop a distance-based formation maneuvering control protocol, based on which the centroid of the formation tracks a desired time-varying velocity. It is assumed that the desired centroid velocity is only available to the leader of the group. The proposed control approach is independent of a global coordinate system and can be applied to arbitrarily oriented local coordinate frames. Furthermore, we prove that the proposed approach naturally reduces potential formation distortions induced by external disturbances, thus preventing convergence to undesired shapes. 
	
	The contributions of this work are summarized as follows\footnote{Our previous work \cite{mehdifar2017distance} allows only connectivity maintenance and collision avoidance among neighboring agents by enforcing constant constraints for the distance errors. Hence, it is not capable of: (i) imposing predefined transient and steady state performance on the closed-loop system, (ii) dealing with external disturbances while achieving the desired formation and (iii) solving the formation maneuvering problem.}:
	\begin{itemize}
		\item To the best of our knowledge, there is no previous work addressing distance-based formation control with guaranteed transient and steady state performance.
		
		\item There are very few previous works addressing distance-based formation  control with external disturbances. In contrast to \cite{bae2018disturbance}, the performance of the proposed scheme does not depend on the upper bound of the external disturbances nor LMI feasibility conditions.
		
		\item To the best our knowledge, there are no previous works on distance-based formation maneuvering control with guaranteed performance, connectivity maintenance and collision avoidance. Moreover, for the first time, this paper solves the distance-based formation maneuvering problem with time-varying reference velocity for arbitrarily oriented local coordinate frames of agents and without requiring any additional measurements on relative orientations. 
	\end{itemize}
	
	The rest of this paper is organized as follows: Section \ref{Sec:Preli} presents preliminary concepts on rigid graph theory as well as some useful technical lemmas. In Section \ref{Sec:Prob}, the problem of robust prescribed performance distance based formation control with connectivity maintenance and collision avoidance of neighboring agents is formulated. The proposed control design along with the stability analysis are provided in Section \ref{designand analyis}. In Section \ref{Centroid_Manue}, a prescribed performance distance-based formation centroid maneuvering controller is proposed. Extensive simulation results are given in Section \ref {Simu}. Finally, Section \ref{Conclu} concludes our work.
	
	\textbf{Notations:}
	The real $n$-dimensional space is denoted by $\mathbb{R}^n$. Also, note that $\mathbb{R}\triangleq (-\infty , \infty)$. $x \in \mathbb{R}^n$ is a $n \times 1$ vector and $x^T$ is its transpose. 
	$\mathrm{dim(.)}$ denotes the dimension of a vector or set. 
	For $i=1,\ldots,n$, $\mathrm{col}(x_i)\triangleq [x_1^T,\ldots,x_n^T]^T$ denotes a $nm \times 1$ vector where $x_i \in \mathbb{R}^m$. $\|x\|$ signifies the standard Euclidean norm and $\|x\|_\infty$ represents the infinity-norm. 
	$|.|$ is the absolute value and for $x \in \mathbb{R}^n$ we have $|x| \triangleq [|x_1|, |x_2|, \ldots , |x_n|]^T$. $\mathbb{R}^{n\times m}$ denotes the real $n \times m$ matrix space. For $A \in \mathbb{R}^{n\times m}$, $A^T$, $\lambda_{\mathrm{min}}(A)$, $\lambda_{\mathrm{max}}(A)$ indicate transpose, minimum and maximum eigenvalues, respectively. 
	$\mathbf{1}_n$ and $\mathbf{0}_n$ stand for $n \times 1$ vectors of ones and zeros, respectively. 
	$\mathrm{diag}(.)$ and $\mathrm{blockdiag}(.)$ are the diagonal operators making a diagonal and block diagonal matrix by their arguments, respectively. $\otimes$ denotes the Kronecker product. Finally, $\mathrm{dist}(y,\mathcal{M})=\inf_{x \in \mathcal{M}} \|y-x\|$ designates the distance of a point $y \in \mathbb{R}^n$ from a set $\mathcal{M} \in \mathbb{R}^n$.
	
	\section{Preliminaries}
	\label{Sec:Preli}
	
	\subsection{Basic Concepts on Graphs and Rigidity Theory}
	\label{SubSec:RigidGraph}
	
\begin{figure}[t]
		\centering
		\begin{subfigure}[t]{0.17\textwidth}
			\centering
			\includegraphics[width=\textwidth]{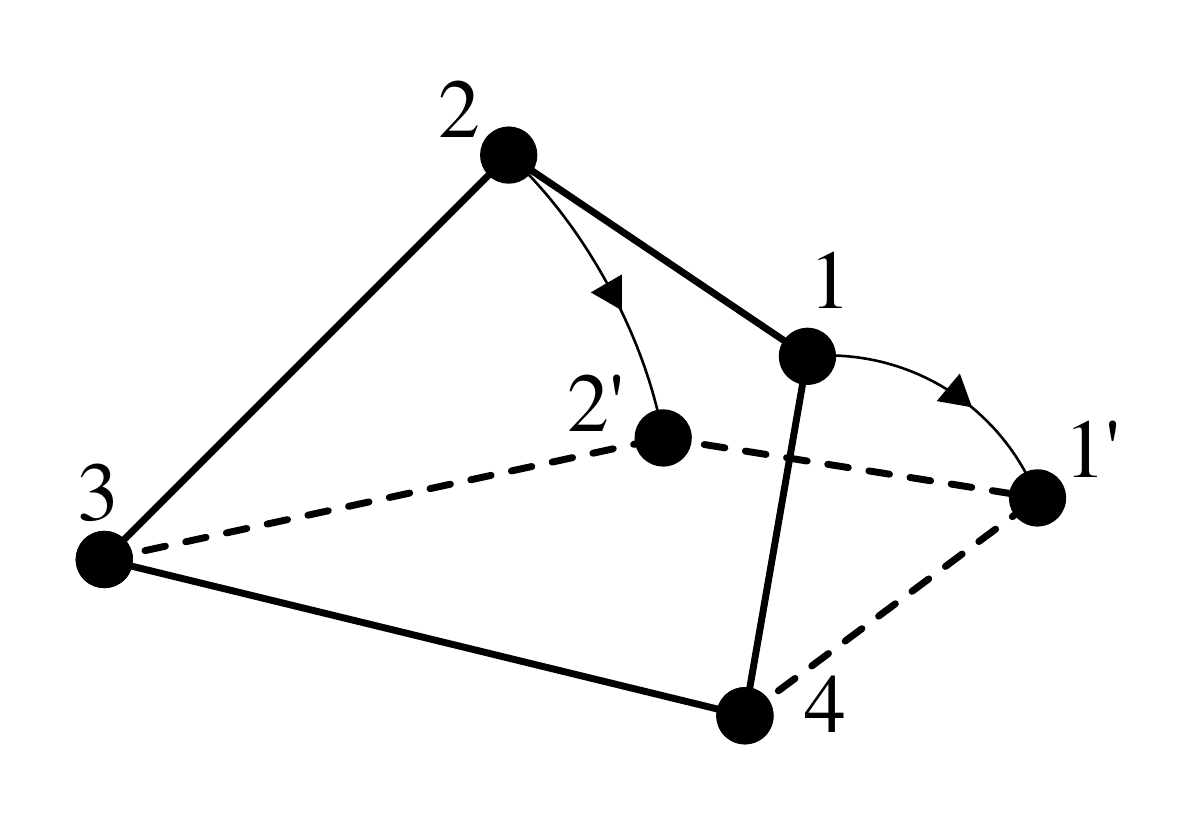}
			\caption{}
		\end{subfigure}%
		~
		\begin{subfigure}[t]{0.14\textwidth}
			\centering
			\includegraphics[width=\textwidth]{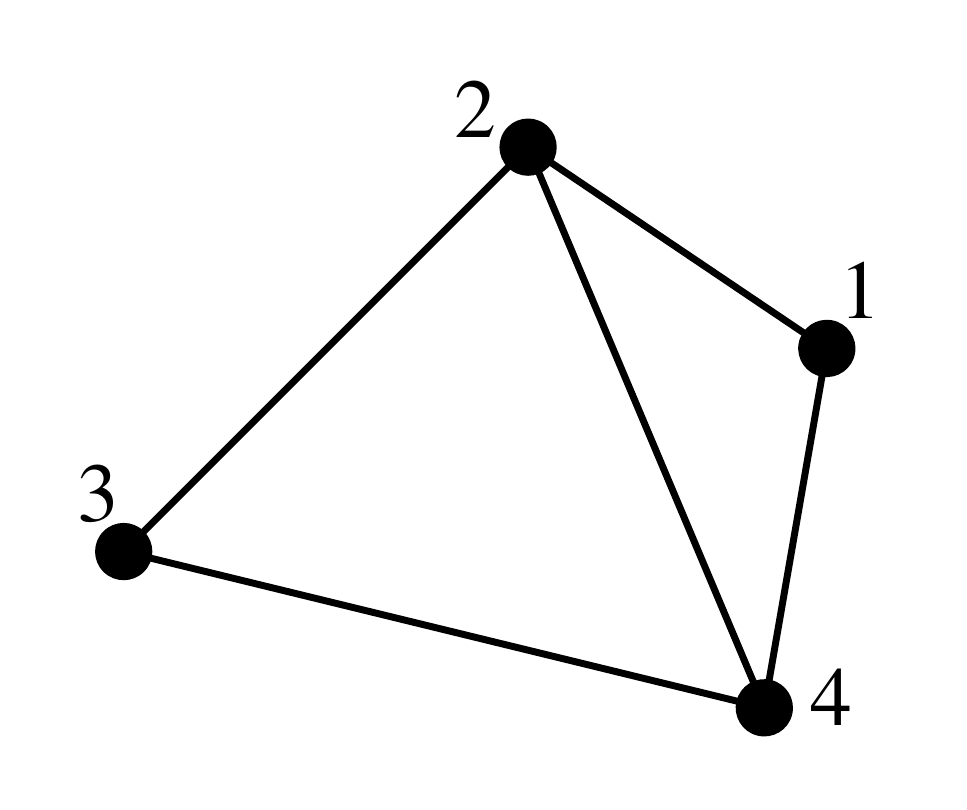}
			\caption{}
		\end{subfigure}
		~
		\begin{subfigure}[t]{0.14\textwidth}
			\centering
			\includegraphics[width=\textwidth]{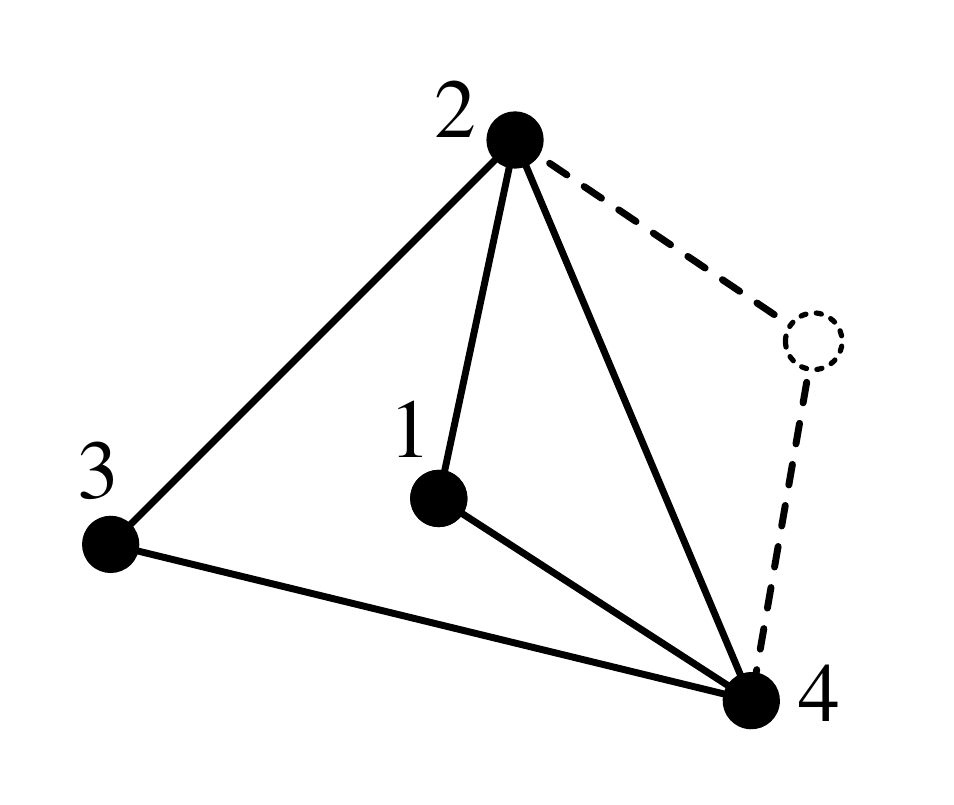}
			\caption{}
		\end{subfigure}
		\caption{Graph (a) is a non-rigid (flexible) graph. Graph (b) is a minimally and infinitesimally rigid graph. Graph (c) belongs to the Amb set of graph (b).}
		\label{fig:rigi_con}
	\end{figure}

	Consider an undirected graph with $l$ edges and $n$ vertices, denoted by $\mathcal{G}\triangleq(\mathcal{V},\mathcal{E})$ where $\mathcal{V}=\{1,2,\ldots,n\}$ is the set of vertices and $\mathcal{E}=\{(i,j)\lvert i,j \in \mathcal{V}, i \neq j\}$ is the set of undirected edges in the sense that there is no distinction between $(i,j)$ and $(j,i)$. The neighbor set of vertex $i$ is defined as $\mathcal{N}_i(\mathcal{E})=\{j\in \mathcal{V} \mid (i,j)\in \mathcal{E}\}$.
	The incidence matrix $H=\{h_{ij}\} \in \mathbb{R}^{l \times n}$, relates the edges of $\mathcal{G}$ with its vertices. Assuming arbitrary edge orientation, the entries of $H$ are defined as:
	\begin{numcases}{h_{ij}=}
	1,& the $i$-th edge sinks at node $j$ \nonumber \\
	-1,& the $i$-th edge leaves node $j$ \nonumber\\
	0,& otherwise \nonumber
	\end{numcases}
	For any connected graph it is known that $\mathrm{ker}(H)=\mathrm{span}\{\mathbf{1}_n\}$ \cite{sun2015finite}.
	Let $p_i \in \mathbb{R}^m$, where $m \in\lbrace 2,3 \rbrace$, denotes a point that is assigned to vertex $i \in \mathcal{V}$. The stacked vector $p=\mathrm{col}(p_i) \in \mathbb{R}^{mn}$, represents the realization of $\mathcal{G}$ in $\mathbb{R}^m$. The pair $\mathcal{F}\triangleq(\mathcal{G},p)$ is said to be a framework of $\mathcal{G}$ in $\mathbb{R}^m$. By introducing the matrix $\bar{H}= H \otimes \mathrm{I}_m \in \mathbb{R}^{ml \times mn}$, the relative position vectors corresponding to the edges can be constructed as:
	\begin{equation}
	\widetilde{p}=\bar{H}p,
	\end{equation}
	where $\widetilde{p}=\mathrm{col}(\widetilde{p}_{ij}) \in \mathbb{R}^{ml}$ with $\widetilde{p}_{ij}=p_i-p_j \in  \mathbb{R}^{m}$ being the relative position vector defined for any pair $(i,j)\in \mathcal{E}$.
	Given an arbitrary ordering of edges in $\mathcal{E}$, an edge function (rigidity function) $\Phi_\mathcal{G} : \mathbb{R}^{mn} \rightarrow \mathbb{R}^{l}$ associated with $(\mathcal{G},p)$ is given as:
	\begin{equation}
	\label{eq:edgefun}
	\Phi_\mathcal{G}(p)=\left[ \ldots,\|p_i-p_j\|^2,\ldots\right] ^T , \quad (i,j)\in \mathcal{E}.
	\end{equation}
	such that its $k$-th component, i.e., $\|p_i-p_j\|^2$, relates to the $k$-th edge of $\mathcal{E}$ connecting the $i$-th and $j$-th vertices. 
	\begin{definition} \label{def:rigid} \cite{oh2011formation}
		A framework $\mathcal{F}=(\mathcal{G},p)$ is rigid if there exists a neighborhood $\mathcal{U}_p$ of $p \in \mathbb{R}^{mn}$ such that $\Phi_\mathcal{G}^{-1}\left[\Phi_\mathcal{G}(p) \right] \cap \mathcal{U}_p = \Phi_{\mathcal{H}}^{-1}\left[\Phi_{\mathcal{H}} (p)\right]\cap \mathcal{U}_p$, where ${\mathcal{H}}$ is a complete graph of $n$ vertices and $\Phi^{-1}_{\star}$ is a set of positions $q\in \mathbb{R}^{mn}$ satisfying $\Phi_\star(p)=\Phi_\star(q)$ for any graph $\star$.
	\end{definition}
	\noindent
	Definition \ref{def:rigid} implies that in a rigid framework, preserving the length of the graph edges guarantees that all distances among all vertices of the graph remain unaltered (i.e., the shape is preserved). The rigidity matrix $R : \mathbb{R}^{mn} \rightarrow \mathbb{R}^{l \times mn}$ of $\mathcal{F}=(\mathcal{G},p)$ is then defined as:
	\begin{equation}
	\label{eq:rigidmat}
	R(p)=\dfrac{1}{2}\dfrac{\partial \Phi_\mathcal{G}(p)}{\partial p}=\dfrac{1}{2}\dfrac{\partial \Phi_\mathcal{G}(p)}{\partial\widetilde{p}} \dfrac{\partial \widetilde{p}}{\partial p}= P^T \bar{H},
	\end{equation}
	where $P=\mathrm{blockdiag}(\widetilde{p}_{ij})$ for the same ordering of edges as in \eqref{eq:edgefun}. Note that, each row of the rigidity matrix $R(p)$ takes the following form:
	\begin{equation}
	\label{eq:rigid_stru}
	\left[\mathbf{0}_{1 \times m}^T  \ldots \, \widetilde{p}_{ij}^T  \ldots \, \mathbf{0}_{1 \times m}^T \, \ldots \, -\widetilde{p}_{ij}^T  \ldots  \mathbf{0}_{1 \times m}^T\right]
	\end{equation}
	Hence, the rigidity matrix depends solely on the relative positions and can be written as $R(\widetilde{p}\,)$. It is know that $\mathrm{rank}[R(p)]\leq 2n-3$ in $\mathbb{R}^{2}$, and $\mathrm{rank}[R(p)]\leq 3n-6$ in $\mathbb{R}^{3}$ \cite{asimow1979rigidity}.
	\begin{definition}\cite{anderson2008rigid}  \label{def:mini}
		A rigid framework is said to be minimally rigid if no single inter-agent distance constraint can be removed without causing the graph to lose its rigidity. In $\mathbb{R}^2$ ( $\mathbb{R}^{3}$) a rigid framework $(G,p)$ is minimally rigid if $l=2n-3$ ($l=3n-6$) .
	\end{definition}
	\begin{definition}\cite{hendrickson1992conditions} \label{def:infini} 
		A framework $\mathcal{F}=(\mathcal{G},p)$ is infinitesimally rigid in $m$-dimensional space if:
		\begin{equation}
		\mathrm{rank}[R(p)]= mn-\dfrac{m(m+1)}{2}.
		\end{equation}
	\end{definition}
	\noindent
	If a framework is infinitesimally rigid in $\mathbb{R}^{2}$ ($\mathbb{R}^{3}$) and its underlying graph has exactly $2n-3$ ($3n-6$) edges, then it is called a minimally and infinitesimally rigid framework. If $\Phi_G(p)=\Phi_G(q)$ holds for two frameworks $\mathcal{F}_p=(\mathcal{G},p)$ and $\mathcal{F}_q=(\mathcal{G},q)$, they are said to be equivalent. Also, if $\|p_i - p_j\| = \|q_i - q_j\|$, $\forall i,j\in \mathcal{V}$, then the two frameworks are congruent. An isometry of $\mathbb{R}^m$ is a bijective map $\mathcal{Q} : \mathbb{R}^{m} \rightarrow \mathbb{R}^{m}$ satisfying \cite{izmestiev2009infinitesimal} $\|x-y\|=\|\mathcal{Q}(x)-\mathcal{Q}(y)\|,  \forall x,y \in \mathbb{R}^{m}$,  
	where $\mathcal{Q}$ accounts for rotation, translation, and reflection of the vector $x-y \in \mathbb{R}^{m}$. Let $\mathrm{Iso}(\mathcal{F})$ denotes the set of all isometric frameworks of $\mathcal{F}$. Note that, \eqref{eq:edgefun} is invariant under isometric motions of $\mathcal{F}$. Two infinitesimally rigid frameworks $\mathcal{F}_p=(\mathcal{G},p)$ and $\mathcal{F}_q=(\mathcal{G},q)$ are said to be ambiguous, if they are equivalent but not congruent \cite{anderson2008rigid}. We denote the set of all ambiguities of an infinitesimally rigid framework $\mathcal{F}$ and its isometries by $\mathrm{Amb}(\mathcal{F})$. It can be reasonably assumed that all frameworks in $\mathrm{Amb}(\mathcal{F})$ are infinitesimally rigid \cite{cai2015formation,cai2014rigidity} (according to \cite{anderson2008rigid} and \cite[Theorem 3]{asimow1979rigidity}, this assumption holds almost everywhere; therefore, it
	is not restrictive). Figure \ref{fig:rigi_con} illustrates the aforementioned concepts.
	
	\begin{lemma}\cite{cai2015formation} \label{co:infini_col}
		If $\mathcal{F}_p=(\mathcal{G},p)$ is infinitesimally rigid then there exists a small positive constant $\vartheta$ such that all frameworks ${\mathcal{F}_q=(\mathcal{G},q)}$ that satisfy:
		\begin{equation}
		\label{eq:Psi}
		\Psi ({\mathcal{F}_q},\mathcal{F}_p) \triangleq \sum_{(i,j)\in \mathcal{E}} (\|q_i - q_j\|-\|p_i - p_j\|)^2 \leq \vartheta,
		\end{equation}
		are also infinitesimally rigid.
	\end{lemma}
	\begin{remark}
		The requirement that $\vartheta$ should be sufficiently small in Lemma \ref{co:infini_col} is a conservative estimate of how far $\mathcal{F}_q$ can be from $\mathcal{F}_p$ (in the sense of its shape) in order to remain infinitesimally rigid. One example is that three or more neighboring agents (nodes) cannot be collinear (coplanar) in $\mathbb{R}^2$ ($\mathbb{R}^3$) because $\mathrm{rank}[R(q)]\neq 2n-3$ ($\neq 3n-6$) and $\mathcal{F}_q$ would not be infinitesimally rigid. In other words, keeping $\vartheta$ sufficiently small is a sufficient condition for avoiding such degenerate cases.
	\end{remark}
	\begin{lemma}\cite{sun2015finite}
		\label{lem:PD_rigid}
		If the framework $\mathcal{F}=(\mathcal{G},p)$ is minimally and infinitesimally rigid in $m$-dimensional space, then the matrix $R(p)R(p)^T$ is positive definite. 
	\end{lemma}

	\subsection{Some Useful Technical Lemmas}
	\label{SubSec:ineq}
	
	\begin{definition} \cite{khalil2002noninear}
		For system
		\begin{equation}
		\label{eq:sys}
		\dot{x}=f(t,x,u)
		\end{equation} 
		where $x\in\mathbb{R}^n$ , $u\in\mathbb{R}^m$ is a bounded input for all $t\geq0$, and $f : [0,\infty) \times \mathbb{R}^{n} \times \mathbb{R}^{m} \rightarrow \mathbb{R}^{n}$ is piece-wise continuous in $t$ and locally Lipschitz in $x$ and $u$, if there exists a class $\mathcal{KL}$ function $\beta_1$ and a class $\mathcal{K}$ function $\beta_2$ such that :
		\begin{equation}
		\|x(t)\|\leq \beta_1(\|x(0)\|,t)+\beta_2(\|u\|_{\infty}), \qquad \forall t>0	\end{equation}
		then it is input to state stable (ISS).
	\end{definition}
	\begin{lemma}  \cite{chen2015stabilization}
		\label{lem:ISS_lyap}
		For system \eqref{eq:sys}, if there exist a continuously differentiable function $V(t,x) : [0,\infty) \times \mathbb{R}^n\rightarrow \mathbb{R}^+$  such that:
		\begin{subequations}
			\begin{align}
			&\alpha_1(\|x\|)\leq V(t,x) \leq\alpha_2(\|x\|) \\
			&\dot{V}(t,x) \leq -\alpha_3(\|x\|) + \alpha_4(\|u\|),
			\end{align}
		\end{subequations}
		where $\alpha_i, i=1,2,3$ are class $\mathcal{K}_{\infty}$ functions and $\alpha_4$ is a class $\mathcal{K}$ function, then it is ISS.
	\end{lemma}
	
	\begin{lemma} \cite[Lemma 11.3]{chen2015stabilization} \label{lem: kapa_inf_ineq}
		Let $\alpha_1$ and $\alpha_2$ be class $\mathcal{K}_{\infty}$ functions. There exists a class $\mathcal{K}_{\infty}$ function $\beta$ such that
		\begin{equation} \label{eq:kapa_inf_ineq}
		\beta(\|x\|) \geq \alpha_1(\|x_1\|)+\alpha_2(\|x_2\|), 
		\end{equation}
		for any $x= \mathrm{col}(x_1,x_2)$ with $x_1 \in \mathbb{R}^{n_1}$ and $x_2 \in \mathbb{R}^{n_2}$.
	\end{lemma}
	
	\begin{lemma}[Young's inequality]
		\label{lem:youngineq}
		For any vectors $x,y\in\mathbb{R}^n$ and any scalar $\epsilon\in\mathbb{R}^+$, the following inequality holds:
		\begin{equation}
		x^Ty\leq\dfrac{\epsilon^p}{p}\|x\|^p+\dfrac{1}{q\epsilon^q}\|y\|^q,
		\end{equation}
		with $p>1$ and $q=p/(p-1)$.
	\end{lemma}
	
	\section{Problem Statement}
	\label{Sec:Prob}
	
	
	Consider $n$ interacting agents in an $m$-dimensional space, with $m \in\lbrace 2,3 \rbrace$, governed by: 
	\begin{equation}
	\label{eq:singledyn}
	\dot{q}_i=u_i + \delta_i(t), \quad i=1,\ldots,n
	\end{equation}
	where $q_i \in \mathbb{R}^m$ is the position, $u_i \in \mathbb{R}^m$ is the velocity control input of agent $i$ with respect to a fixed coordinate frame, and $\delta_i(t) \in \mathbb{R}^m$ is an unknown, bounded and piece-wise continuous external disturbance vector.
	Let the desired formation be defined by a minimally and infinitesimally rigid framework $\mathcal{F}^\ast = (\mathcal{G}^\ast,q^\ast)$ where $\mathcal{G}^\ast=(\mathcal{V}^\ast,\mathcal{E}^\ast)$, $\mathrm{dim}(\mathcal{V}^\ast)=n$, $\mathrm{dim}(\mathcal{E}^\ast)=l$, and $q^\ast = \mathrm{col}({q_i^\ast}) \in \mathbb{R}^{mn}$.  Moreover, assume that the actual framework (actual formation) of the agents, which shares the same graph with $\mathcal{F^*}$, is represented by $\mathcal{F}(t)=(\mathcal{G}^\ast , q(t))$, where $q(t)=\mathrm{col}(q_i(t)) \in \mathbb{R}^{mn}$. Let the desired distances between agent $i$ and its neighboring and non-neighboring agents in the rigid framework be given by:
	\begin{equation}
	d_{ij}=\|q_i^\ast - q_j^\ast\| > 0, \quad i,j \in \mathcal{V}^\ast,
	\end{equation} 
	and the relative positions between neighboring agents as:
	\begin{equation}\label{eq:relpo}
	\widetilde{q}_{ij}=q_i - q_j, \quad (i,j)\in \mathcal{E}^\ast.
	\end{equation}
	Thus, for each edge of the rigid graph, the distance error is given by:
	\begin{equation}
	\label{eq:eij}
	e_{ij}=\|\widetilde{q}_{ij}\|-d_{ij}, \quad (i,j)\in \mathcal{E^\ast},
	\end{equation}
	where $\|\widetilde{q}_{ij}\|$ is the actual distance between agents $i$ and $j$. From \eqref{eq:eij} it is clear that $e_{ij} \in [-d_{ij},\infty)$. In what follows, collision avoidance and connectivity maintenance among neighboring agents are formulated with respect to $e_{ij}$.
	
	\textit{\textbf{Collision Avoidance of Neighboring Agents:}}
	The agents should not collide  during their motion towards the desired formation. In general, to cope with this issue, a circular safety region around each agent is assumed and the controller is designed to ensure that the aforementioned safety regions do not overlap during the operation. In this paper, we assume that all agents have spherical shapes. Let $r_{si}\in \mathbb{R}^+$ , $r_{sj}\in \mathbb{R}^+$ be the geometrical radius of the neighboring agents $i$ and $j$, respectively. Furthermore, define $r_{sij}=(r_{si}+r_{sj})>0 ,(i,j)\in \mathcal{E}^\ast$. Obviously, it is necessary that $d_{ij}>r_{sij}$, otherwise, the formation is not feasible. Thus, in order to ensure collision avoidance between neighboring agents, it is required that $\|\widetilde{q}_{ij}(t)\|>r_{sij}$ for all $t\geq0$, which can be restated in terms of the distance errors as:
	\begin{equation} \label{eq:coll_condit}
	r_{sij}-d_{ij}<e_{ij}(t), \quad (i,j)\in \mathcal{E}^\ast, \, \forall t\geq0,
	\end{equation}
	where $r_{sij}-d_{ij}$ is a negative value.
	
\begin{figure}[tbp]
		\centering
		\includegraphics[scale=0.32]{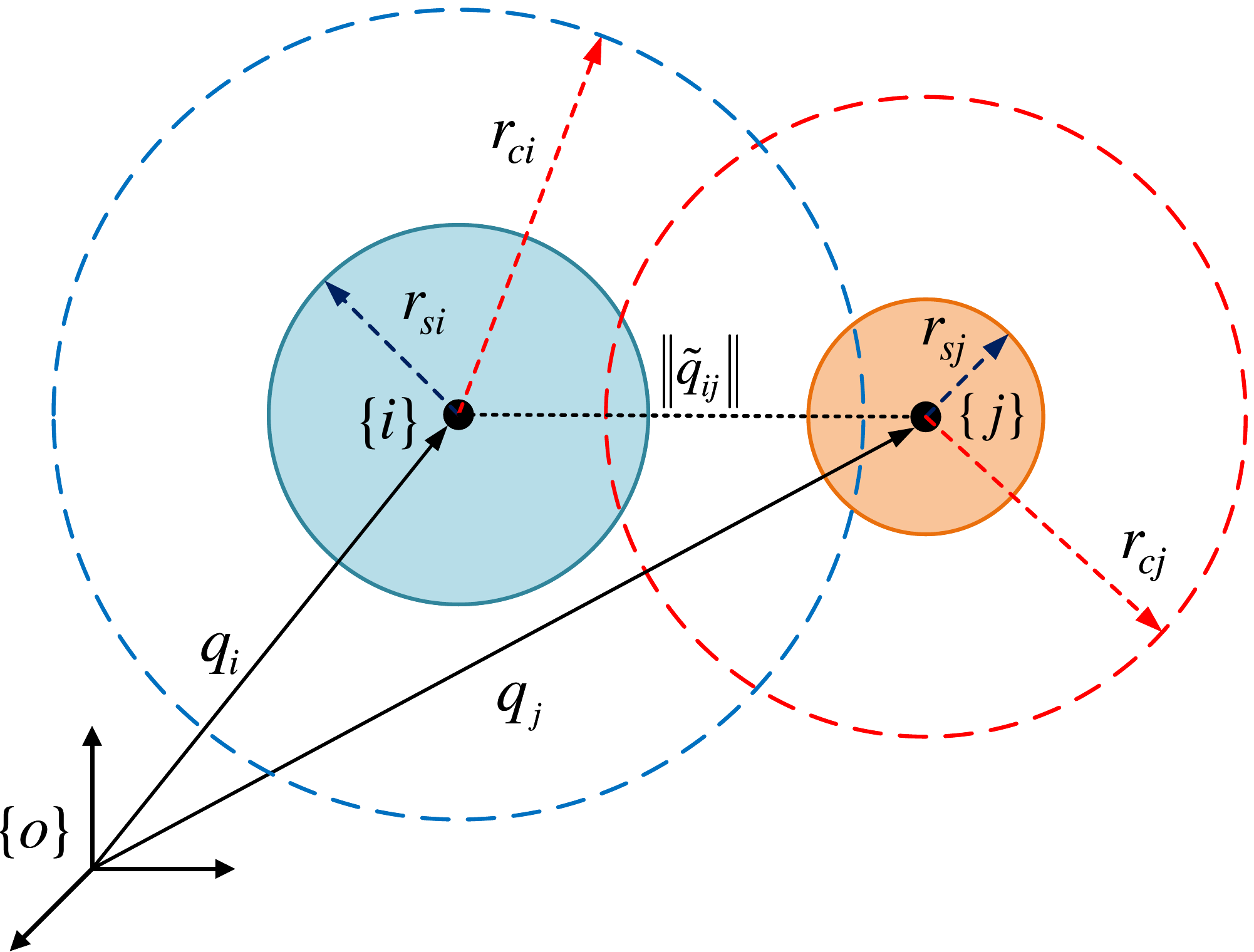}
		\caption{The sensing and safety regions of two neighboring agents.} 
		\label{pic:radius}
	\end{figure}

	\textit{\textbf{Connectivity Maintenance:}}
	In practice, since each agent has a limited sensing capability it is also necessary for neighboring agents to remain within their common sensing area during the operation, otherwise, the whole system might become unstable or inactive. Hence, it is important to design the formation controller in a way that ensures connectivity of neighboring agents. In this paper, connectivity maintenance is equivalent to not losing any edge in the actual undirected rigid formation graph during the operation. Let $r_{ci}\in \mathbb{R}^+$ and $r_{cj} \in \mathbb{R}^+$ be the sensing radius of agents $i$ and $j$, respectively. Furthermore, let us define $r_{cij}=\mathrm{min}\lbrace r_{ci}+r_{sj}\, ,\,r_{cj}+r_{si}\rbrace, (i,j)\in \mathcal{E}^\ast$. Notice that $d_{ij}$ should be less than $r_{cij}$ so that we seek for a feasible target formation. Therefore, securing $\|\widetilde{q}_{ij}(t)\|<r_{cij}$ for all $t\geq0$ is sufficient to ensure connectivity maintenance, which can be formulated in terms of the distance errors as:
	\begin{equation} \label{eq:con_condit}
	e_{ij}(t)<r_{cij}-d_{ij}, \quad (i,j)\in \mathcal{E}^\ast, \, \forall t\geq0.
	\end{equation}
	Apparently, \eqref{eq:con_condit} indicates that the initial graph should be connected at $t=0$ as well. Notice that, always $r_{ci}>r_{si}, \forall i \in \mathcal{V}^\ast$ holds. Moreover, \eqref{eq:con_condit} and \eqref{eq:coll_condit} indicate that  $r_{cij}>r_{sij}, (i,j)\in \mathcal{E}^\ast$. Figure \ref{pic:radius} depicts two neighboring agents.
	
	\textit{\textbf{Control Objective:}} Under the assumption that initially the actual formation graph is minimally and infinitesimally rigid, which is common in distance-based formation control literature \cite{cai2015formation,sun2015finite, sun2016exponential, rozenheck2015proportional}, the objective is to design a decentralized robust control protocol such that:
	\begin{subequations}\label{eq:goal}
		\begin{align}
		\mathcal{F}(t) &\rightarrow \mathrm{Iso}(\mathcal{F}^\ast) \quad \mathrm{as} \quad  t\rightarrow \infty \label{eq:objec} \\ 
		-\underline{e}_{ij}(t) < &e_{ij}(t) < \overline{e}_{ij}(t), \quad (i,j)\in \mathcal{E}^\ast, \, \forall t\geq0, \label{eq:e_bound}
		\end{align} 
	\end{subequations}
	where $\underline{e}_{ij}(t)>0$ and $\overline{e}_{ij}(t)>0$ are decreasing performance bounds satisfying $\lim_{t \rightarrow \infty} \underline{e}_{ij}(t)>0$ and $\lim_{t \rightarrow \infty} \overline{e}_{ij}(t)>0$. Notice further that if the performance bounds are selected appropriately to satisfy $\overline{e}_{ij}(0) \leq r_{cij}-d_{ij}$ and $\underline{e}_{ij}(0) \leq d_{ij}-r_{sij}$, then connectivity maintenance and collision avoidance among neighboring agents will be ensured for all $t\geq0$. 
	
	\begin{remark}
		The aforementioned robust distance-based formation acquisition problem is defined for agents modeled by \eqref{eq:singledyn}. In this paper, distance-based formation centroid maneuvering will also be studied in Section \ref{Centroid_Manue} for nominal single integrator agents.
	\end{remark}
	
	\section{Controller Design and Stability Analysis}
	\label{designand analyis}
	
	The following lemma provides a sufficient condition to establish infinitesimal rigidity of the actual formation $\mathcal{F}(t)$ based on the distance error bounds \eqref{eq:e_bound}.
	\begin{lemma} \label{lem:infi_my}
		If $\mathcal{F}^\ast$ is infinitesimally rigid and $\bar{\vartheta}$ is a sufficiently small positive constant satisfying:
		\begin{equation} \label{eq:psi_bar}
		\overline{\Psi} ({\mathcal{F}},\mathcal{F^\ast}) \triangleq \sum_{(i,j)\in \mathcal{E}^\ast} \mathrm{max}\{|\underline{e}_{ij}(0)| , |\overline{e}_{ij}(0)|\} \leq \bar{\vartheta},
		\end{equation}
		then securing \eqref{eq:e_bound} guarantees that $\mathcal{F}$ is also infinitesimally rigid for all time.
	\end{lemma}
	\begin{pf}
		Using \eqref{eq:relpo} and \eqref{eq:eij}, \eqref{eq:Psi} can be represented as:
		\begin{equation}
		\label{eq:psi_prof}
		\Psi ({\mathcal{F}},\mathcal{F^\ast})=\sum_{(i,j)\in \mathcal{E}^\ast} (\|\widetilde{q}_{ij}\|-d_{ij})^2 = \sum_{(i,j)\in \mathcal{E}^\ast} e_{ij}^2 \leq \vartheta.
		\end{equation}
		for a small positive constant $\vartheta$. The following inequality is a sufficient condition for \eqref{eq:psi_prof}:
		\begin{equation}
		\label{eq:e_bounded}
		\sum_{(i,j)\in \mathcal{E}^\ast} |{e}_{ij}| < \bar{\vartheta}
		\end{equation}
		where $\bar{\vartheta}$ is a sufficiently small positive constant such that $\bar{\vartheta}^ {\, 2} \leq \vartheta$. Moreover, if we select $\underline{e}_{ij}(0), \overline{e}_{ij}(0)$ and guarantee \eqref{eq:e_bound} for decreasing performance functions $\underline{e}_{ij}(t), \overline{e}_{ij}(t)$ we arrive at: 
		\begin{equation}
		\label{eq:e_bounded2}
		\sum_{(i,j)\in \mathcal{E}^\ast} |{e}_{ij}(t)| < \sum_{(i,j)\in \mathcal{E}^\ast} \mathrm{max}\{|\underline{e}_{ij}(0)| , |\overline{e}_{ij}(0)|\}, \forall t\geq0
		\end{equation}
		Clearly, establishing \eqref{eq:psi_bar} is sufficient for \eqref{eq:e_bounded2}, and thus it is sufficient for \eqref{eq:e_bounded} and \eqref{eq:psi_prof} as well. Finally, Lemma \ref{co:infini_col} yields that $\mathcal{F}$ is infinitesimally rigid, which concludes the proof.
	\end{pf}
	
	For an infinitesimally rigid target formation $\mathcal{F^*}$, Lemma \ref{lem:infi_my} indicates that if $\overline{e}_{ij}(0)$ and $\underline{e}_{ij}(0)$ are properly selected and  \eqref{eq:e_bound} is satisfied, then the infinitesimal rigidity of the actual formation graph $\mathcal{F}(t)$ is ensured for all time. Note that, the assumption on  infinitesimal rigidity of the actual rigid graph at $t=0$ can be demonstrated by $e(0) \in \Omega_I$ where
	\begin{equation} \label{eq:Omega_I}
	\Omega_I=\left\lbrace e\in \mathbb{R}^l \mid \overline{\Psi}(\mathcal{F},\mathcal{F}^\ast)\leq\bar{\vartheta} \right\rbrace,
	\end{equation}
	and $e=\mathrm{col}(e_{ij}) \in \mathbb{R}^l$ for the same ordering as in the edge function \eqref{eq:edgefun}. Moreover, since rigid graphs are connected, the aforementioned assumption also secures the initial connectivity specification of neighboring agents.
	
	\begin{remark}
		Notice that $\mathcal{F^\ast}$ can be replaced by $\mathrm{Iso}(\mathcal{F^\ast})$ (or even $\mathrm{Amb}(\mathcal{F^\ast})$) in Lemma \ref{lem:infi_my} since $\mathrm{Iso}(\mathcal{F^\ast})$ ($\mathrm{Amb}(\mathcal{F^\ast})$) is infinitesimally rigid and has the same edges lengths as $\mathcal{F^\ast}$.
	\end{remark}
	
	\subsection{Proper Selection of $\overline{e}_{ij}(0)$ and $\underline{e}_{ij}(0)$}
	\label{sub:algo}
	
	As mentioned earlier, since $\overline{e}_{ij}(t)$ and $\underline{e}_{ij}(t)$ are assumed to be decreasing, by selecting $\overline{e}_{ij}(0)$ and $\underline{e}_{ij}(0)$ appropriately, one can obtain the distance error bounds which ensure infinitesimal rigidity of the actual formation as well as connectivity maintenance and collision avoidance among neighboring agents. Note that, when $e(0) \in \Omega_I$ and $|e_{ij}(t)|<|e_{ij}(0)|+\mu$, where $\mu$ is a sufficiently small positive constant, then from Lemma \ref{lem:infi_my} it is reasonable to set $\underline{e}_{ij}(0)=\overline{e}_{ij}(0)=|e_{ij}(0)|+\mu$ as the maximum allowable bounds for $e_{ij}$ in order to guarantee infinitesimal rigidity of the actual formation graph. It is clear that increasing $\mu$ reduces the conservativeness in choosing the bounds of $e_{ij}$ that ensure this property. Notice that since $e_{ij} \in [-d_{ij},\infty)$, if $e_{ij}(0)>0$ and $-(|e_{ij}(0)|+\mu)<-d_{ij}$ then we should take $\underline{e}_{ij}(0)=d_{ij}$. We also define the \textit{formation robustness constants} $\overline{\mu}_{ij}$ and $\underline{\mu}_{ij}$ with respect to each edge of the rigid graph where $0<\overline{\mu}_{ij} \leq r_{cij}-d_{ij}$ and $0<\underline{\mu}_{ij} \leq  d_{ij} - r_{sij}, (i,j)\in \mathcal{E}^\ast$. Adjusting these constants is useful for improving robustness against formation distortion caused by external disturbances, which will be discussed later.
	
	We propose Algorithm \ref{algo} for selecting the allowable upper and lower bounds of $e_{ij}$ in \eqref{eq:e_bound}  ($\overline{e}_{ij}(0)$ and $\underline{e}_{ij}(0)$) to ensure infinitesimal rigidity, connectivity maintenance and collision avoidance among neighboring agents. Algorithm \ref{algo} is applied in a distributed fashion and each agent calculates its corresponding edges' maximum allowable distance error bounds. Agents should sense the initial distance errors $e_{ij}(0)$ with their neighbors and use the sensing and the geometric radius. Note that, when the sensing and geometric radius are not identical for all agents, then agent $i$ may receive scalars $r_{cj}$ and $r_{sj}$ through communication with its neighbors at $t=0$.
	\begin{algorithm}[htb]
		\caption{Distributed selection of $\overline{e}_{ij}(0)$ and $\underline{e}_{ij}(0)$, $j \in \mathcal{N}_i(\mathcal{E}^\ast)$ for $i$-th agent at $t=0$.}
		\begin{algorithmic}[1] \label{algo}
			\renewcommand{\algorithmicrequire}{\textbf{Input:}}
			\renewcommand{\algorithmicensure}{\textbf{Output:}}
			\REQUIRE $e_{ij}(0)$, $d_{ij}$, $r_{ci}$, $r_{cj}$, $r_{si}$, $r_{sj}$, $\mu$ 
			\STATE 
			$r_{sij}=(r_{si}+r_{sj})$ and $r_{cij}=\mathrm{min}\lbrace r_{ci}\, ,\,r_{cj}\rbrace$
			\STATE \textbf{Choose} \; $\overline{\mu}_{ij}$ and $\underline{\mu}_{ij}$ such that $0<\overline{\mu}_{ij} \leq r_{cij}-d_{ij}$ and $0<\underline{\mu}_{ij} \leq  d_{ij} - r_{sij}$
			\IF {($e_{ij}(0) \geq 0$)}
			\STATE 
			$\overline{e}_{ij}(0)=\mathrm{min}\lbrace |e_{ij}(0)|+\mu \, , \, r_{cij}-d_{ij} \rbrace$
			\STATE
			$\underline{e}_{ij}(0)=\mathrm{min}\lbrace |e_{ij}(0)|+\mu \, , \, \underline{\mu}_{ij} \rbrace$
			\ELSIF{($e_{ij}(0) < 0$)}
			\STATE
			$\overline{e}_{ij}(0)=\mathrm{min}\lbrace |e_{ij}(0)|+\mu \, , \, \overline{\mu}_{ij} \rbrace$
			\STATE
			$\underline{e}_{ij}(0)=\mathrm{min}\lbrace |e_{ij}(0)|+\mu \, , \, d_{ij}-r_{sij} \rbrace$
			\ENDIF
			\RETURN $\overline{e}_{ij}(0)$, $\underline{e}_{ij}(0)$
		\end{algorithmic}
	\end{algorithm} 
	\subsection{Prescribed Performance Bounds}
	\label{sub:ppb_con}
	
	Define the following squared distance error
	\begin{equation}
	\label{eq:etaij}
	\eta_{ij}=\|\widetilde{q}_{ij}\|^2 - d_{ij}^2, \quad (i,j)\in \mathcal{E}^\ast,
	\end{equation}
	then it follows from \eqref{eq:eij} that:
	\begin{equation}
	\label{eq:eta_and_e}
	\eta_{ij}=e_{ij}\left(\|\widetilde{q}_{ij}\|+d_{ij}\right).
	\end{equation} 
	Since $e_{ij} \in [-d_{ij},\infty)$, from \eqref{eq:eta_and_e}, it is easy to see $\eta_{ij}=0$ if and only if $e_{ij}=0$. Therefore, instead of using $e_{ij}$ directly, we will use the squared distance error \eqref{eq:etaij} to impose specific prescribed performance bounds for all times. Similar to the results developed in \cite{bechlioulis2008robust,wang2010adaptive,karayiannidis2012multi,macellari2017multi}, for each edge of the rigid framework, first a smooth decreasing performance function $\rho_{ij}(t): [0,\infty) \rightarrow \mathbb{R}^+ , (i,j)\in \mathcal{E}^\ast$ with $\lim_{t \rightarrow \infty} \rho_{ij}(t)=\rho_{ij\infty}>0$ is chosen. In this paper, we adopt the following exponential performance function:
	\begin{equation}
	\label{eq:rho_ij}
	\rho_{ij}(t)=(\rho_{ij0}-\rho_{ij\infty}) \exp(-a_{ij}t)+ \rho_{ij\infty},  (i,j)\in \mathcal{E}^\ast,
	\end{equation}
	where $\rho_{ij0}>\rho_{ij\infty}$ and $a_{ij}>0$. The objective of guaranteeing transient performance can be achieved if the following condition is always satisfied
	\begin{equation}
	\label{eq:eta_bound}
	-\underline{\eta}_{ij}(t) < \eta_{ij}(t) < \overline{\eta}_{ij}(t), \quad (i,j)\in \mathcal{E}^\ast,
	\end{equation}
	where
	\begin{subequations}\label{eq:ppb}
		\begin{align}
		&\overline{\eta}_{ij}(t)=\overline{b}_{ij} \rho_{ij}(t), \label{eq:eta_up}	\\
		&\underline{\eta}_{ij}(t)=\underline{b}_{ij} \rho_{ij}(t),
		\label{eq:eta_dow}
		\end{align}	
	\end{subequations}
	and $\overline{b}_{ij} , \underline{b}_{ij}>0$ are positive scalars. It is clear that $\overline{\eta}_{ij}(0)=\overline{b}_{ij} \rho_{ij}(0)$ and $\underline{\eta}_{ij}(0)=\underline{b}_{ij} \rho_{ij}(0)$ are the maximum overshoot and the maximum undershoot (i.e. negative overshoot) of $\eta_{ij}(t)$, respectively. Furthermore, the decreasing rate of $\rho_{ij}(t)$ introduces a lower bound on the convergence speed of $\eta_{ij}(t)$. Moreover, based on \eqref{eq:eta_and_e}, $\overline{e}_{ij}(0)$ and $\underline{e}_{ij}(0)$ can be also related to $\overline{\eta}_{ij}(0)=\overline{b}_{ij} \rho_{ij0}$ and $\underline{\eta}_{ij}(0)=\underline{b}_{ij} \rho_{ij0}$ as:
	\begin{subequations}\label{eq:bebeta}
		\begin{align} 
		&\overline{e}_{ij}^2(0)+2d_{ij}\overline{e}_{ij}(0)=\overline{b}_{ij} \rho_{ij0}, \label{eq:ove_ovet} \\
		&2d_{ij}\underline{e}_{ij}(0)-\underline{e}_{ij}^2(0)=\underline{b}_{ij} \rho_{ij0}. \label{eq:loe_loet}
		\end{align} 
	\end{subequations}
	Consequently, if $\overline{e}_{ij}(0)$ and $\underline{e}_{ij}(0)$ are given from Algorithm\ref{algo}, then one can always assume $\rho_{ij0}=1$ and find $\overline{b}_{ij}$ and $\underline{b}_{ij}$ to encapsulate the requirements of infinitesimal rigidity, connectivity maintenance and collision avoidance between neighboring agents within the performance bounds defined in \eqref{eq:eta_bound}. 
	
	\begin{remark} \label{rem:eta&e}
		Note that, by using \eqref{eq:eta_and_e} one can write $\overline{\eta}_{ij}=\overline{e}_{ij}(\overline{e}_{ij}+2d_{ij})$ and $-\underline{\eta}_{ij}=-\underline{e}_{ij}(-\underline{e}_{ij}+2d_{ij})$. Since $\overline{e}_{ij}(t)>0$ and $0< \underline{e}_{ij}(t)<d_{ij}$, then we can obtain the following relations:
		\begin{subequations}\label{ppb_e}
			\begin{align}
			\overline{e}_{ij}(t)&=-d_{ij}+\sqrt{d_{ij}^2+\overline{b}_{ij} \rho_{ij}(t)}, \label{eq:e_pres_e_over} \\
			\underline{e}_{ij}(t)&=d_{ij}-\sqrt{d_{ij}^2-\underline{b}_{ij} \rho_{ij}(t)}. \label{eq:e_pres_e_under}
			\end{align} 
		\end{subequations}
		between performance bounds of $\eta_{ij}(t)$ and $e_{ij}(t)$. Therefore, satisfaction of \eqref{eq:eta_bound}, fulfills the condition in \eqref{eq:e_bound} with the aforementioned decreasing prescribed performance bounds in \eqref{ppb_e}.
	\end{remark}
	\subsection{Transformed Error System}
	\label{sub:trans_err}
	
	The distance error dynamics corresponding to each edge of the rigid framework can be obtained invoking \eqref{eq:singledyn} and \eqref{eq:eij} as follows:
	\begin{equation}
	\label{eq:eij_dot}
	\dot{e}_{ij}=\dfrac{d}{dt}\left(\sqrt{\widetilde{q}_{ij}^T\widetilde{q}_{ij}} \, \right) = \dfrac{\widetilde{q}_{ij}^T (u_i- u_j +\delta_i-\delta_j)}{e_{ij}+d_{ij}}, \; (i,j)\in \mathcal{E^\ast}.
	\end{equation}
	Using \eqref{eq:eij_dot} and \eqref{eq:eta_and_e}, the time derivative of $\eta_{ij}$ is given by
	\begin{equation}
	\label{eq:etaij_dot}
	\dot{\eta}_{ij}=2e_{ij}\dot{e}_{ij}+2\dot{e}_{ij}d_{ij} 
	=2\widetilde{q}_{ij}^T(u_i - u_j +\delta_i -\delta_j), \; (i,j)\in \mathcal{E^\ast}.
	\end{equation}
	Based on the structure of rigidity matrix in \eqref{eq:rigid_stru}, \eqref{eq:etaij_dot} can be written in compact form as:
	\begin{equation}
	\label{eq:eta_dot}
	\dot{\eta}=2R(q)(u+\delta),
	\end{equation}
	where $u=\mathrm{col}(u_i)\in \mathbb{R}^{mn}$, $\delta=\mathrm{col}(\delta_i)\in \mathbb{R}^{mn}$, and $\eta=\mathrm{col}(\eta_{ij}) \in \mathbb{R}^l$, $(i,j)\in \mathcal{E}^\ast$ for the same ordering as in \eqref{eq:edgefun}. 
	
	The problem of designing a controller that meets error constraints in \eqref{eq:eta_bound} can be simply transformed into establishing the boundedness of certain error signals via the prescribed performance control methodology \cite{bechlioulis2008robust}. Specifically, to handle the time-varying constraints in \eqref{eq:eta_bound}, an error transformation technique is used to convert the original dynamical error system \eqref{eq:etaij_dot} with constraints \eqref{eq:eta_bound} into a new equivalent unconstrained one, whose stability ensures satisfaction of the restrictions given in \eqref{eq:eta_bound}. First, consider the following modulated squared distance errors:
	\begin{equation}
	\label{eq:modu}
	\widehat{\eta}_{ij}(t)\triangleq \dfrac{\eta_{ij}(t)}{\rho_{ij}(t)}, \quad (i,j)\in \mathcal{E}^\ast.
	\end{equation}
	To transform the constrained error system (in the sense of \eqref{eq:eta_bound}) into an equivalent unconstrained one, we introduce the following error transformation \cite{wang2010adaptive} corresponding to each edge in the rigid framework:
	\begin{equation}
	\label{eq:sigmaij}
	\sigma_{ij}=T_{ij}(\widehat{\eta}_{ij})=\dfrac{1}{2} \ln \left( \dfrac{\overline{b}_{ij} \widehat{\eta}_{ij} + \overline{b}_{ij} \underline{b}_{ij}}{\overline{b}_{ij} \underline{b}_{ij} - \underline{b}_{ij} \widehat{\eta}_{ij}} \right), (i,j)\in \mathcal{E}^\ast,
	\end{equation}
	where $\sigma_{ij}$ is the transformed error corresponding to $\eta_{ij}$ and $T_{ij}(.) : (- \underline{b}_{ij} , \overline{b}_{ij}) \rightarrow (- \infty , + \infty)$ is a smooth, strictly increasing bijective mapping satisfying $T_{ij}(0)=0$. It is clear that $\eta_{ij} \rightarrow 0$ if and only if $\sigma_{ij} \rightarrow 0$. Finally, notice that maintaining the boundedness of $\sigma_{ij} \in \mathbb{R}$, enforces $-\underline{b}_{ij}<\widehat{\eta}_{ij}(t)<\overline{b}_{ij}$ or equivalently \eqref{eq:eta_bound}.
	
	Now, taking the time derivative of $\sigma_{ij}$, yields:
	\begin{align}
	\label{eq:sigmaij_dot}
	\dot{\sigma}_{ij} &= \dfrac{d T_{ij}}{d \widehat{\eta}_{ij}}\dot{\widehat{\eta}}_{ij} \nonumber \\
	&=\dfrac{1}{2} \left[ \dfrac{1}{\widehat{\eta}_{ij} + \underline{b}_{ij}} - \dfrac{1}{\widehat{\eta}_{ij} - \overline{b}_{ij}} \right] \left(\dfrac{\dot{\eta}_{ij}}{\rho_{ij}}-\dfrac{\eta_{ij} \dot{\rho}_{ij}}{\rho_{ij}^2}\right) \nonumber \\
	&= \dfrac{1}{2} \xi_{ij} (\dot{\eta}_{ij}-\widehat{\eta}_{ij} \dot{\rho}_{ij}), 
	\end{align}
	where
	\begin{equation}
	\label{eq:xi_ij}
	\xi_{ij}\triangleq \dfrac{1}{\rho_{ij}} \left[\dfrac{1}{\widehat{\eta}_{ij} + \underline{b}_{ij}} - \dfrac{1}{\widehat{\eta}_{ij} - \overline{b}_{ij}}\right], \quad (i,j)\in \mathcal{E^\ast}.
	\end{equation}
	Then \eqref{eq:sigmaij_dot} is given in compact form as:
	\begin{equation}
	\label{eq:sigma_dot}
	\dot{\sigma}=  \dfrac{1}{2} \xi (\dot{\eta}-\dot{\rho}\widehat{\eta}),
	\end{equation}
	where $\sigma=\mathrm{col}(\sigma_{ij}) \in \mathbb{R}^l$, $\xi=\mathrm{diag}(\xi_{ij}) \in \mathbb{R}^{l\times l}$, $\dot{\rho}=\mathrm{diag}(\dot{\rho}_{ij}) \in \mathbb{R}^{l\times l}$ and $\widehat{\eta}=\mathrm{col}(\widehat{\eta}_{ij}) \in \mathbb{R}^l$,  for $(i,j)\in \mathcal{E}^\ast$ and the same ordering as in \eqref{eq:edgefun}. In the next subsection $\sigma$ is employed in the design of a distance-based formation controller that stabilizes the transformed error dynamics \eqref{eq:sigma_dot}, thus satisfying the error transformation in \eqref{eq:eta_bound} for all time.
	
	\begin{remark}[\textbf{PPC Design Philosophy}] 
		\noindent \\ When $-\underline{\eta}_{ij}(0) < \eta_{ij}(0) < \overline{\eta}_{ij}(0)$, based on the properties of the error transformation \eqref{eq:sigmaij}, prescribed performance in the sense of \eqref{eq:eta_bound} is achieved, if  $\sigma_{ij}, (i,j)\in \mathcal{E^\ast}$ is kept bounded. Notice that, although for $\sigma_{ij} \in \mathbb{R}$ the Prescribed Performance Bounds (PPB) in \eqref{eq:eta_bound} are satisfied, the boundedness of $\sigma_{ij}$ is required to gurantee internal stability of the closed-loop system (bounded control inputs). Moreover, it is important to note that the specific bounds of $\sigma_{ij}(t)$ (no matter how large they are, which is the key property of the adopted error transformation) do not affect the achieved transient and steady state performance on $\eta_{ij}(t)$, which is solely determined by \eqref{eq:eta_bound} and thus by the selection of the performance functions $\rho_{ij}(t)$ as well as $\overline{b}_{ij}$ and $\underline{b}_{ij}, (i,j)\in\mathcal{E^\ast}$.
	\end{remark}
	\subsection{Main Result}
	\label{singleinteg_main}
	
	The following theorem summarize the main results of this section.
	\begin{theorem} \label{th:single}
		Consider $n$ agents with dynamics \eqref{eq:singledyn} in an $m$-dimensional space ($m \in \{2,3\}$) with the actual formation $\mathcal{F}(t)=(\mathcal{G}^\ast , q(t))$. Let the initial conditions be such that $e(0) \in \Omega_I$, which implies infinitesimal rigidity of the actual rigid graph at $t=0$. If the desired formation $\mathcal{F}^\ast$ is minimally and infinitesimally rigid and also $\overline{e}_{ij}(0)$, $\underline{e}_{ij}(0)$ are selected according to Algorithm \ref{algo}, then the following control law:
		\begin{equation}
		\label{eq:u_single}
		u= - R^T \xi K \sigma,
		\end{equation}
		where $K=\mathrm{diag}(k_{ij}) > 0$, $R$ is the rigidity matrix, $\sigma=\mathrm{col}(\sigma_{ij})$, and $\xi=\mathrm{diag}(\xi_{ij})$ with $\xi_{ij}, \sigma_{ij}$ defined in \eqref{eq:sigmaij} and \eqref{eq:xi_ij}, respectively, ensures prescribed performance in the sense of \eqref{eq:eta_bound} and leads to connectivity maintenance and collision avoidance among neighboring agents.
	\end{theorem}
	\begin{pf}
		The proof of Theorem \ref{th:single} proceeds in three phases. First, we show that all $\widehat{\eta}_{ij}(t)$ remain within $(-\underline{b}_{ij}, \overline{b}_{ij})$ for a specific time interval $[0, \tau_{\mathrm{max}})$ (i.e., the existence and uniqueness of a maximal solution). Next, we prove that the proposed control scheme guarantees, for all $[0, \tau_{\mathrm{max}})$: a) the boundedness of all closed loop signals as well as b) that $\widehat{\eta}_{ij}(t)$ remains strictly in a compact subset of $(-\underline{b}_{ij}, \overline{b}_{ij})$, which by contradiction leads to
		$\tau_{\mathrm{max}}=\infty$ (i.e., forward completeness) in the last phase, thus completing the proof. 
		
		In particular, the compact form of the modulated errors in \eqref{eq:modu} is given by $\widehat{\eta}=\rho(t)^{-1} \eta$ where $\rho(t)^{-1}=\mathrm{diag}({\rho}_{ij}(t)^{-1}) \in \mathbb{R}^{l\times l}$. Differentiating $\widehat{\eta}$ as well as employing \eqref{eq:eta_dot} and \eqref{eq:u_single}, yields:
		\begin{align} 
		\dot{\widehat{\eta}}= h_{\widehat{\eta}}(t,\widehat{\eta})&=\rho(t)^{-1} (\dot{\eta}-\dot{\rho}(t) \widehat{\eta}) \nonumber \\
		&= \rho(t)^{-1} (-2RR^T\xi K \sigma(\widehat{\eta})+2R\delta-\dot{\rho}(t) \widehat{\eta}). \label{eq:etahat_dyn}
		\end{align}
		Let us also define the open set $\Omega_{\widehat{\eta}}$ as: 
		\begin{equation}
		\Omega_{\widehat{\eta}}= \prod_{(i,j)\in \mathcal{E}^\ast}(-\underline{b}_{ij}, \overline{b}_{ij}).
		\end{equation}
		
		\textit{Phase I.} Algorithm \ref{algo} and \eqref{eq:bebeta} define $\underline{b}_{ij}$ and $\overline{b}_{ij}$, which guarantee that the set $\Omega_{\widehat{\eta}}$ is nonempty and open. Moreover, Algorithm \ref{algo}, ensures that $\widehat{\eta}(0) \in \Omega_{\widehat{\eta}}$. Additionally, $h_{\widehat{\eta}}$ is continuous on $t$ and locally Lipschitz on $\widehat{\eta}$ over the set $\Omega_{\widehat{\eta}}$. Therefore, the hypotheses of Theorem 54 in \cite[p.~476]{sontag1998mathematical} hold and the existence and uniqueness of a maximal solution $\widehat{\eta}(t)$ of \eqref{eq:etahat_dyn} for a time interval $[0, \tau_{\mathrm{max}})$ such that $\widehat{\eta}(t) \in \Omega_{\widehat{\eta}}, \forall t \in [0, \tau_{\mathrm{max}})$ is guaranteed. Equivalently, we infer that:
		\begin{equation} \label{eq:eta_hat_satisfied}
		\widehat{\eta}_{ij}(t) \in (-\underline{b}_{ij}, \overline{b}_{ij}),\; (i,j)\in \mathcal{E}^\ast,\; \forall t \in [0, \tau_{\mathrm{max}}),
		\end{equation}
		from which we obtain that $\eta_{ij}(t)$ are absolutely bounded as in \eqref{eq:eta_bound} for all $t \in [0, \tau_{\mathrm{max}})$. Accordingly, from Remark \ref{rem:eta&e}, it can be deduced that: 
		\begin{equation} \label{eq:e_bound_maximal}
		-\underline{e}_{ij}(t) < e_{ij}(t) < \overline{e}_{ij}(t), \; (i,j)\in \mathcal{E}^\ast, \, \forall t \in [0, \tau_{\mathrm{max}}). 
		\end{equation}
		Recall that $\underline{e}_{ij}(t)$ and $\overline{e}_{ij}(t)$ are decreasing functions of time. As a result, \eqref{eq:e_bound_maximal} ensures: a) connectivity maintenance and collision avoidance of neighboring agents in the actual formation graph $\mathcal{F}(t)$ as well as b) infinitesimal rigidity of $\mathcal{F}(t)$ (due to $e(0) \in \Omega_I$), for all $t \in [0, \tau_{\mathrm{max}})$.
		
		\textit{Phase II.} From \textit{Phase I} we know that infinitesimal rigidity and connectivity of the neighboring agents is guaranteed for all $t \in [0, \tau_{\mathrm{max}})$. Note that, $\mathcal{F}^\ast$ and $\mathcal{F}(t)$ have the same number of edges, whenever connectivity maintenance is ensured for $\mathcal{F}(t)$. Thus, since  $\mathcal{F}^\ast$ is minimally rigid, $\mathcal{F}(t)$ is also minimally rigid for all $t \in [0, \tau_{\mathrm{max}})$. Consequently, it can be deduced that $\mathcal{F}(t)$ is \textit{minimally and infinitesimally rigid} for all $t \in [0, \tau_{\mathrm{max}})$. Furthermore, owing to \eqref{eq:eta_hat_satisfied}, the error signals $\sigma_{ij}$ in \eqref{eq:sigmaij} are well defined. Therefore, consider the following potential function:
		\begin{equation}
		\label{eq:vij}
		V_{ij}=\dfrac{k_{ij}}{2} \sigma_{ij}^2, \quad (i,j)\in \mathcal{E}^\ast,
		\end{equation}
		and define the overall candidate Lyapunov function:
		\begin{equation}
		\label{eq:lyap1}
		V=\sum_{(i,j)\in \mathcal{E}^\ast} V_{ij}(\sigma_{ij}) = \dfrac{1}{2} \sigma^T K \sigma.
		\end{equation}
		Employing \eqref{eq:eta_dot} and \eqref{eq:sigma_dot}, we arrive at:
		\begin{equation}
		\label{eq:lyap1_dot}
		\dot{V}= \sigma^T K \xi R u + \sigma^T K \xi R \delta -\dfrac{1}{2} \sigma^T K \xi \dot{\rho} \widehat{\eta}.
		\end{equation}
		By Lemma \ref{lem:youngineq} (Young's inequality) we get:
		\begin{equation}
		\label{eq:distur_young}
		\sigma^T K \xi R d \leq \dfrac{1}{2}(\sigma^T K \xi R R^T \xi K \sigma)+ \dfrac{1}{2} \|\delta\|^2.
		\end{equation}
		Substituting \eqref{eq:u_single} into \eqref{eq:lyap1_dot} and using \eqref{eq:distur_young} and Lemma \ref{lem:PD_rigid}, we arrive at:
		\begin{align}
		\dot{V} &\leq - \dfrac{1}{2}(\sigma^T K \xi R R^T \xi K \sigma) + \dfrac{1}{2} \|\delta\|^2 -\dfrac{1}{2} \sigma^T K \xi \dot{\rho} \widehat{\eta} \nonumber \\
		&\leq -\dfrac{\lambda_{\mathrm{min}}(RR^T)}{2}  \sigma^T K \xi \xi K \sigma + \dfrac{1}{2} \|\delta\|^2 -\dfrac{1}{2} \sigma^T K \xi \dot{\rho} \widehat{\eta}.& \label{eq:lyap1_dotcalc}
		\end{align}
		Let $\epsilon$ be an arbitrarily small positive constant satisfying  $\lambda_{\mathrm{min}}(RR^T)>\epsilon^2/2$. Invoking Young's inequality on $-(1/2) \sigma^T K \xi \dot{\rho} \widehat{\eta}$ and exploiting the diagonality of $K$, $\xi$, $\rho$ matrices we obtain:
		\begin{align}
		\label{eq:lyap1_dot_young}
		\dot{V} \leq &-\left[\dfrac{\lambda_{\mathrm{min}}(RR^T)}{2}-\dfrac{\epsilon^2}{4}\right] \sigma^T \! K \xi \xi K \sigma \!+\! \dfrac{1}{2} \|\delta\|^2 \! + \! \dfrac{1}{4\epsilon^2} \widehat{\eta}^T \dot{\rho}^2 \widehat{\eta} \nonumber \\
		\leq &-\left[\dfrac{\lambda_{\mathrm{min}}(RR^T)}{2}-\dfrac{\epsilon^2}{4}\right] \lambda_{\mathrm{min}}(\xi^2) \lambda_{\mathrm{min}}(K^2) \|\sigma\|^2 + \dfrac{1}{2} \|\delta\|^2 \nonumber \\
		&+\dfrac{1}{4\epsilon^2} \lambda_{\mathrm{max}}(\dot{\rho}^2) \|\widehat{\eta}\|^2, \quad \forall t \in [0, \tau_{\mathrm{max}}),
		\end{align}
		where $\|\delta\|^2$, $\lambda_{\mathrm{max}}(\dot{\rho}^2)$, and $\|\widehat{\eta}\|^2$ are bounded. Moreover, by \eqref{eq:eta_hat_satisfied}, we get:
		\begin{equation}
		\label{eq:norm_etahat}
		\|\widehat{\eta}\|^2=\sum_{(i,j)\in \mathcal{E}^\ast} \widehat{\eta}_{ij}^2 < \sum_{(i,j)\in \mathcal{E}^\ast} \mathrm{max}\{\underline{b}_{ij}^2 , \overline{b}_{ij}^2\} = \gamma,
		\end{equation}
		Now let $\lambda\triangleq[\lambda_{\mathrm{min}}(RR^T)/2-\epsilon^2/4] \lambda_{\mathrm{min}}(\xi^2) \lambda_{\mathrm{min}}(K^2) > 0$. Then from \eqref{eq:lyap1_dot_young} and \eqref{eq:norm_etahat} we arrive at:
		\begin{equation}\label{eq:lyap_dot_final}
		\dot{V} \leq -\lambda \|\sigma\|^2 + \dfrac{1}{2} \|\delta\|^2 
		+\dfrac{\gamma}{4\epsilon^2} \lambda_{\mathrm{max}}(\dot{\rho}^2), \;\forall t \in [0, \tau_{\mathrm{max}}).
		\end{equation}
		Assume that $\phi=\mathrm{col}(\delta,\lambda_{\mathrm{max}}(\dot{\rho}^2)) \in \mathbb{R}^{n+1}$. Using Lemma \ref{lem: kapa_inf_ineq} yields:
		\begin{equation} \label{eq:ISS_Lyap_Final}
		\dot{V} \leq -\lambda \|\sigma\|^2 + \beta(\|\phi\|), \quad \forall t \in [0, \tau_{\mathrm{max}}).
		\end{equation}
		where $\beta$ is a $\mathcal{K}_{\infty}$-function. Clearly \eqref{eq:ISS_Lyap_Final} satisfies the conditions of Lemma \ref{lem:ISS_lyap} (note that, all $\mathcal{K}_{\infty}$ -functions are indeed class $\mathcal{K}$ functions). Therefore, $\sigma$ is ISS with respect to input $\phi$ for all $t \in [0, \tau_{\mathrm{max}})$. Due to ISS and the boundedness of $\phi$ for all $t\geq 0$, there exists an ultimate bound $\overline\sigma>0$ independent of $\tau_{\text{max}}$ such that $\|\sigma\|\leq \overline\sigma$.
		Based on this result it is clear that \eqref{eq:u_single} is also bounded for all $t \in [0, \tau_{\mathrm{max}})$. Now using $\overline\sigma$ and taking the inverse logarithmic function in \eqref{eq:sigmaij}, leads to:
		\begin{equation}\label{eq:eta_hat_in_bound}
		\begin{aligned}
		-\underline{b}_{ij}<-&\dfrac{\overline{b}_{ij}(1-\exp(-2\overline{\sigma}))}{\underline{b}_{ij} \exp(-2\overline{\sigma})+\overline{b}_{ij}}\, \underline{b}_{ij}=\underline{\widehat{\eta}}_{ij} \leq \widehat{\eta}_{ij}(t) \\ 
		& \quad \leq \overline{\widehat{\eta}}_{ij} = \dfrac{\underline{b}_{ij}(\exp(2\overline{\sigma})-1)}{\underline{b}_{ij} \exp(2\overline{\sigma})+\overline{b}_{ij}} \, \overline{b}_{ij}< \overline{b}_{ij},
		\end{aligned}
		\end{equation}
		for all $t \in [0, \tau_{\mathrm{max}})$ and $(i,j)\in \mathcal{E}^\ast$.
		
		\textit{Phase III.} Up to this point, what remains to be shown is that $\tau_{\mathrm{max}}$ can be extended to $\infty$. Towards this direction, notice by \eqref{eq:eta_hat_in_bound} that $\widehat{\eta}(t) \in \Omega_{\widehat{\eta}}^{\prime}, \forall t \in [0, \tau_{\mathrm{max}})$ where
		\begin{equation}
		\Omega_{\widehat{\eta}}^{\prime}= \prod_{(i,j)\in \mathcal{E}^\ast}[\underline{\widehat{\eta}}_{ij},\overline{\widehat{\eta}}_{ij}],
		\end{equation}
		is a nonempty compact subset of $\Omega_{\widehat{\eta}}$. Hence, assuming $\tau_{\mathrm{max}} \leq \infty$ and since $\Omega_{\widehat{\eta}}^{\prime} \subset \Omega_{\widehat{\eta}}$, Proposition C.3.6 in \cite[p.~481]{sontag1998mathematical} dictates the existence of a time instant $t^{\prime} \in [0,\tau_{\mathrm{max}})$ such that $\widehat{\eta}(t^{\prime}) \notin \Omega_{\widehat{\eta}}^{\prime}$, which is a clear contradiction. Therefore, $\tau_{\mathrm{max}} = \infty$. Thus, all closed loop signals remain bounded and moreover $\widehat{\eta}(t) \in \Omega_{\widehat{\eta}}^{\prime} \subset \Omega_{\widehat{\eta}}, \forall t \geq 0$. Finally, multiplying \eqref{eq:eta_hat_in_bound} by $\rho_{ij}(t)$ results in:
		\begin{flalign}
		&-\underline{b}_{ij}\rho_{ij}(t) < \underline{\widehat{\eta}}_{ij} \rho_{ij}(t) \leq {\eta}_{ij}(t) \leq \overline{\widehat{\eta}}_{ij} \rho_{ij}(t) < \overline{b}_{ij} \rho_{ij}(t),&
		\end{flalign}
		for $(i,j)\in \mathcal{E}^\ast$ and $t \geq 0$, which further guarantees  \eqref{eq:e_bound} for all $t \geq 0$. 	
	\end{pf}
	
	\begin{remark}
		In the absence of external disturbances (i.e. $\delta_i = 0, \; i=1, \ldots,n$) or even when $\delta_i(t)$ are vanishing with time (i.e. $\delta_i(t) \rightarrow 0$ as $t \rightarrow \infty$), the proposed controller guarantees exact zero convergence of the errors with prescribed performance. Note that, since $\rho_{ij}(t)$ is smooth and $\lim_{t \rightarrow \infty} \rho_{ij}(t)=\rho_{ij\infty}>0$ then $\lim_{t\rightarrow \infty} \dot{\rho}_{ij}=0$ and so $\lambda_{\mathrm{max}}(\dot{\rho}^2)$ converges to zero as $t\rightarrow \infty$. Accordingly, whenever $\delta$ is zero or vanishing with time we have $\phi \rightarrow 0$ as $t \rightarrow \infty$. 
		Proof of Theorem \ref{th:single} clearly shows \eqref{eq:ISS_Lyap_Final} and \eqref{eq:e_bound} hold for all $t \geq 0$. 
		Therefore, based on the ISS property for vanishing input $\phi$, we have $\sigma(t) \rightarrow 0$ which implies $\eta(t) \rightarrow 0$, and hence $e(t) \rightarrow 0$ as $t\rightarrow \infty$. 
	\end{remark}
	
	\begin{remark}
		Notice that the speed of convergence is affected by the constants $a_{ij}$ in \eqref{eq:rho_ij}, which introduce a lower bound on the convergence speed of $e_{ij}(t)$. On the other hand for non-vanishing disturbances note that the squared distance errors $\eta_{ij}$ converge exponentially to the set $\Omega_f=(-\underline{b}_{ij} \rho_{ij\infty}, \overline{b}_{ij} \rho_{ij\infty})$. Since $-\underline{b}_{ij}$ and $\overline{b}_{ij}$ are determined by Algorithm \ref{algo}, one can always narrow the ultimate bound around zero by reducing $\rho_{ij\infty}$.
	\end{remark}
	
	\begin{remark} \label{rem:iso}
		Note that \eqref{eq:objec} is met when 
		\begin{equation}\label{eq:equval}
		\|q_i(t) - q_j(t)\| \rightarrow d_{ij} \quad \text{as} \quad t\rightarrow \infty, \quad i,j \in \mathcal{V}^\ast
		\end{equation}
		In general, the following is a necessary condition for \eqref{eq:objec}:
		\vspace{-0.5cm}
		\begin{equation}\label{eq:necce}
		e_{ij}(t)\rightarrow 0 \quad \text{as} \quad t\rightarrow \infty, \quad (i,j)\in \mathcal{E}^\ast.
		\end{equation}
		However, notice that \eqref{eq:necce} is equivalent to \eqref{eq:objec} only when $\mathrm{Amb}(\mathcal{F}^\ast)$ is an empty set (i.e., when $\mathcal{F}^\ast$ is a complete graph, also known as global rigidity).
		Recall that $\mathrm{Amb}(\mathcal{F}^\ast)$ is the set of undesired shapes with correct distance constraints between neighboring agents. Notice that, since $\mathcal{F}^\ast$ is assumed to be minimally rigid, $\mathrm{Amb}(\mathcal{F}^\ast)$ is not necessarily an empty set. Hence, $e(t) \rightarrow 0$ implies either $\mathcal{F}(t) \rightarrow \mathrm{Iso}(\mathcal{F}^\ast)$ or $\mathcal{F}(t) \rightarrow \mathrm{Amb}(\mathcal{F}^\ast)$. Now consider the following set
		\begin{equation} \label{eq:Omega_2}
		\Omega_F=\left\lbrace e\in \mathbb{R}^l \mid \mathrm{dist}(q , \mathrm{Iso}(\mathcal{F}^\ast)) < \mathrm{dist}(q , \mathrm{Amb}(\mathcal{F}^\ast)) \right\rbrace.
		\end{equation}
		Since stability of $e=0$ is ensured in Theorem \ref{th:single}, in order to guarantee \eqref{eq:objec}, the initial formation shape should be closer to $\mathrm{Iso}(\mathcal{F}^\ast)$. Thus, convergence to the correct formation can be achieved by considering $e(0)\in\Omega_I \cap \Omega_F$ in Theorem \ref{th:single}. This condition for converging to the right shape is an inherent issue in distance-based formation control problems and is a common requirement among various related works such as \cite{cai2015formation,krick2009stabilisation,ramazani2017rigidity,sun2015finite,sun2016exponential}.
	\end{remark}
	
	\begin{remark}\label{rem:robus_to_Distur}
		Notice that in the conventional approaches (e.g., \cite{cai2015adaptive,cai2014rigidity,cai2015formation,krick2009stabilisation,sun2016exponential,sun2015finite}) when external disturbances affect the agents' dynamics as in \eqref{eq:singledyn}, the condition $e(0)\in\Omega_I \cap \Omega_F$, discussed in Remark \ref{rem:iso}, is not sufficient to secure convergence to the correct shape (formation), since unpredictable transient behavior in agents' motion may lead the actual formation closer to an undesired shape ($\mathcal{F}(t) \rightarrow \mathrm{Amb}(\mathcal{F}^\ast)$). On the other hand, in our approach the proposed control scheme prevents such phenomena since it establishes a predetermined transient response. Moreover, according to Algorithm \ref{algo}, it is clear that one can limit the initial bounds of $e_{ij}$ (i.e. $\underline{e}_{ij}(0)$ and $\overline{e}_{ij}(0)$) to further increase the robustness against formation distortions as much as possible by decreasing $\underline{\mu}_{ij}$, $\overline{\mu}_{ij}$ (\textit{formation robustness constants}) and/or $\mu$. Note that, decreasing $\mu$ results in more conservativeness for choosing the proper bounds that ensure infinitesimally rigidity of the actual formation graph.
	\end{remark}
	
	\begin{remark}
		The proposed control protocol \eqref{eq:u_single} is in a similar form with the conceptual schemes of \cite{sun2016exponential,sun2015finite,cai2015formation,krick2009stabilisation,cai2014rigidity}, but it further encapsulates guaranteed transient and steady state performance with increased robustness against any bounded external disturbances and ensures connectivity maintenance and collision avoidance among neighboring agents.
	\end{remark}
	
	\begin{remark}\label{rem:decent_single}
		The control law in \eqref{eq:u_single} can be expressed for each agent as:
		\begin{equation}
		\label{eq:u_single_decen}
		u_i= -\sum_{j \in \mathcal{N}_i(\mathcal{E}^\ast)} k_{ij} \; g_{ij}(\eta_{ij}) \; \widetilde{q}_{ij} 
		\end{equation}
		with $g_{ij}(\eta_{ij})=\xi_{ij} \, \sigma_{ij}$, which is decentralized since each agent is required to sense only the relative position with respect to its neighbors via on-board sensors. Hence, the control law is independent of a global coordinate frame and does not require agents' local coordinate systems to be aligned. Notice that the quantities in \eqref{eq:u_single_decen} are all expressed in a global coordinate frame. However, this control law can be implemented with locally measurable relative positions w.r.t. each agent's local coordinate system. Let $u_i$ be the control law of agent $i$ in the global coordinate frame. Also, let the superscript $i$ indicate a quantity expressed in the local coordinate frame of the $i$-th agent. Furthermore, suppose that $\mathcal{R}_i \in \mathbb{R}^{m \times m}$ is the transformation matrix from the $i$-th local frame to the global frame. Obviously, we have $u_i=\mathcal{R}_i u_i^{i}$ and $\widetilde{q}_{ij}=\mathcal{R}_i \widetilde{q}_{ij}^{\,i}=\mathcal{R}_i(q_i^{i}-q_j^{i})$. The control law \eqref{eq:u_single_decen} in the $i$-th agent's local frame is:
		\begin{align}
		u_i^{i}=\mathcal{R}_i^{-1}u_i &=- \mathcal{R}_i^{-1} \sum_{j \in \mathcal{N}_i(\mathcal{E}^\ast)} k_{ij} \; g_{ij}(\eta_{ij}) \; \widetilde{q}_{ij} \nonumber \\
		&=- \sum_{j \in \mathcal{N}_i(\mathcal{E}^\ast)} k_{ij} \; g_{ij}(\eta_{ij}) \; \mathcal{R}_i^{-1} \widetilde{q}_{ij} \nonumber \\
		&=- \sum_{j \in \mathcal{N}_i(\mathcal{E}^\ast)} k_{ij} \; g_{ij}(\eta_{ij}^{i}) \; \widetilde{q}_{ij}^{\,i} 	\label{eq:u_single_decen_frame}
		\end{align}
		Note that $g_{ij}(\eta_{ij})$ is a scalar function of $\eta_{ij}$ and since $\eta_{ij}$ relates to inter-agent distance, its value does not dependent on the coordinate system. In other words, since $\|q_i - q_j\|=\|q_i^i - q_j^i \|$ the same holds for $\eta_{ij}=\eta_{ij}^i$. It is clear that \eqref{eq:u_single_decen_frame} has the same form as in \eqref{eq:u_single_decen} in which all the quantities are expressed
		with respect to the $i$th local coordinate system. This indicates that the decentralized control law can be implemented in arbitrarily oriented local coordinate frames of the agents merely by measuring the relative positions.
	\end{remark}
	
	\begin{remark}
		Note that, the control law \eqref{eq:u_single_decen} does not require any knowledge on the external disturbances bounds. Moreover, it does not require any estimation of their unknown upper bounds which simplifies implementation. Furthermore, the controller's gain does not depend on the performance specification, thus, making its selection easier.
	\end{remark}
	\section{Extension to Formation Centroid Maneuvering}
	\label{Centroid_Manue}
	
	In this section, we extend the previous results to solve the distance-based formation centroid maneuvering control problem with prescribed performance as well as connectivity maintenance and collision avoidance among neighboring agents. In this respect, a single agent will be considered as the leader that is injected with an external reference velocity command, while the rest should follow the leader and maintain the formation shape such that the centroid of the formation moves with a desired time-varying reference velocity.
	
	Thus, consider $n$ disturbance-free single integrator agents in an $m$-dimensional space modeled by: 
	\begin{equation}
	\label{eq:singledyn_nomi}
	\dot{q}_i=u_i,  \quad i=1,\ldots,n
	\end{equation}
	Let $v_{d}(t) \in \mathbb{R}^m$ be the bounded time-varying desired velocity of the formation's centroid, which is known  only to the leader. The formation's centroid is given by:
	\begin{equation} \label{eq:centeroid}
	q_c(t) = \dfrac{1}{n} \sum_{i=1}^{n} q_i(t) = \dfrac{1}{n} (\mathbf{1}_n^T \otimes \mathrm{I}_m) q(t).	
	\end{equation}
	Now, the problem is to design a proper control protocol that ensures \eqref{eq:goal} as well as $\dot{q}_c(t)=v_d(t)$.
	
	Let the proper formation maneuvering controller be designed as:
	\begin{equation}\label{eq:u_single_maneuever}
	u_{m}= - R^T \xi K \sigma + n M_{L} v_d(t),
	\end{equation}
	where $M_{L} \in \mathbb{R}^{mn \times m}$ denotes the pinning matrix to the leader that has direct access to the desired centroid velocity $v_d$ (i.e., if agent $i$ is the leader, then the $i$th block of $M_{L}$ is $\mathrm{I}_m$).
	
	\begin{theorem}\label{th: single_cent_maneu}
		For the agents' dynamics \eqref{eq:singledyn_nomi}, the control law \eqref{eq:u_single_maneuever} guarantees $\dot{q}_c(t)=v_d(t)$ and \eqref{eq:goal}, which solves the prescribed performance distance-based formation centroid maneuvering control problem with connectivity maintenance and collision avoidance among neighboring agents.
	\end{theorem}
	\begin{pf}
		First, since $\mathrm{ker}(H)=\mathrm{span}\{\mathbf{1}_n\}$, observe from \eqref{eq:rigidmat} that:
		\begin{align}
		(\mathbf{1}_n^T \otimes \mathrm{I}_m)R^T &= (\mathbf{1}_n^T \otimes \mathrm{I}_m) \bar{H}^T Q \nonumber \\
		&=  (\mathbf{1}_n^T \otimes \mathrm{I}_m) (H^T \otimes \mathrm{I}_m) Q = 0
		\end{align}
		where $Q=\mathrm{diag}(\widetilde{q}_{ij})$. Using the aforementioned property along with \eqref{eq:centeroid} and \eqref{eq:u_single_maneuever}, the velocity of the centroid can be written as:
		\begin{align}	
		\dot{q}_c(t) &= \dfrac{1}{n} (\mathbf{1}_n^T \otimes I_m) \dot{q}(t) \nonumber \\ 
		&= \dfrac{1}{n} (\mathbf{1}_n^T \otimes I_m) (- R^T \xi K \sigma + n M_{L} v_d(t)) \nonumber \\
		& =  (\mathbf{1}_n^T \otimes I_m) M_{L} v_d(t).	\label{eq:centeroid_vel}
		\end{align}	
		Since there is only one leader in the group (i.e. $(\mathbf{1}_n^T \otimes I_m) M_{L} = \mathrm{I}_m$) the centroid velocity reduces to $\dot{q}_c(t)=v_d(t)$.
		
		Moreover, it is easy to see that the squared distance errors can be written in compact form as $\dot{\eta}=2R(q)u_m$, where $u_m=u+n M_{L} v_d$ with $u$ as defined in \eqref{eq:u_single}. Thus, invoking the boundedness of $n M_{L} v_d(t)$ instantly leads us to infer that $\delta(t)$ in Theorem \ref{th:single} is equivalent to $n M_{L} v_d(t)$ in Theorem \ref{th: single_cent_maneu}. Hence, the achieved results in Theorem \ref{th:single} (satisfaction of \eqref{eq:e_bound}) also hold for the proposed controller  in \eqref{eq:u_single_maneuever}.
	\end{pf}	
	
	\begin{remark}
		As mentioned in the proof of Theorem \ref{th: single_cent_maneu}, $n M_{L} v_d(t)$ is equivalent to $\delta$ in Theorem \ref{th:single}. In other words, embedding $\delta=n M_{L} v_d(t)$ (i.e., $v_d(t)$ is inferred as external disturbance for the leader) the control law in \eqref{eq:u_single_maneuever} solves the centroid maneuvering problem. Based on this equivalence, if the agents models are affected by external disturbances $\delta_e(t)$, we will have $\delta(t)=n M_{L} v_d(t)+\delta_e(t)$ In this case, following similar analysis we may conclude that the centroid velocity will be $\dot{q}_c(t)=(1/n)(\mathbf{1}_n^T \otimes I_m) (nM_{L} v_d(t) + \delta_e)$. Notice that, in practice, the amplitude of the desired centroid velocity in $nM_{L} v_d(t)$ is more than the amplitude of the external disturbances $\delta_e$. Hence, when the external disturbances are periodic (which is common in practical applications), the proposed controller will form the desired shape and track the desired centroid velocity with certain small fluctuations caused by the external disturbances. Consequently, in case the external disturbances $\delta_e(t)$ are vanishing with time, the velocity of the formation's centroid will eventually converge to $v_d(t)$ (i.e., $\dot{q}_c(t) \rightarrow v_d(t)$ when $\delta_e(t) \rightarrow 0$).
	\end{remark}
	
	\begin{remark}
		The control law \eqref{eq:u_single_maneuever} can be expressed component wise for each agent exactly the same as \eqref{eq:u_single_decen} except for the leader has an extra term $nv_d(t)$. The structure of the leader's controller indicates that in addition to the desired centroid velocity, the leader should also know the total number of agents in the formation. Moreover, since the leader is the only agent that has access to the desired centroid velocity, $v_d(t)$ is indeed available w.r.t. to the leader's local coordinate system. Due to these facts, all the points stated in Remark \ref{rem:decent_single} hold for \eqref{eq:u_single_maneuever} as well. Therefore, the control law \eqref{eq:u_single_maneuever} is decentralized and also it is independent of a global coordinate frame.
	\end{remark}
	
	\begin{remark}
		In contrast to \cite{rozenheck2015proportional}, where the distance-based formation centroid maneuvering problem for undirected rigid graphs with constant reference velocity is solved, in this paper we extend the results for time-varying desired reference velocities of the formation centroid and guaranteed further connectivity maintenance and collision avoidance as well as predefined transient performance. Moreover, in \cite{cai2015formation} the proposed distance-based formation maneuvering controller is applicable whenever the agent's local coordinate systems are aligned and the reference velocity is available to all of the agents. In \cite{mehdifar2018finite,khaledyan2019flocking} and \cite{yang2018distributed} distance-based formation maneuvering and formation centroid tracking are studied for undirected rigid graphs, respectively. However, the proposed schemes require agents to be capable of extracting their relative orientation with respect to their neighbors. To the best of the author's knowledge this is the first time that the distance-based formation maneuvering problem for undirected rigid graphs with time-varying reference velocity is solved, in which the control law is implementable in arbitrarily oriented local coordinate frames. In addition, none of the previous works have considered collision avoidance nor connectivity maintenance for distance-based formation maneuvering.
	\end{remark}

	\section{Simulation Results}
	\label{Simu}
	In this section, three simulation examples are presented to demonstrate the effectiveness of the proposed decentralized control protocols\footnote{A short video demonstrating the following simulation results can be found at: https://youtu.be/eIxCKLcVnM8}.
	\subsection{Formation Acquisition}
	\label{Simu:acqu}
\begin{figure}[tbp]
		\centering
		\begin{subfigure}[t]{0.20\textwidth}
			\flushleft
			\includegraphics[width=\textwidth]{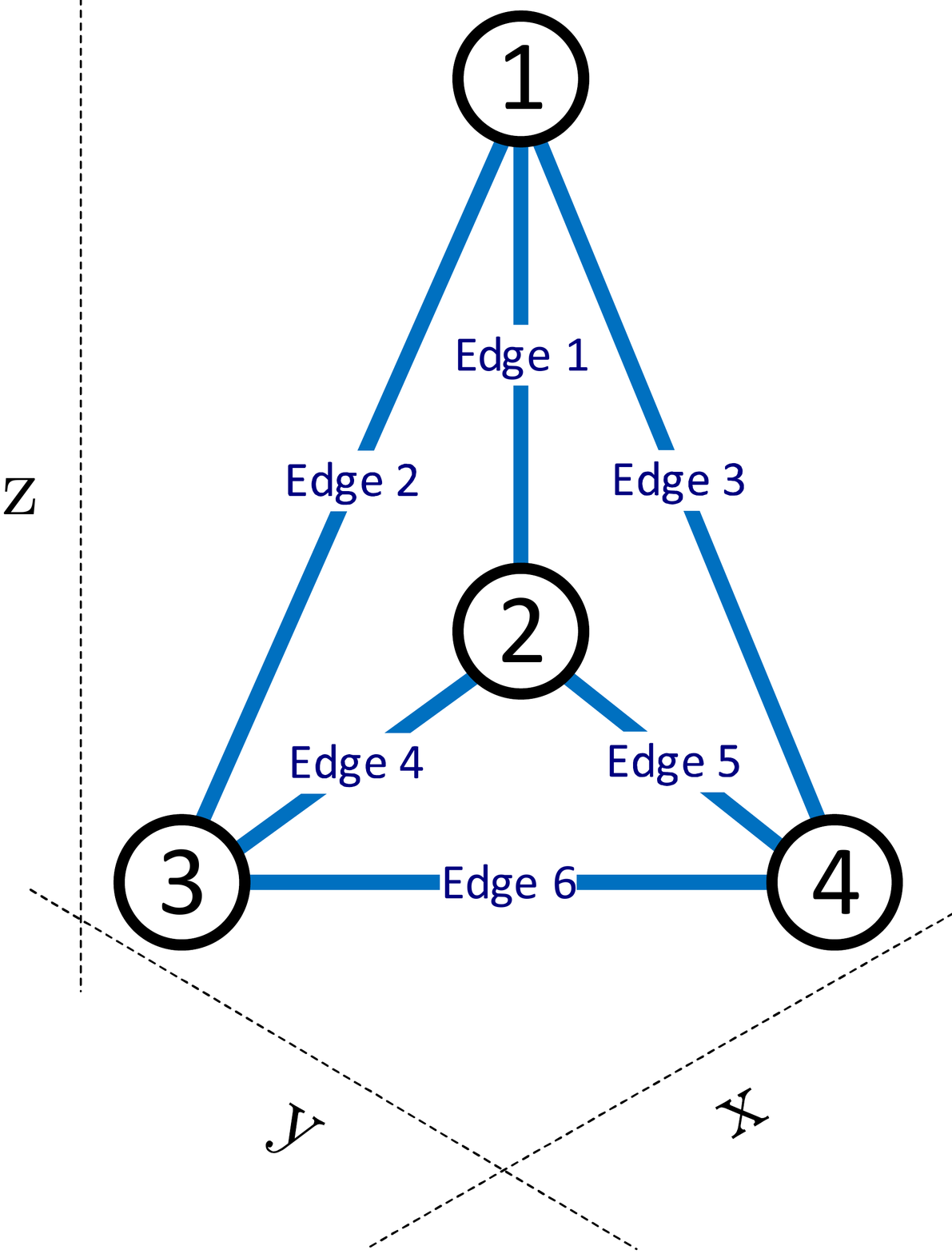}
			\caption{}
			\label{fig:3-D_tetrahedron}
		\end{subfigure}%
		\begin{subfigure}[t]{0.28\textwidth}
			\centering
			\includegraphics[width=\textwidth]{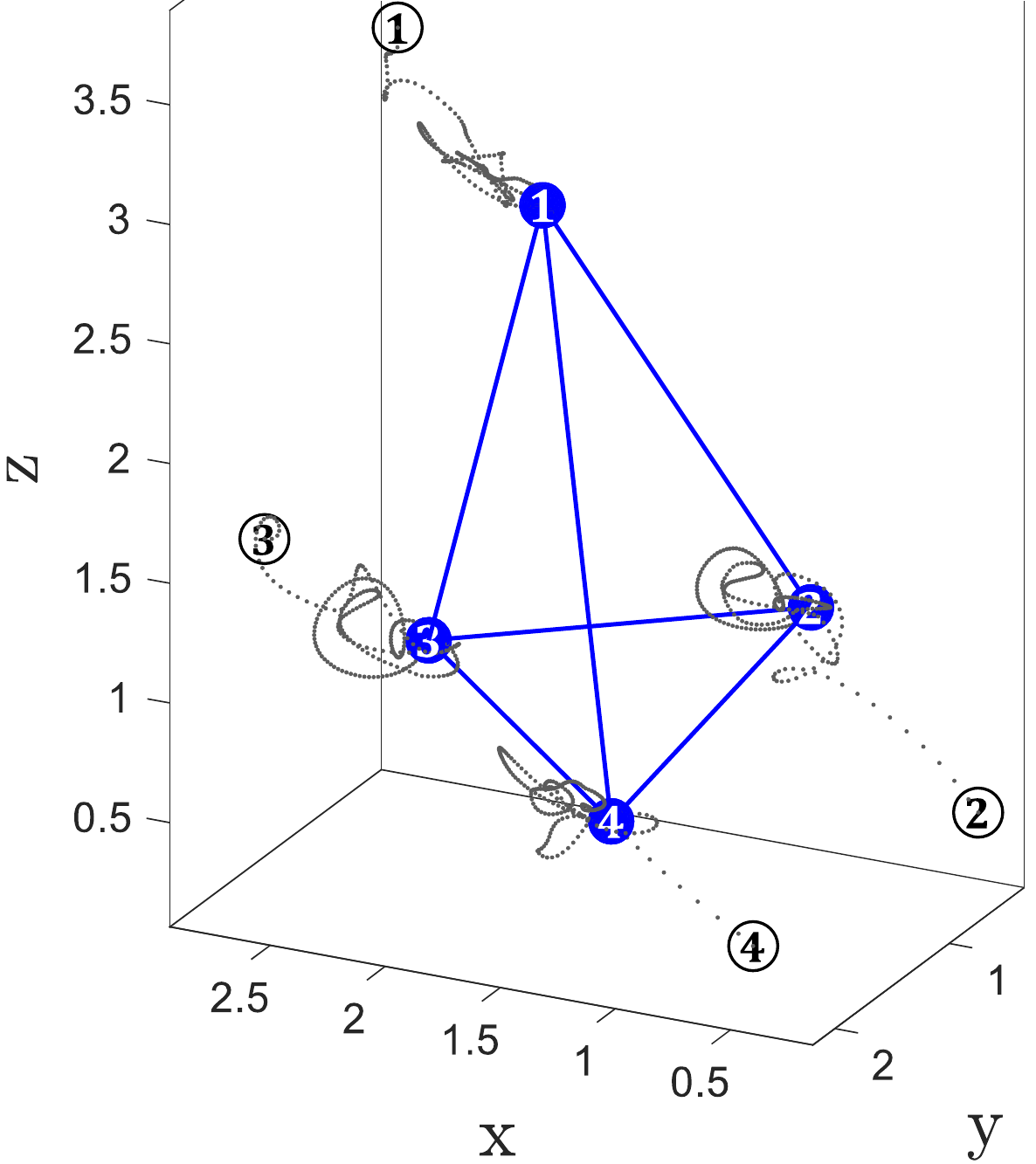}
			\caption{}
			\label{fig:FD_3D}
		\end{subfigure}
		\caption{(a) The desired minimally and infinitesimally rigid framework. (b) Agents' trajectories towards the desired formation.}
	\end{figure}

\begin{figure}[tbp]
		\centering
		\begin{subfigure}[t]{0.235\textwidth}
			\centering
			\includegraphics[width=\textwidth]{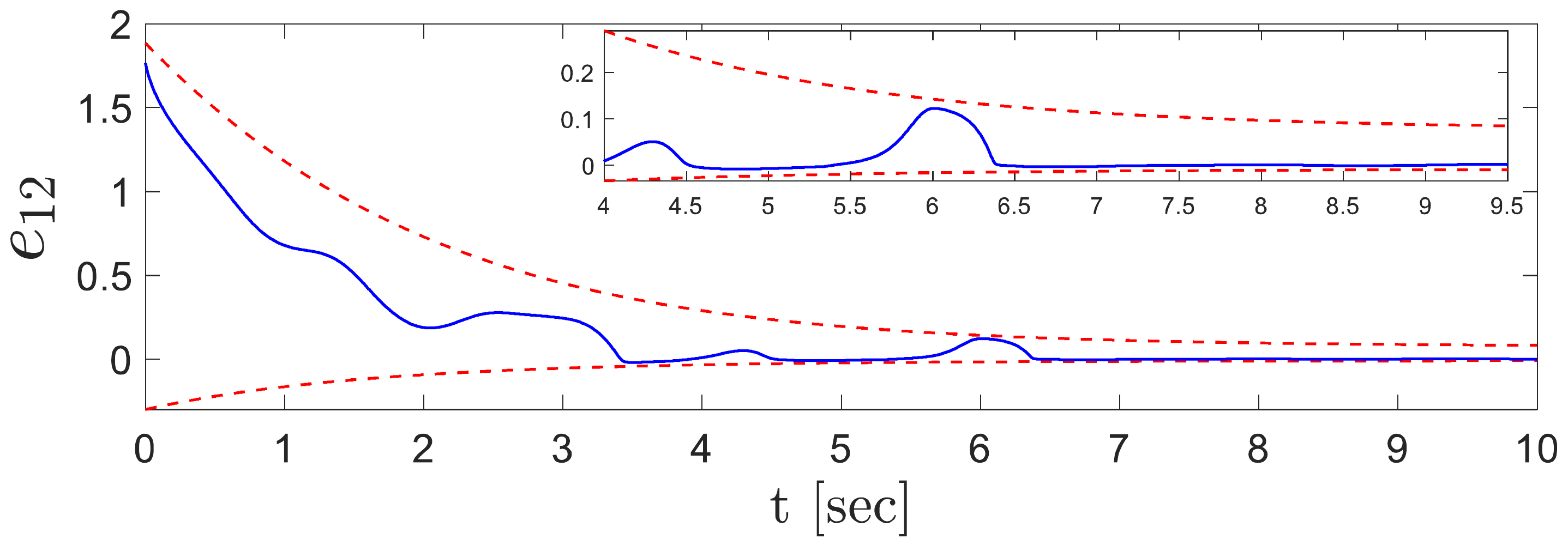}
		\end{subfigure}
		\begin{subfigure}[t]{0.237\textwidth}
			\centering
			\includegraphics[width=\textwidth]{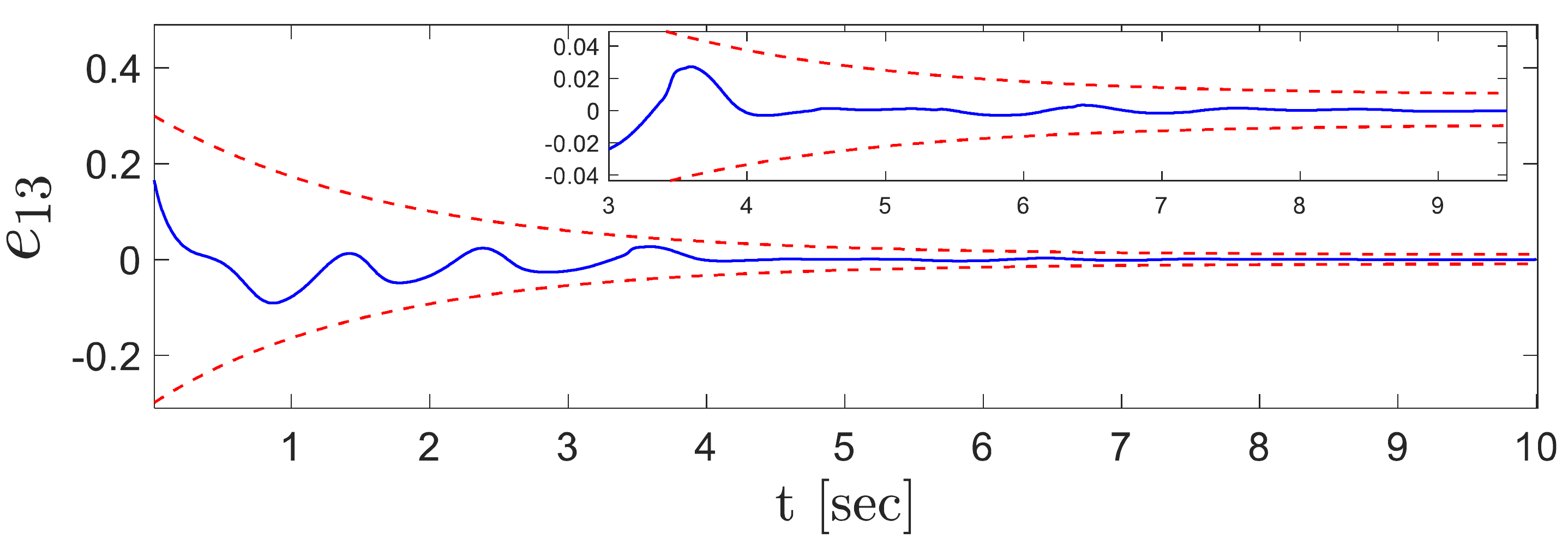}
		\end{subfigure}
		\begin{subfigure}[t]{0.236\textwidth}
			\centering
			\includegraphics[width=\textwidth]{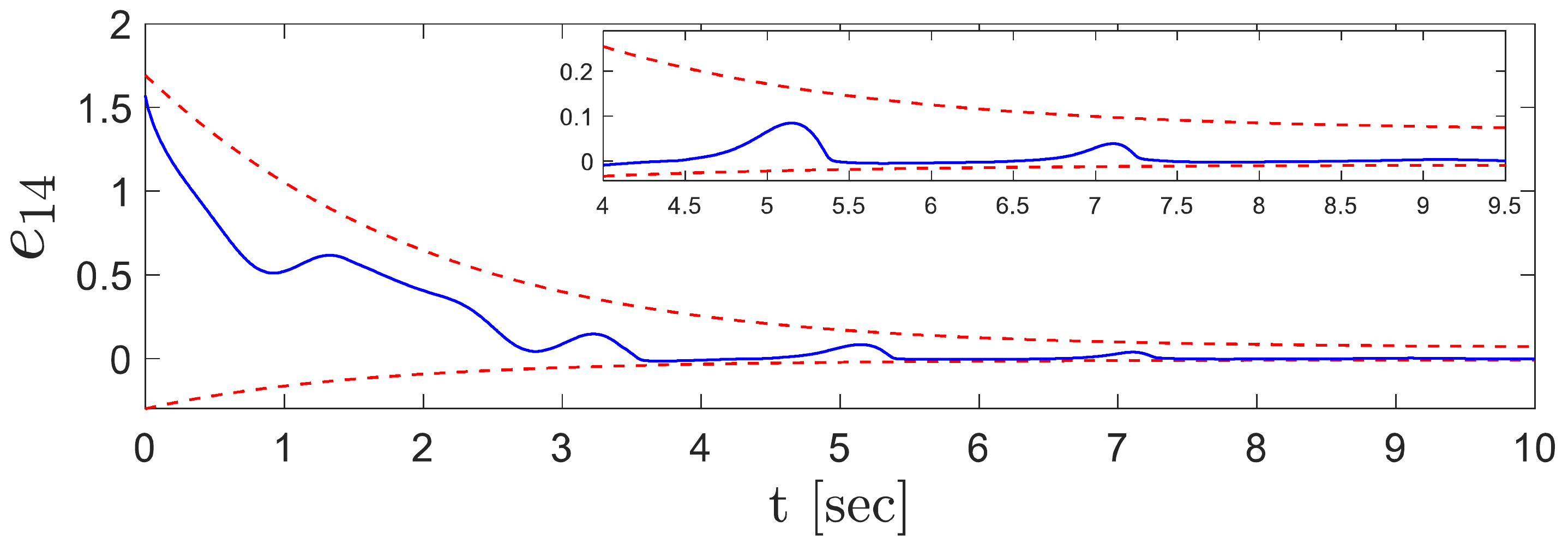}
		\end{subfigure}
		\begin{subfigure}[t]{0.236\textwidth}
			\centering
			\includegraphics[width=\textwidth]{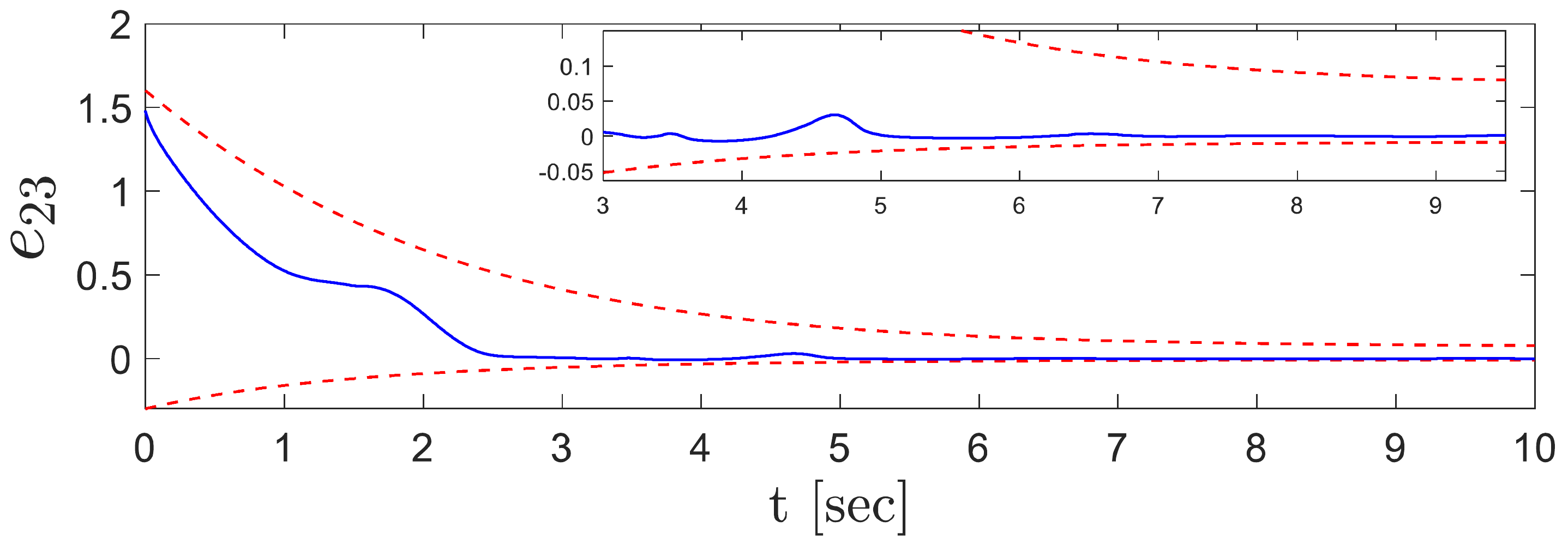}
		\end{subfigure}
		\begin{subfigure}[t]{0.236\textwidth}
			\centering
			\includegraphics[width=\textwidth]{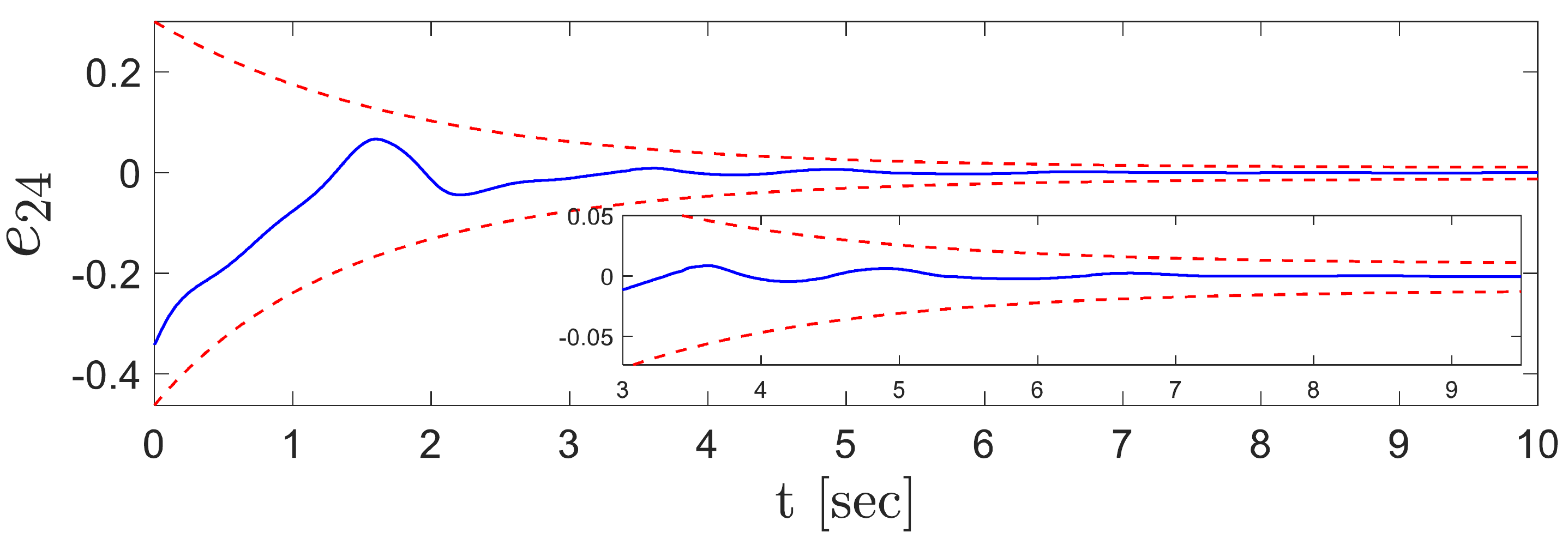}
		\end{subfigure}
		\begin{subfigure}[t]{0.236\textwidth}
			\centering
			\includegraphics[width=\textwidth]{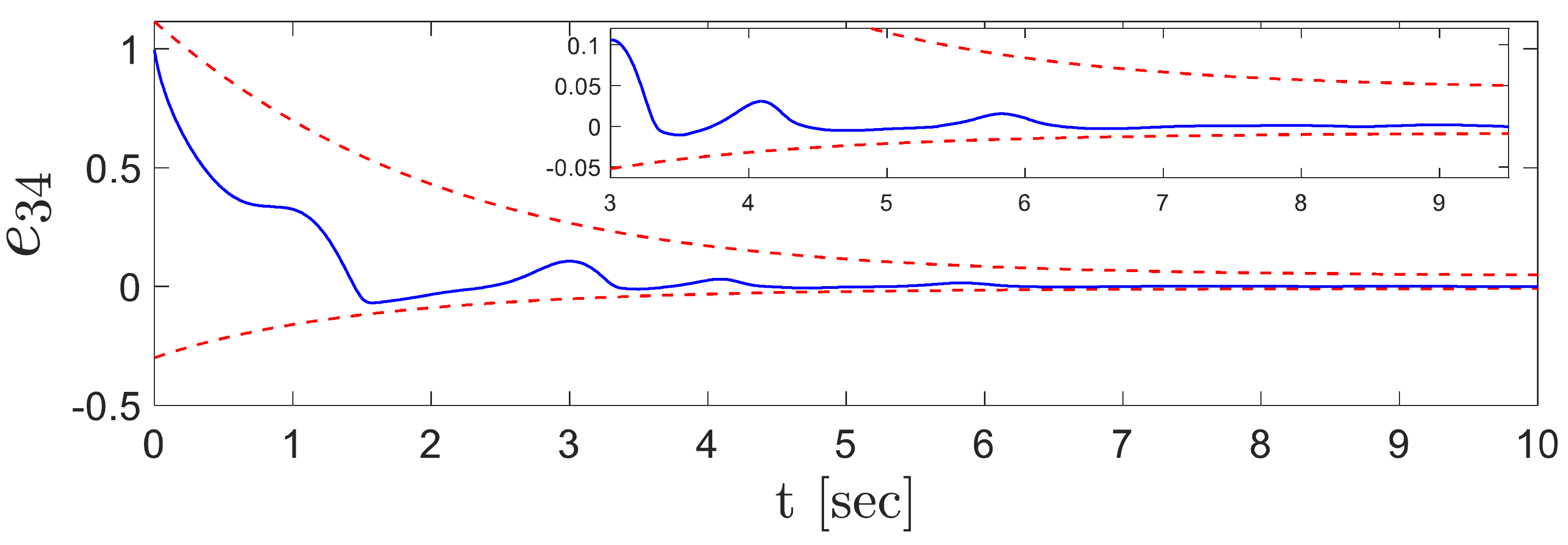}
		\end{subfigure}
		\caption{Inter-agent distance errors of the tetrahedron formation for $(i,j)\in \mathcal{E^\ast}$.}
		\label{fig:error_diag_3D}
	\end{figure}

	Consider a group of four agents modeled by \eqref{eq:singledyn} in a three dimensional space. Assume that the desired formation is a tetrahedron defined by a minimally and infinitesimally rigid graph as depicted in Fig.~\ref{fig:3-D_tetrahedron}. Given the edge ordering as in Fig.~\ref{fig:3-D_tetrahedron}, we obtain the edge set as $\mathcal{E}^\ast=\{(1,2),(1,3),(1,4),(2,3),(2,4),(3,4)\}$. The desired distances between neighboring agents (desired edges lengths) in the rigid framework are assumed to be $d_{12}=d_{13}=d_{14}=2.5$ and $d_{23}=d_{34}=d_{24}=1.5$. The initial positions of the agents are given by  $q_1(0)=[2.0610, \,1.9605, \,3.8940]$, $q_2(0)= [0.3424, \,0.3424, \,0.3424]$, $q_3(0)=[2.9121, \, 1.4121, \,1.4121]$, $q_4(0)=[0.8137, \,1.3627, \,0.0637]$ indicating that $\mathcal{F}(0)$ is  infinitesimally rigid. Moreover, agent's external disturbances are considered to be: $\delta_1(t)= [0.4\sin(0.8\pi t)+0.25\cos(2\pi t),\, 0.5\cos(\pi t),\, 0.4\sin(2\pi t) + 0.2\cos(1.2\pi t)]$, $\delta_2(t)=[0.8\sin(1.2 \pi t) + 0.3\cos(0.5\pi t),\, 0.4\sin(0.8\pi t)+{0.25\cos(2\pi t),\, 0.35\sin(0.6\pi t)+0.6\cos(1.2\pi t)],\, \delta_3(t)}=[0.2\sin(1.2\pi t)+0.4\cos(0.5\pi t),\,0.2\sin(1.2\pi t),\,0.5\sin(1.5\pi t) \allowbreak +0.4\cos(2\pi t)], \, {\delta_4(t)=[0.35\sin(0.6\pi t)+0.6\cos(1.2\pi t),\,} \allowbreak 0.4\sin(0.8\pi t)+0.25\cos(2\pi t) ,\,0.5\sin(0.8\pi t)+0.7\cos(\pi t)]$. Without loss of generality, in the simulations we assumed that all agents have the same geometrical and sensing radius, i.e., $r_{si}=0.2$ and $r_{ci}=5$ for $i=1,\ldots,4$. In Algorithm \ref{algo}, we selected $\mu=0.12$, and $\overline{\mu}_{ij}=\underline{\mu}_{ij}=0.3, \forall (i,j) \in \mathcal{E^\ast}$. Furthermore, the parameters in the performance function \eqref{eq:rho_ij} are considered as $a_{ij}=0.6$, $\rho_{ij0}=1$, $\rho_{ij\infty}=0.03$ for all $(i,j)\in \mathcal{E^\ast}$. Moreover, the controller gains are set to $k_{ij}=0.1, (i,j) \in \mathcal{E^\ast}$. Fig.~\ref{fig:FD_3D} depicts a snapshot of the agents' trajectories  towards $\mathrm{Iso}(\mathcal{F^*})$ after 10 seconds under the proposed controller \eqref{eq:u_single}. The evolution of distance errors among neighboring agents along with the Prescribed Performance Bounds (PPB) obtained from \eqref{ppb_e} (dashed lines) are illustrated in Fig.~\ref{fig:error_diag_3D}. The results clearly indicate that the proposed formation acquisition controller is capable of handling the problem of robust distance-based formation control with prescribed performance, connectivity maintenance and collision avoidance among neighboring agents in the presence of external disturbances.
	\subsection{Robustness to Formation Distortions and Undesired Shapes}
	\label{subsec:simu_NoAmb}
	
\begin{figure}[tbp]
		\centering
		\begin{subfigure}[t]{0.187\textwidth}
			\centering
			\includegraphics[width=\textwidth]{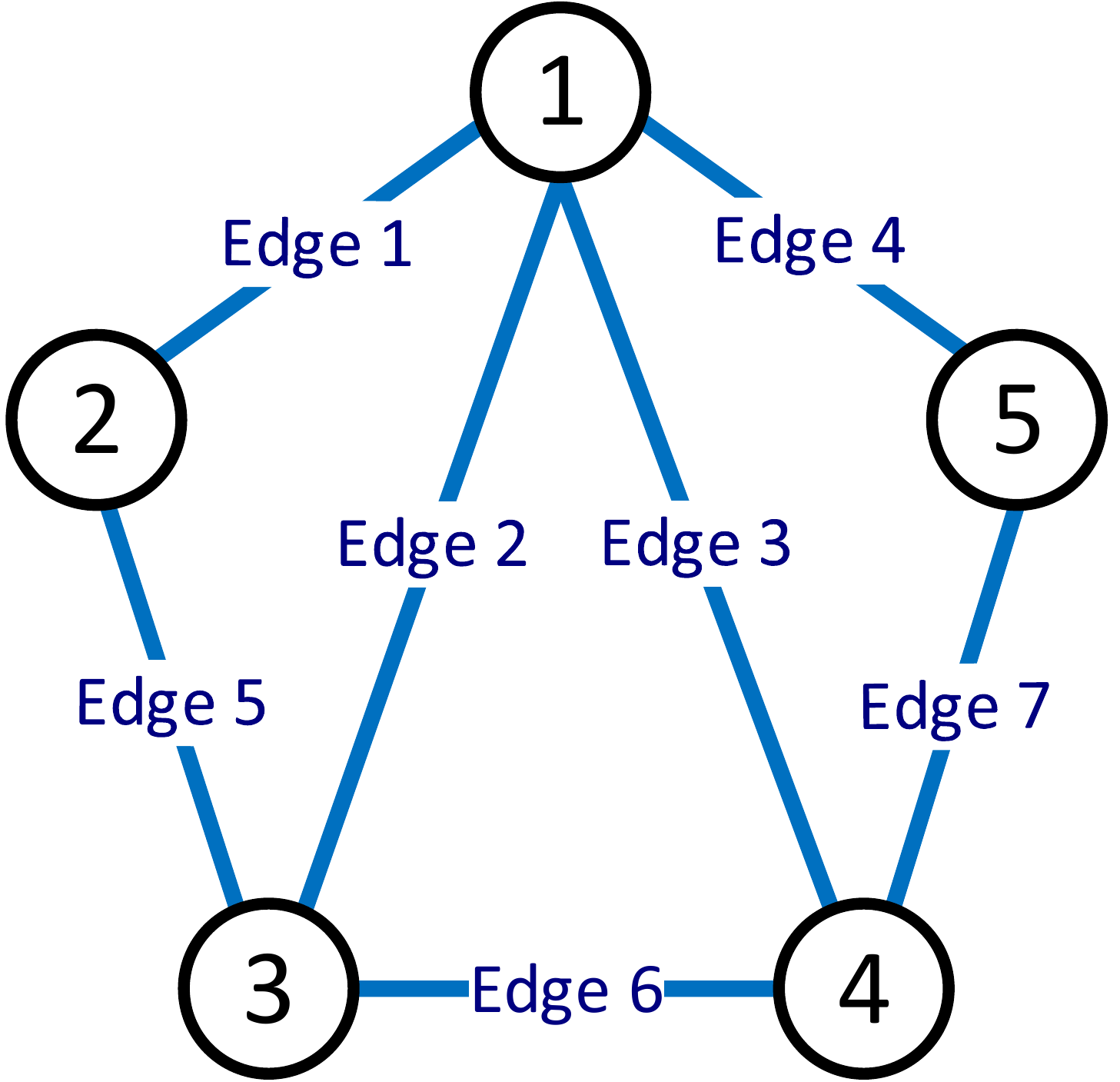}
			\caption{}
			\label{fig:penta_Rigid}
		\end{subfigure}
		\begin{subfigure}[t]{0.28\textwidth}
			\centering
			\includegraphics[width=\textwidth]{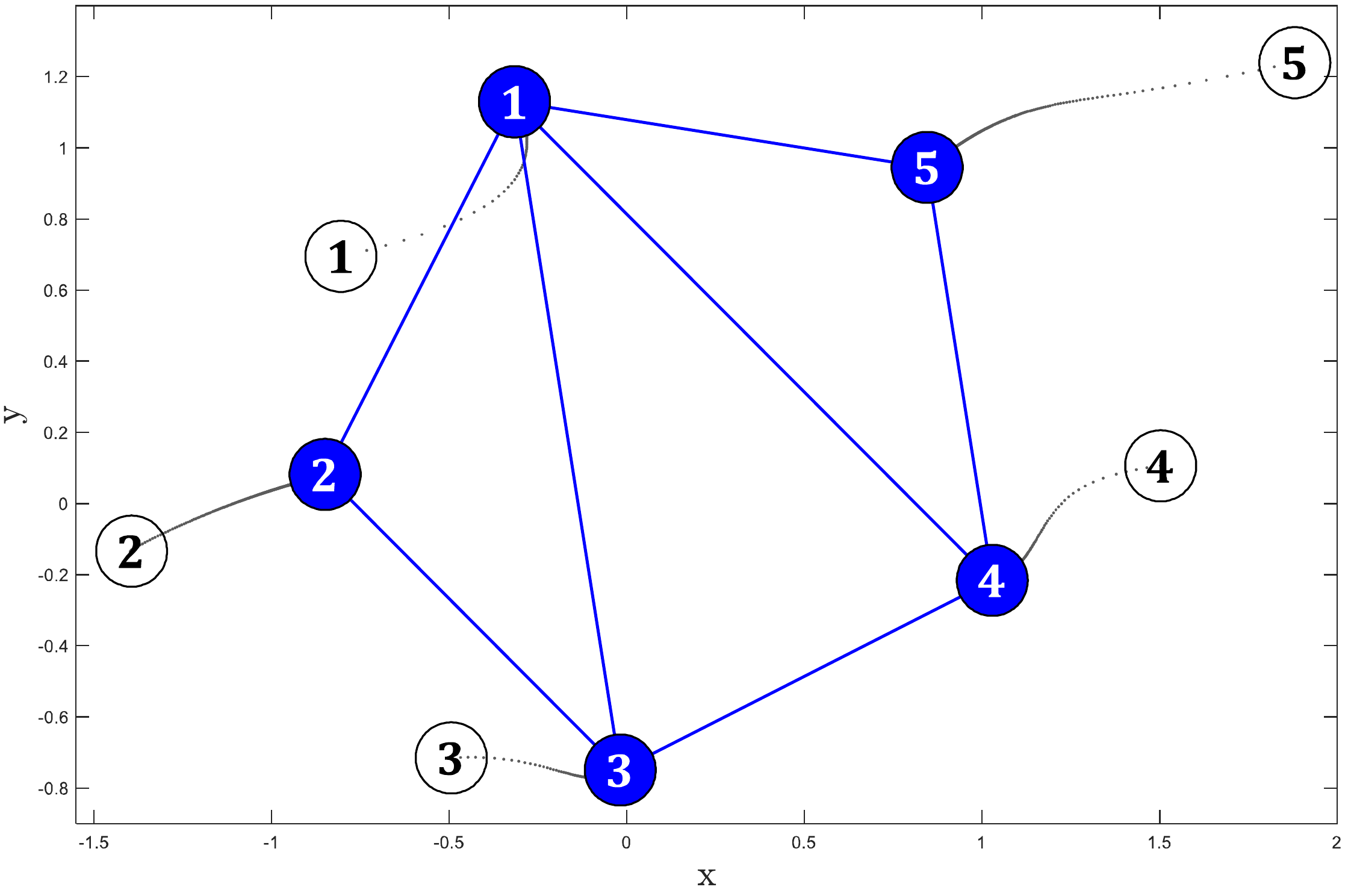}
			\caption{}
			\label{fig:penta_Normal}
		\end{subfigure}
		\caption{(a) The desired minimally and infinitesimally rigid framework of pentagon. (b) Agents' trajectories toward the desired formation using conventional distance-based formation control law with no external disturbances.}
		\label{fig:penta_rigid+normalControl}
	\end{figure}

	In this subsection, we provide a comparative simulation study to show that the proposed formation control law \eqref{eq:u_single} can effectively prevent formation distortions that lead to undesired shapes when external disturbances affect the agents dynamics (see Remark \ref{rem:robus_to_Distur}). Consider a group of five agents with a pentagon as the desired formation defined by a minimally and infinitesimally rigid graph as in Fig.~\ref{fig:penta_Rigid}. Based on the edge ordering in Fig.~\ref{fig:penta_Rigid}, the edge set is $\mathcal{E}^\ast=\{(1,2), (1,3), (1,4), (1,5), (2,3), (3,4), (4,5)\}$. The desired distances between neighboring agents are $d_{12}=d_{23}=d_{34}=d_{45}=\sqrt{2(1-\cos(2\pi/5))}$ and $d_{13}=d_{14}=\sqrt{2(1+\cos(\pi/5))}$. Moreover, the initial positions of the agents are given by $q_1(0)=[-0.8049, \, 0.6951]$, $q_2(0)= [-1.3941,\,-0.1340]$, $q_3(0)=[-0.4940,\,-0.7153]$, $q_4(0)=[1.5028, \, 0.1060]$, $q_5(0)=[1.8808,\,1.2388]$ ensuring $e(0)\in\Omega_I \cap \Omega_F$ (see Remark \ref{rem:iso}). Fig~\ref{fig:penta_Normal} shows that in the absence of external disturbances, the conventional distance-based formation control law $u=-kR^T\eta$, ($k > 0$) as proposed in \cite{cai2014rigidity}, establishes the desired formation. Alternatively, in the presence of external disturbances, in order to provide reliable comparisons, we introduce a decentralized and robustly modified distance-based formation control law: 
	\begin{equation}
	\begin{aligned}
	u &= -k R^T \eta - \mathrm{diag}(\widehat{D}) \tanh (\dfrac{\mathrm{diag}(\widehat{D}) k R^T \eta}{\varepsilon}) \\
	\dot{\widehat{D}} &= W |k R^T \eta| - \theta \widehat{D}
	\end{aligned}	
	\label{eq:contr_normal_robus}
	\end{equation} 
	where $k, \varepsilon, \theta$ are positive scalars and $W=\mathrm{diag}(w_{ij}) \in \mathbb{R}^{l\times l}$ with $w_{ij} > 0, \forall (i,j)\in \mathcal{E}^\ast$. Notice that, $\widehat{D}=\mathrm{col}(\widehat{D}_i)$ is an estimate of the upper bounds  of the agents' external disturbances. One can prove that the modified conventional control law \eqref{eq:contr_normal_robus} is capable of stabilizing inter agent distance errors in a sufficiently small neighborhood of $e=0$ by utilizing the uniform ultimate boundedness notion. In the simulations, it is assumed that $k=0.3, \varepsilon=0.01, \theta=0.01,\text{ and } w_{ij}=1.5 \text{ for } (i,j)\in \mathcal{E}^\ast$. Moreover, for the prescribed performance based control law \eqref{eq:u_single} it is assumed that the controller gains are set to $k_{ij}=0.3$ and $a_{ij}=1, (i,j) \in \mathcal{E^\ast}$ in the performance functions. All other parameters for the performance functions as well as Algorithm \ref{algo} are considered the same as in Subsection \ref{Simu:acqu}. Without loss of generality, consider that only the second and third agents are subject to external disturbances. To this end, assume $\delta_1(t) = \delta_4(t) = \delta_5(t)=[0 , 0]$, $\delta_2(t)=[0.6 \sin (1.2 \pi t) -0.3 \sin (0.6 \pi t), 0.5 \sin(\pi t)]$, and $\delta_3(t)=[0.3 \sin(0.6 \pi t) - 0.6 \sin (1.2 \pi t), -0.5 \sin(\pi t)]$. Recall that $\delta(t)=\mathrm{col}(\delta_i(t))$. In what follows, the performance of both control laws \eqref{eq:u_single} and \eqref{eq:contr_normal_robus}, are compared in three cases of disturbance level.
	
\begin{figure}[tbp]
		\flushleft
		\begin{subfigure}[t]{0.234\textwidth}
			\centering
			\includegraphics[width=\textwidth]{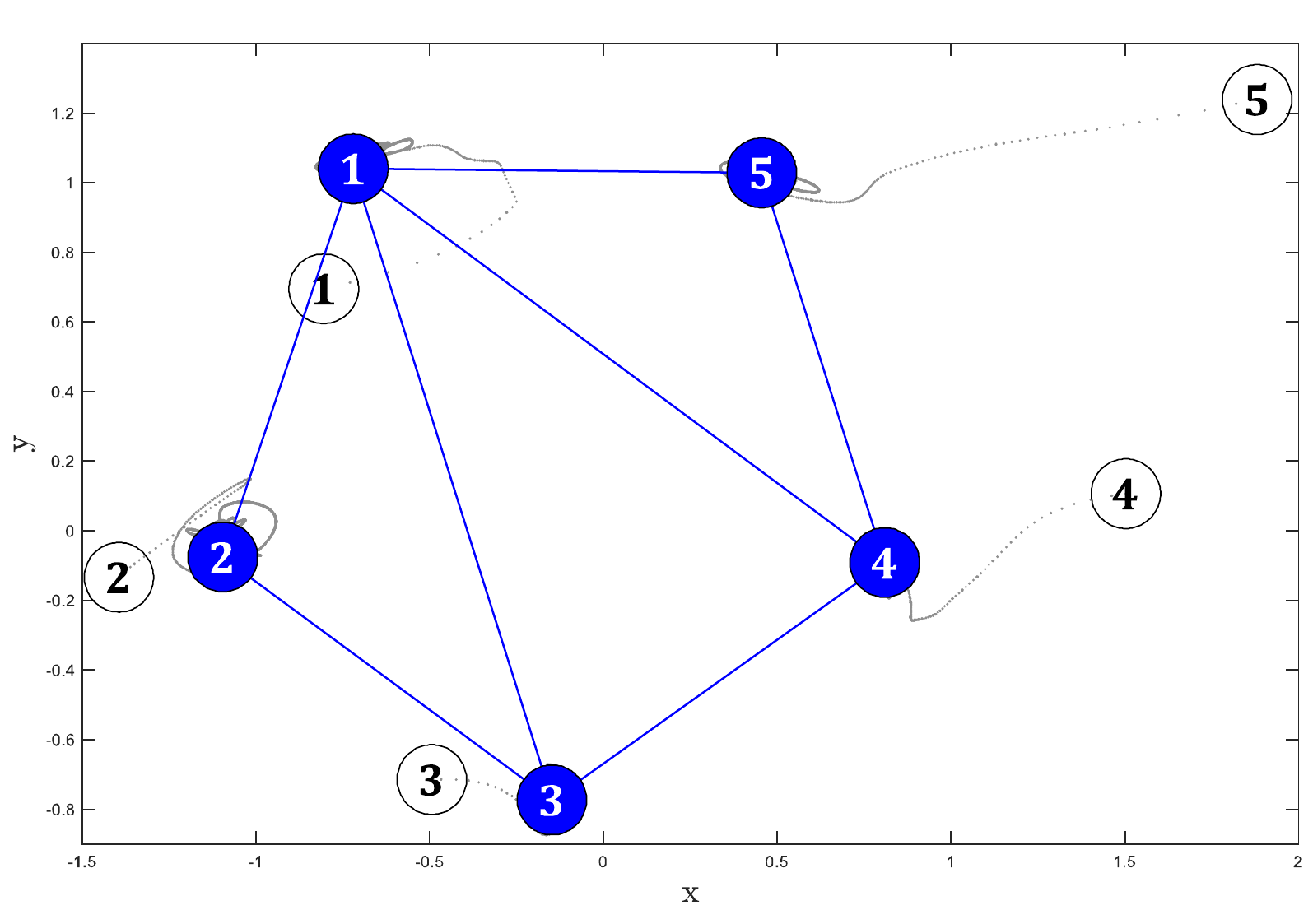}
		\end{subfigure}
		\begin{subfigure}[t]{0.237\textwidth}
			\centering
			\includegraphics[width=\textwidth]{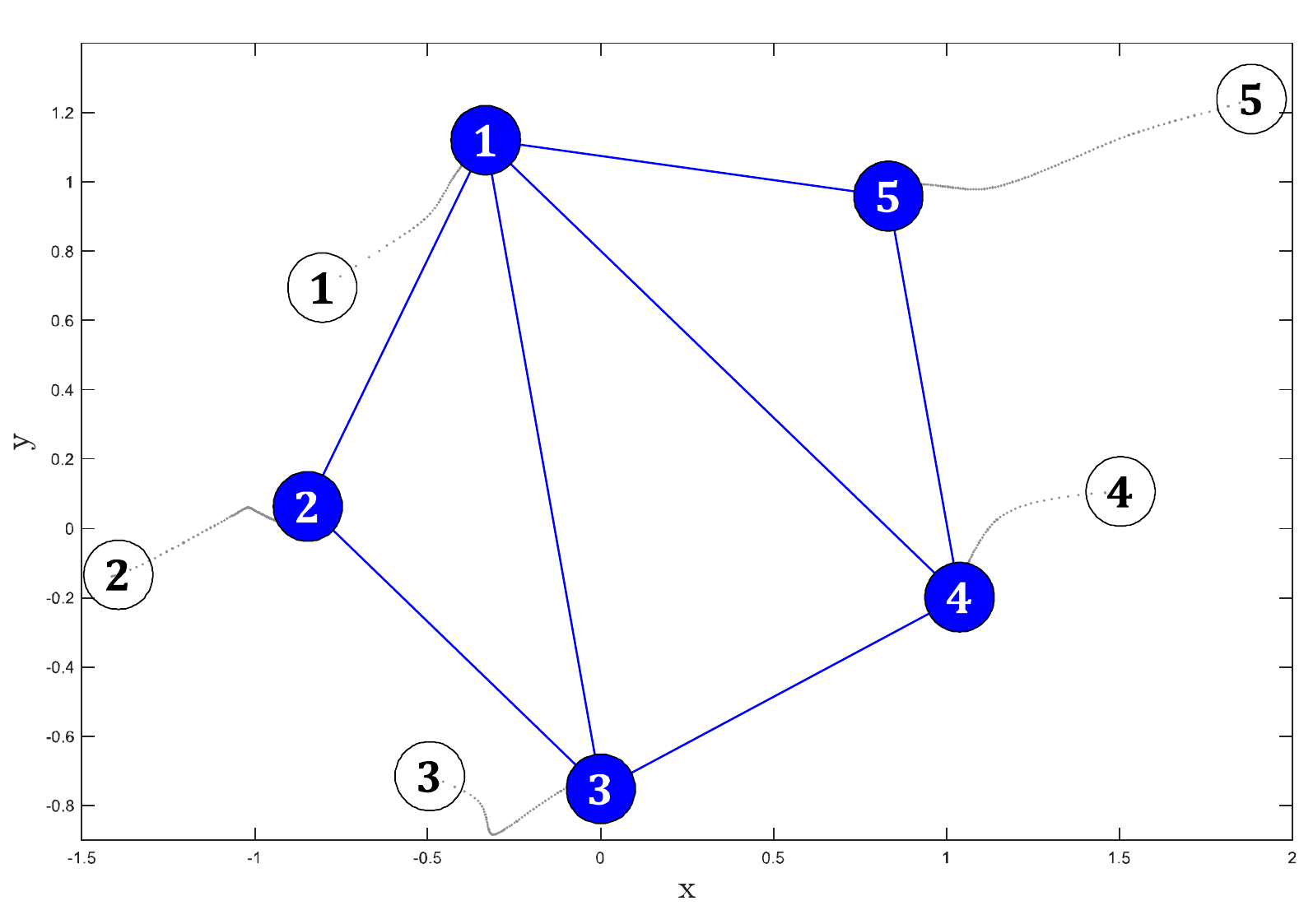}
		\end{subfigure}
		\begin{subfigure}[t]{0.236\textwidth}
			\centering
			\includegraphics[width=\textwidth]{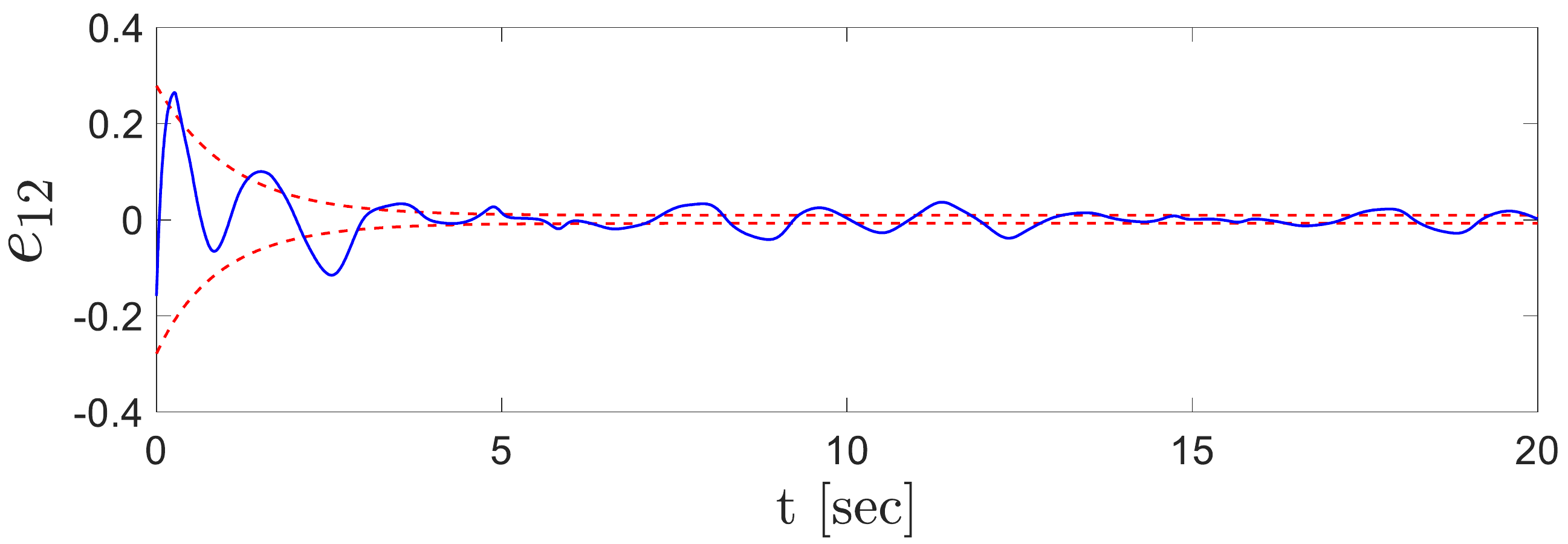}
		\end{subfigure}
		\begin{subfigure}[t]{0.236\textwidth}
			\centering
			\includegraphics[width=\textwidth]{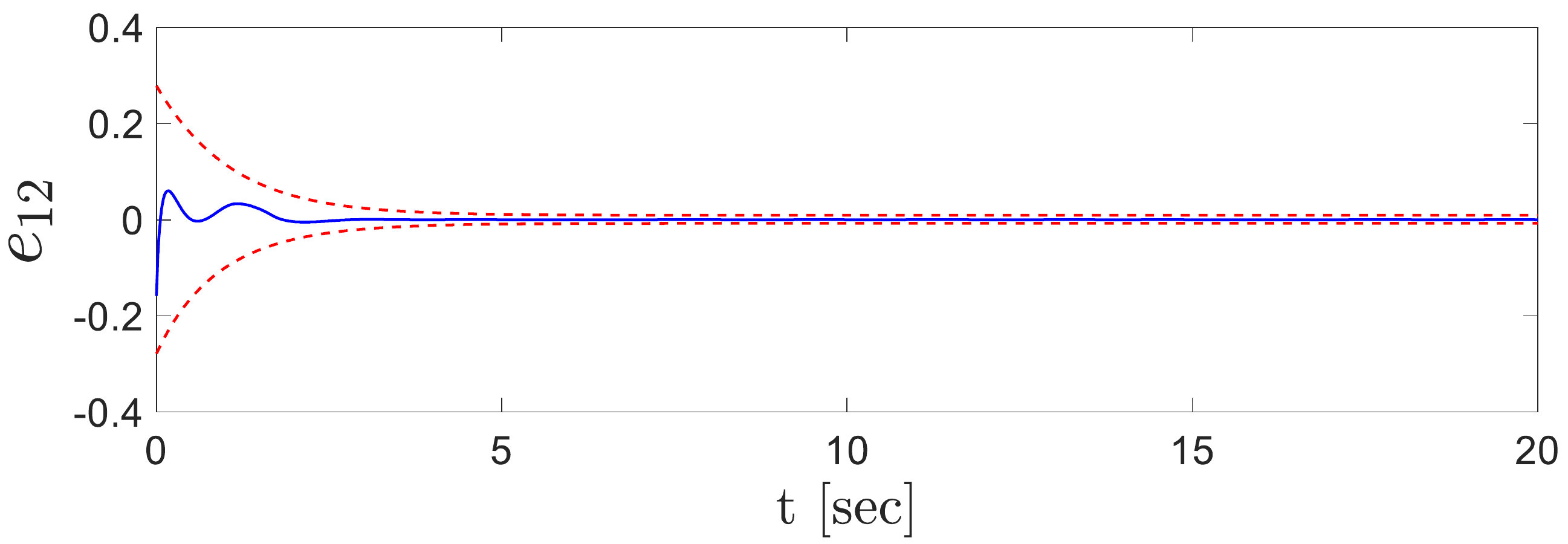}
		\end{subfigure}
		\begin{subfigure}[t]{0.236\textwidth}
			\centering
			\includegraphics[width=\textwidth]{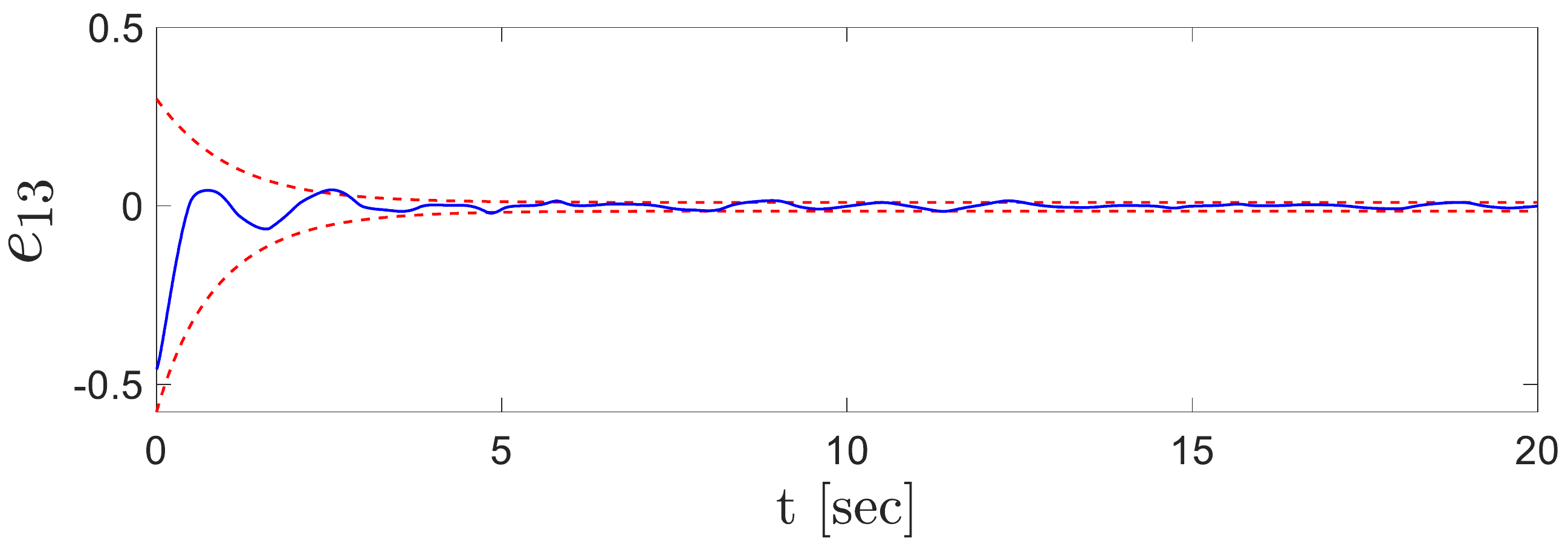}
		\end{subfigure}
		\begin{subfigure}[t]{0.236\textwidth}
			\centering
			\includegraphics[width=\textwidth]{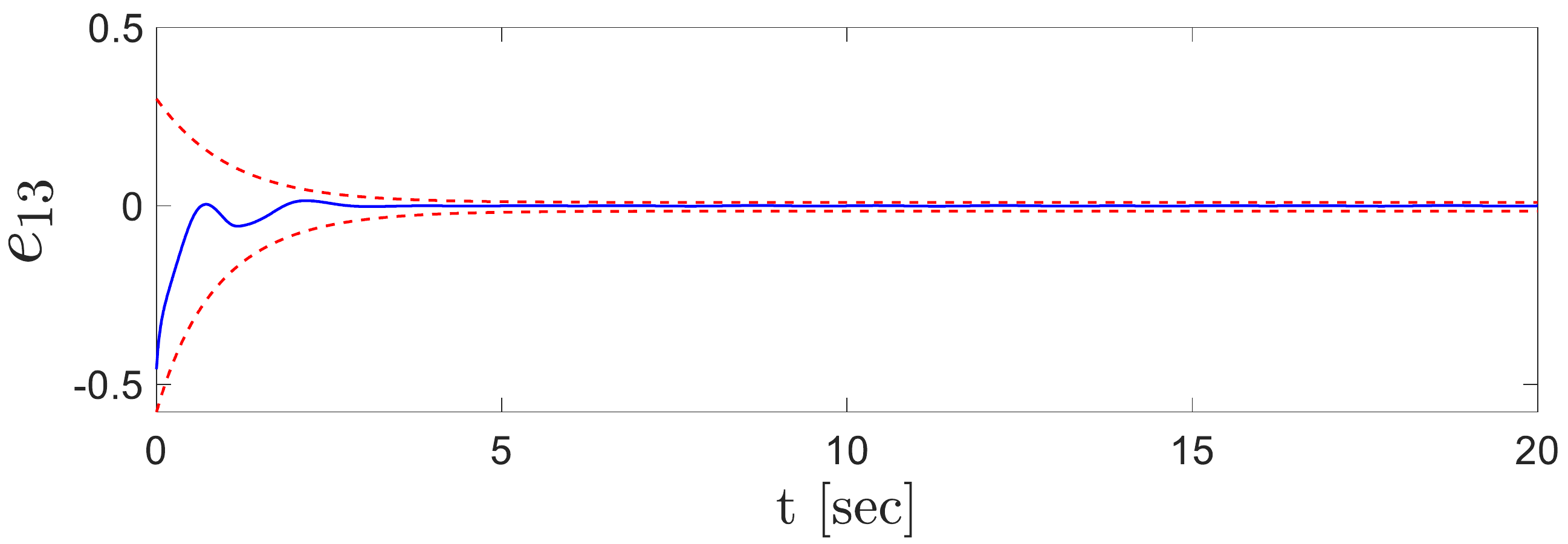}
		\end{subfigure}
		\begin{subfigure}[t]{0.236\textwidth}
			\centering
			\includegraphics[width=\textwidth]{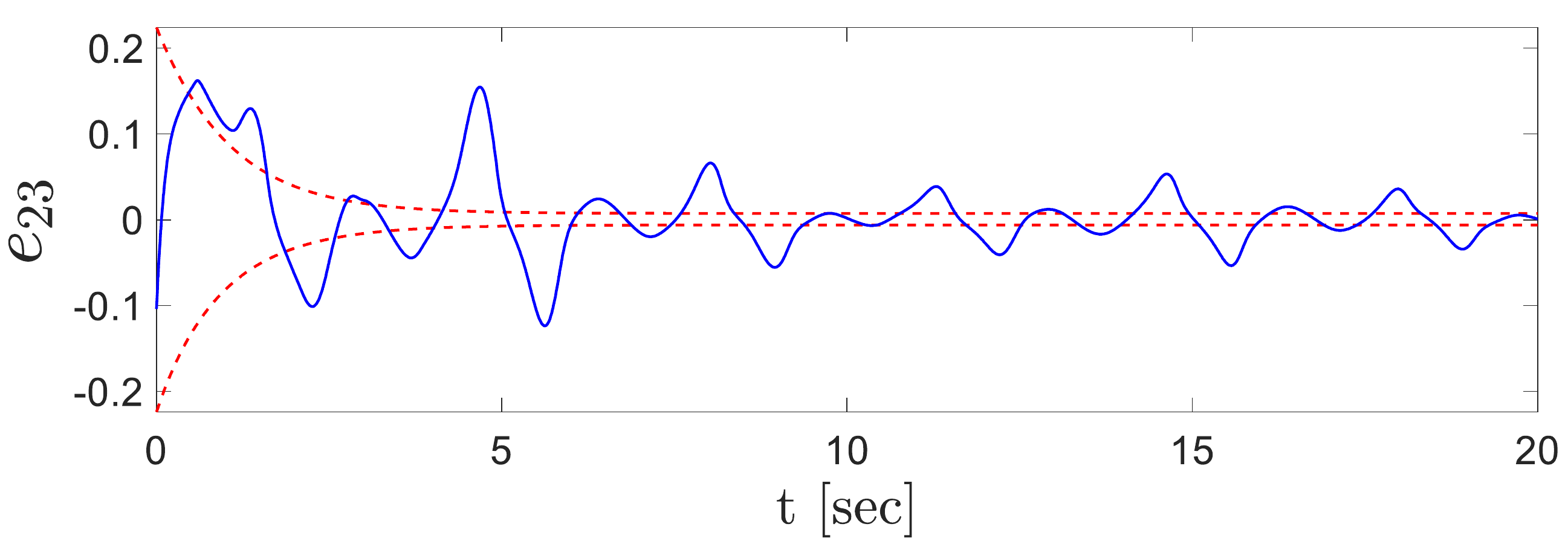}
		\end{subfigure}
		\begin{subfigure}[t]{0.236\textwidth}
			\centering
			\includegraphics[width=\textwidth]{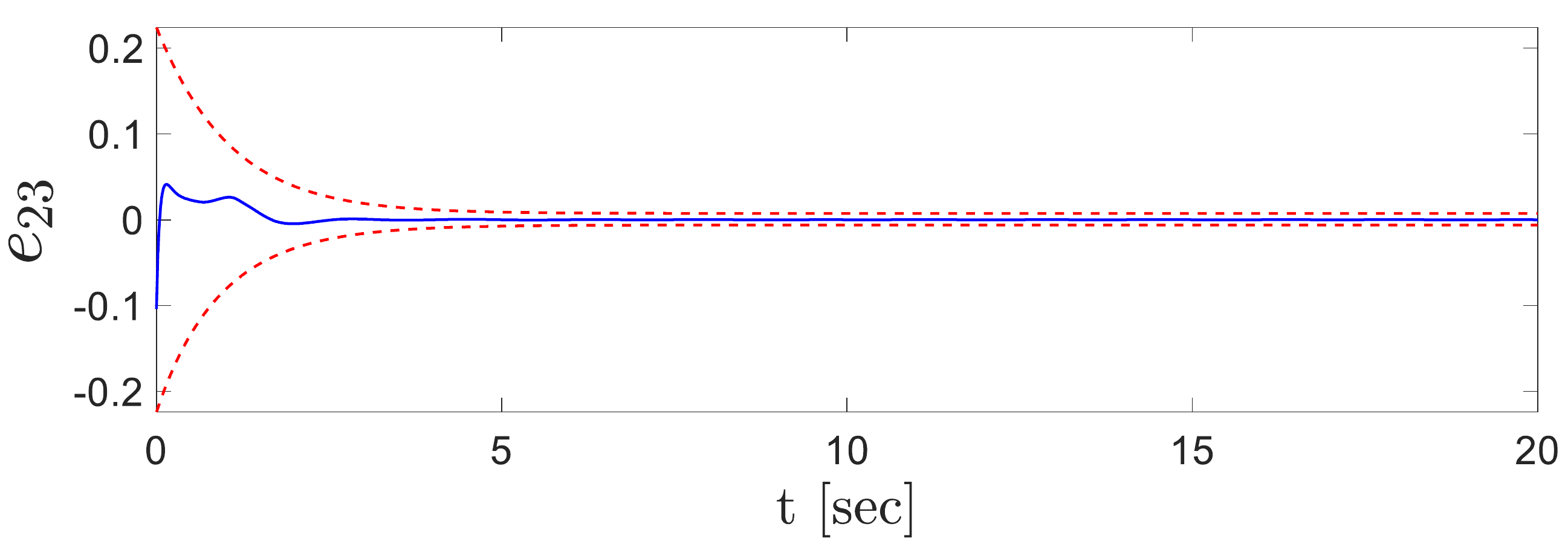}
		\end{subfigure}
		\begin{subfigure}[t]{0.236\textwidth}
			\centering
			\includegraphics[width=\textwidth]{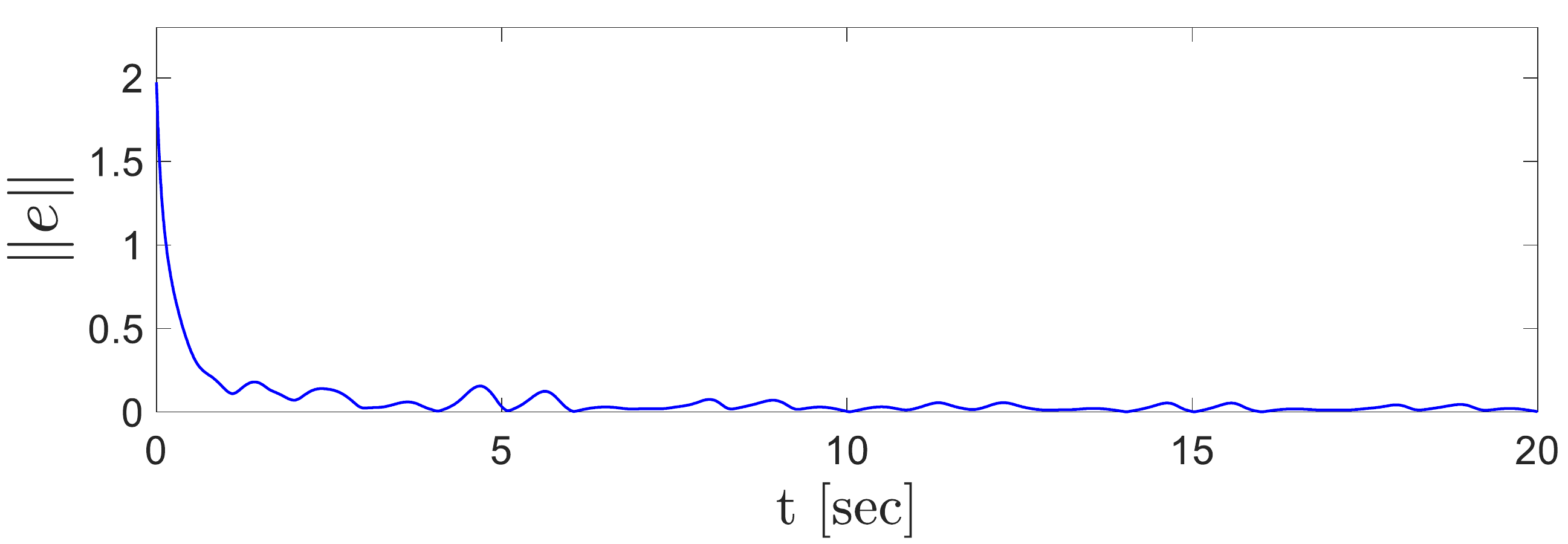}
			\caption{}
			\label{fig:robust_connv_d}
		\end{subfigure}
		\begin{subfigure}[t]{0.236\textwidth}
			\centering
			\includegraphics[width=\textwidth]{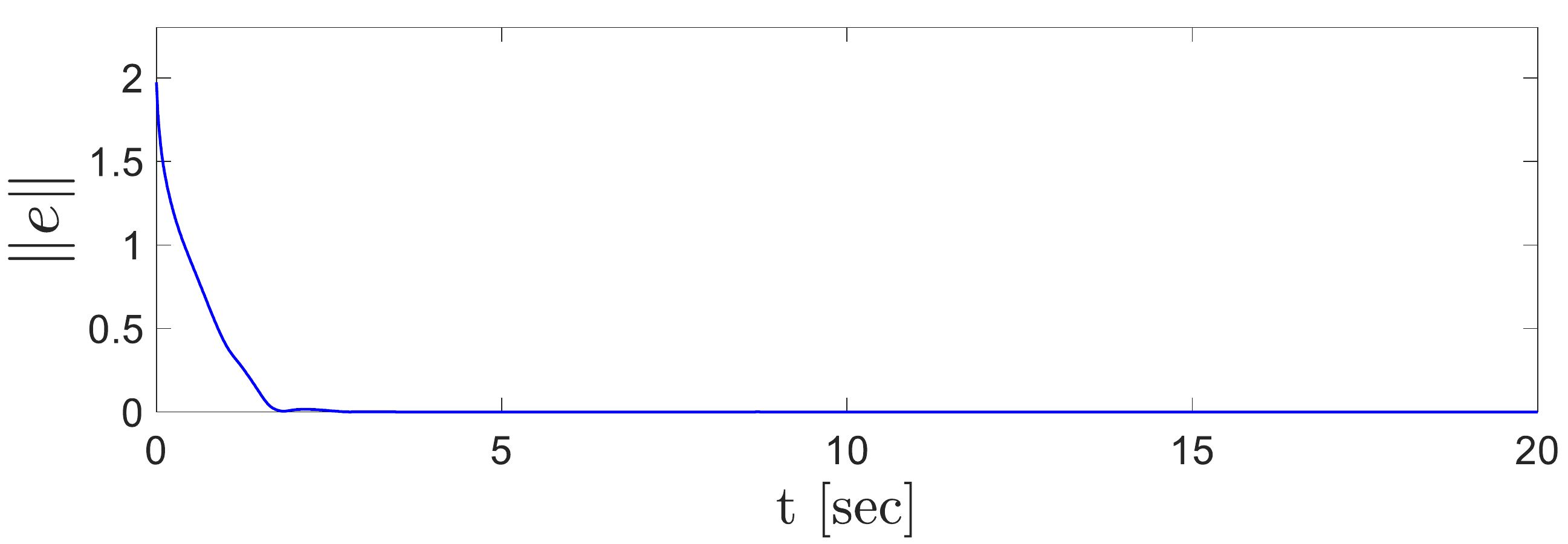}
			\caption{}
			\label{fig:PPC_d}
		\end{subfigure}
		\caption{Comparison of results using control protocols: (a) \eqref{eq:contr_normal_robus}, and (b) \eqref{eq:u_single} when $\delta(t)$ is considered as the external disturbance of the agents. The dashed lines represent the prescribed performance bounds.}
		\label{fig:all_d}
	\end{figure}

	\textbf{\textit{Case I)}}
	In this simulation, the external disturbance vector $\delta(t)$ is applied to the agents' dynamics for both control schemes \eqref{eq:contr_normal_robus} and \eqref{eq:u_single}. Fig~\ref{fig:robust_connv_d} and \ref{fig:PPC_d} show the agents' trajectories towards the desired formation after 20 seconds using  \eqref{eq:contr_normal_robus} and the proposed control protocol \eqref{eq:u_single}, respectively. Fig~\ref{fig:all_d} also depicts  some of the inter-agent distance errors evolution (namely, $e_{12}$, $e_{13}$, $e_{23}$) as well as the whole system's inter-agent distance error norm $\|e\|$. According to Fig~\ref{fig:all_d}, in this case both controllers were able to ensure $\mathcal{F}(t) \rightarrow \mathrm{Iso}(\mathcal{F}^\ast)$ in the presence of external disturbances. However, as Fig~\ref{fig:robust_connv_d} depicts, \eqref{eq:contr_normal_robus} cannot guarantee the errors to remain within the PPB and as a result it is not as robust as \eqref{eq:u_single} with respect to formation distortions. 
	This can be clearly understood by noticing the evolution of $e_{23}$ and $\|e\|$ in Fig~\ref{fig:robust_connv_d}.
	
\begin{figure}[tbp]
		\flushleft
		\begin{subfigure}[t]{0.235\textwidth}
			\centering
			\includegraphics[width=\textwidth]{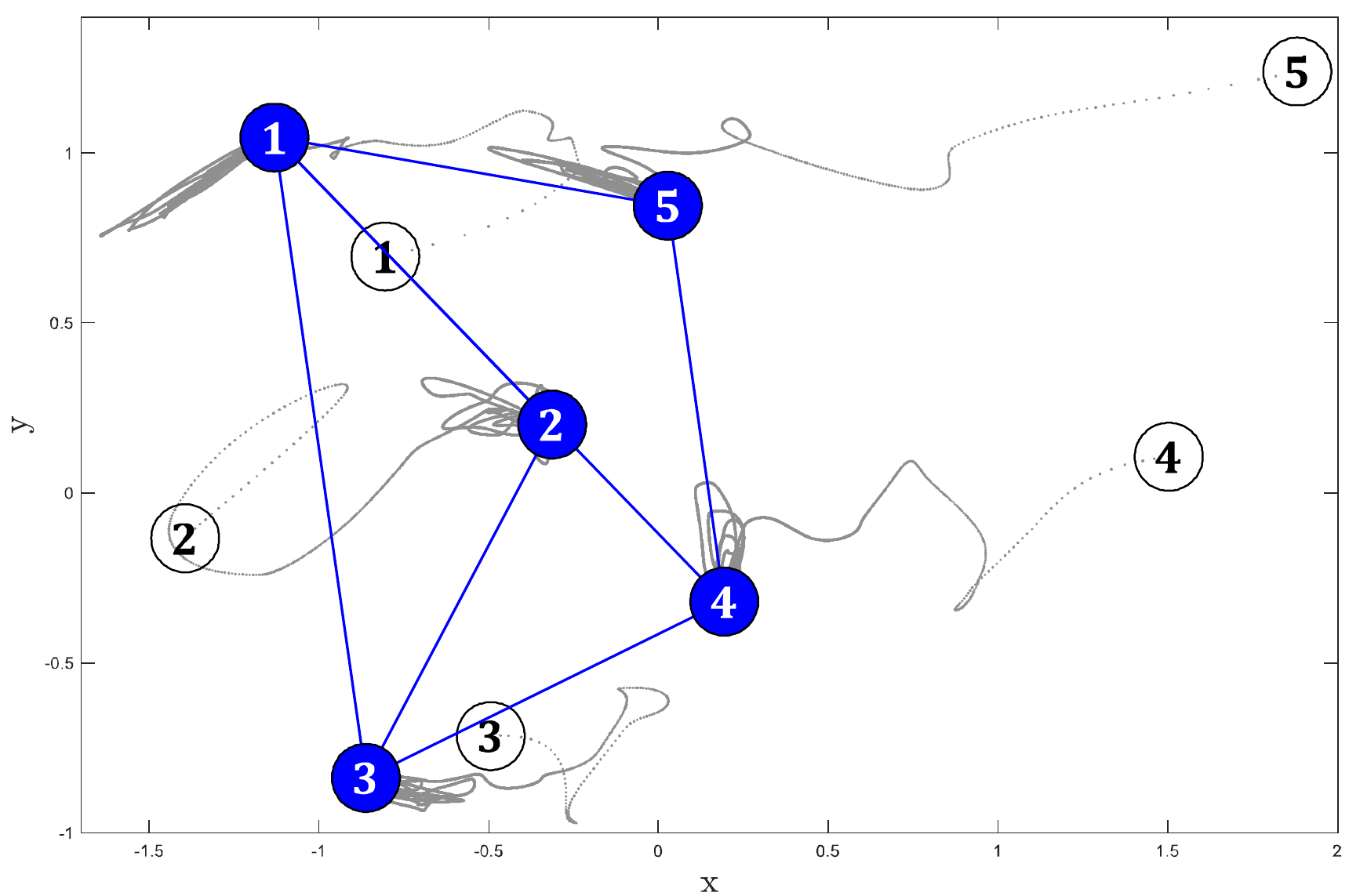}
		\end{subfigure}
		\begin{subfigure}[t]{0.235\textwidth}
			\centering
			\includegraphics[width=\textwidth]{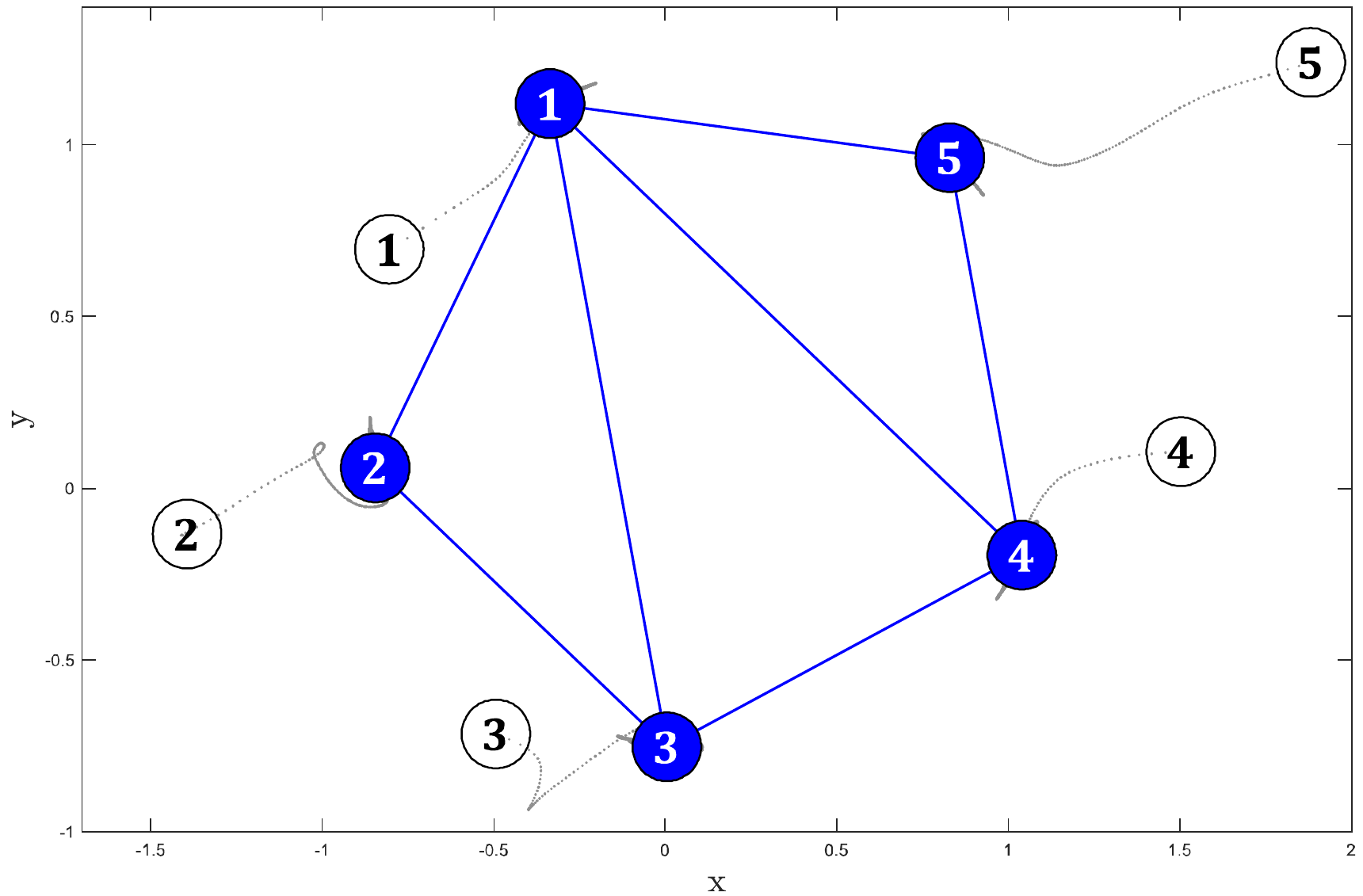}
		\end{subfigure}
		\begin{subfigure}[t]{0.236\textwidth}
			\centering
			\includegraphics[width=\textwidth]{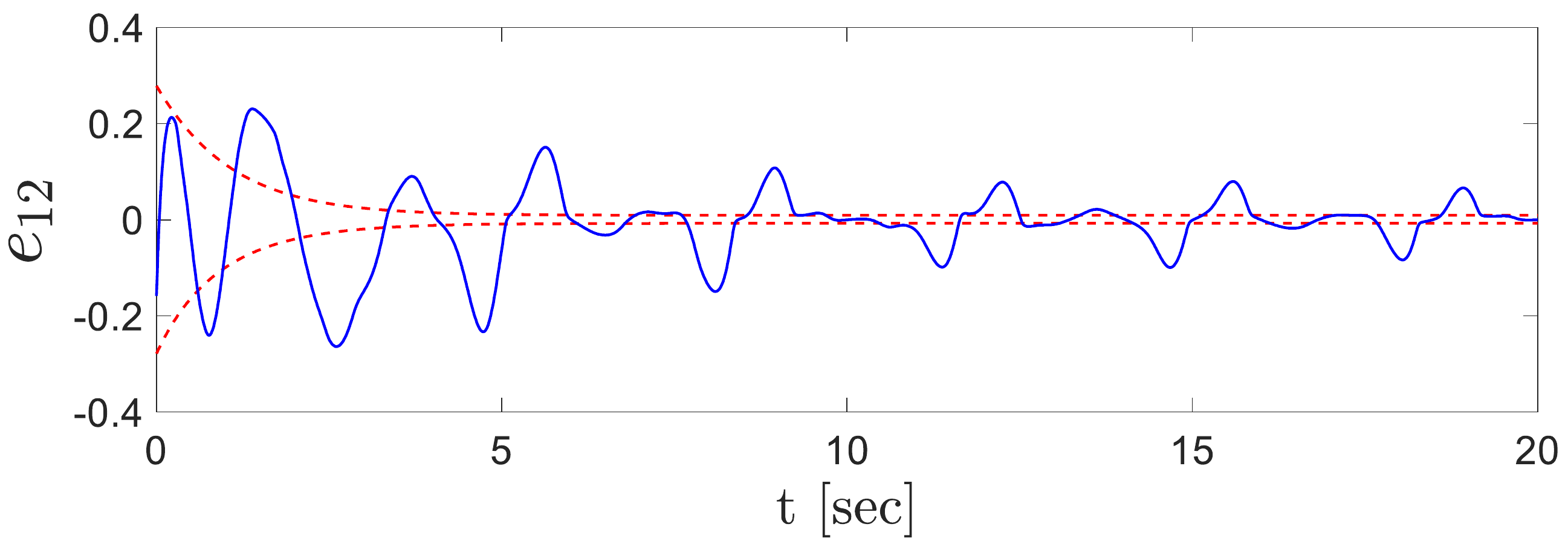}
		\end{subfigure}
		\begin{subfigure}[t]{0.236\textwidth}
			\centering
			\includegraphics[width=\textwidth]{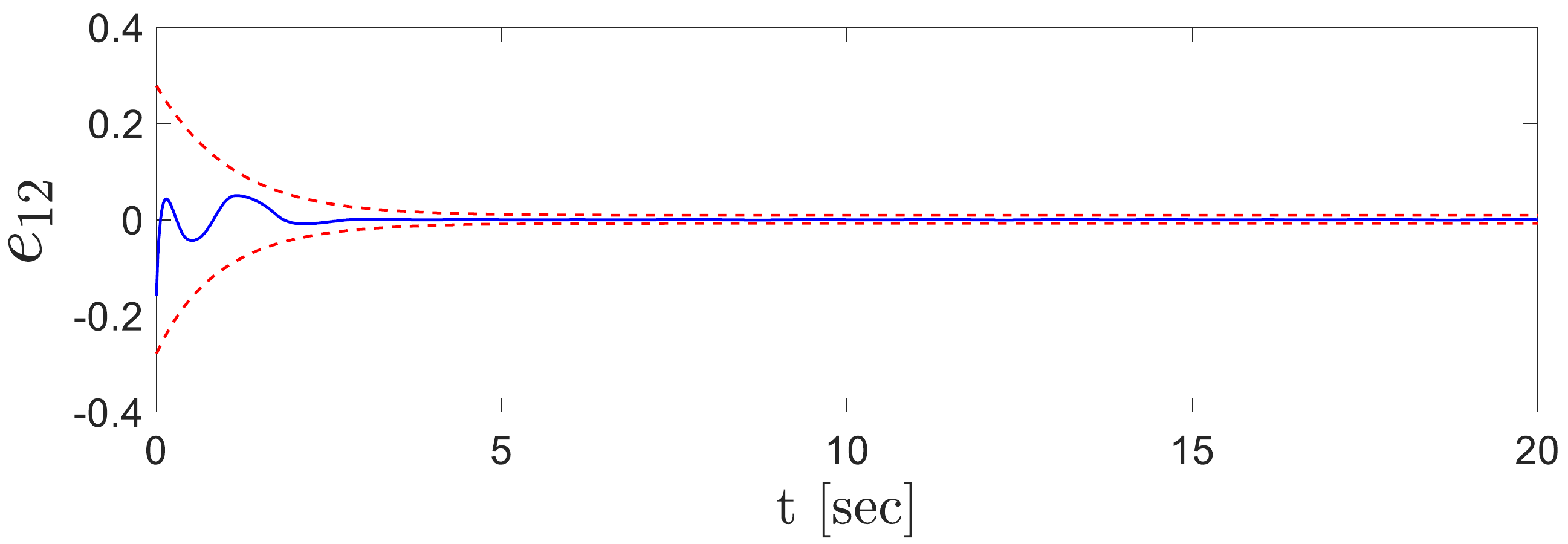}
		\end{subfigure}
		\begin{subfigure}[t]{0.236\textwidth}
			\centering
			\includegraphics[width=\textwidth]{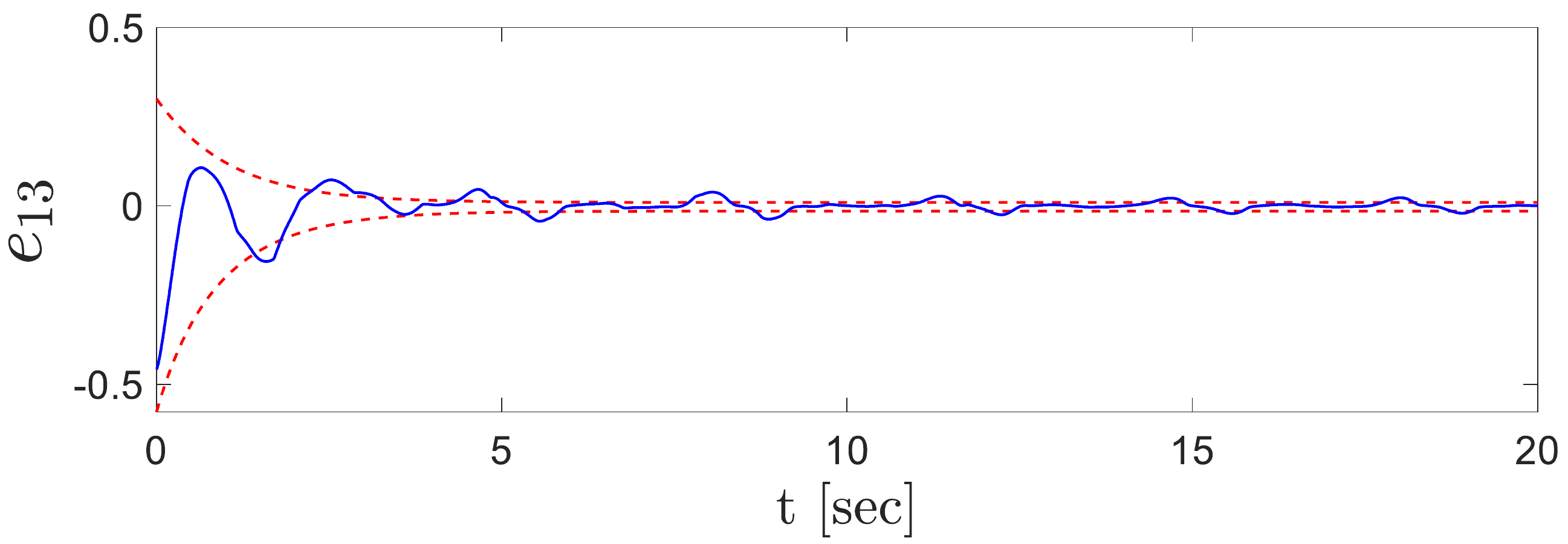}
		\end{subfigure}
		\begin{subfigure}[t]{0.236\textwidth}
			\centering
			\includegraphics[width=\textwidth]{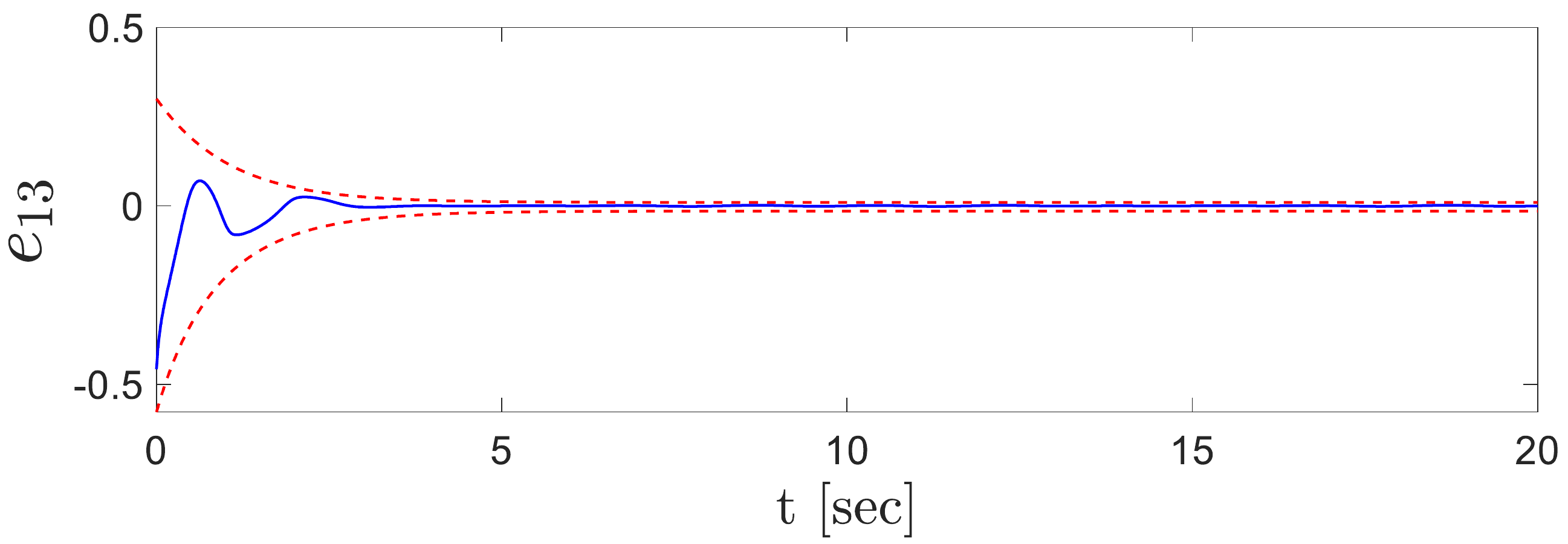}
		\end{subfigure}
		\begin{subfigure}[t]{0.236\textwidth}
			\centering
			\includegraphics[width=\textwidth]{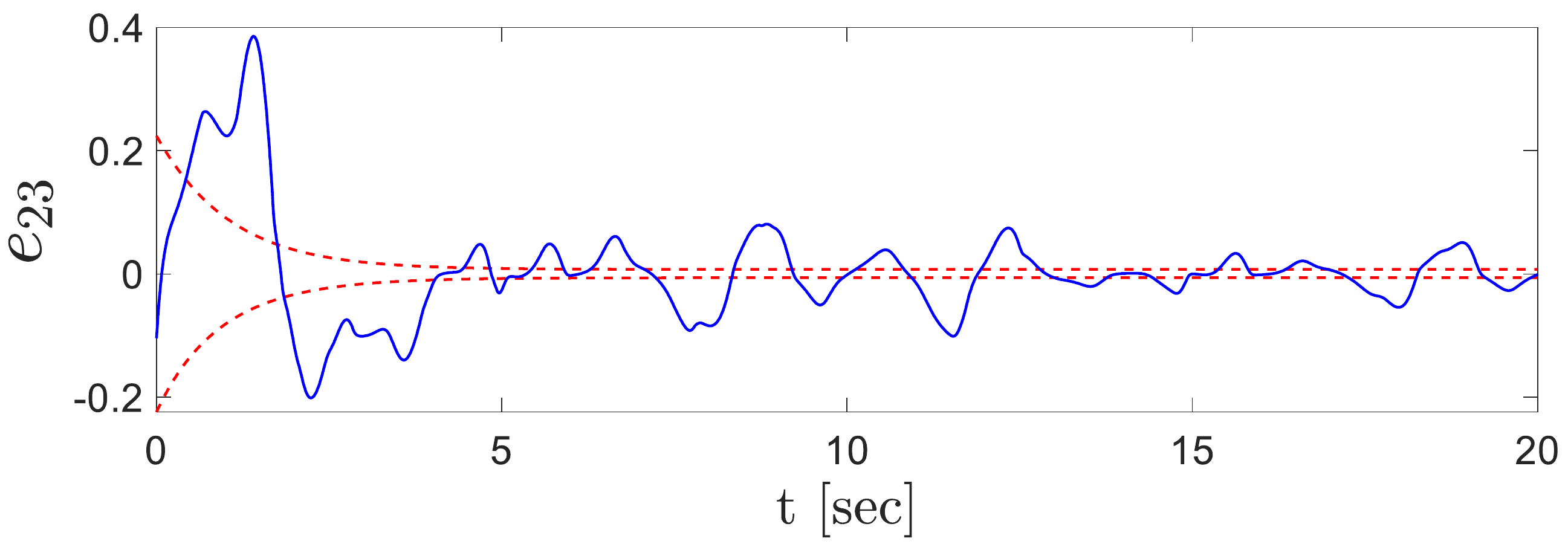}
		\end{subfigure}
		\begin{subfigure}[t]{0.236\textwidth}
			\centering
			\includegraphics[width=\textwidth]{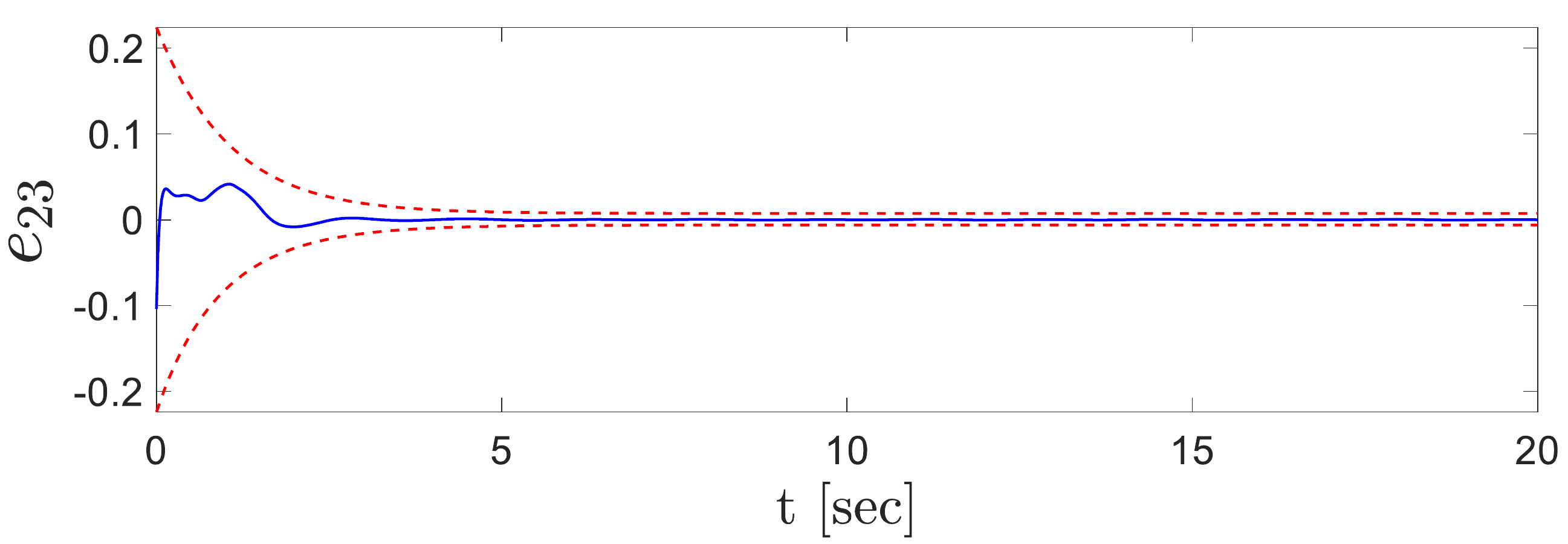}
		\end{subfigure}
		\begin{subfigure}[t]{0.236\textwidth}
			\centering
			\includegraphics[width=\textwidth]{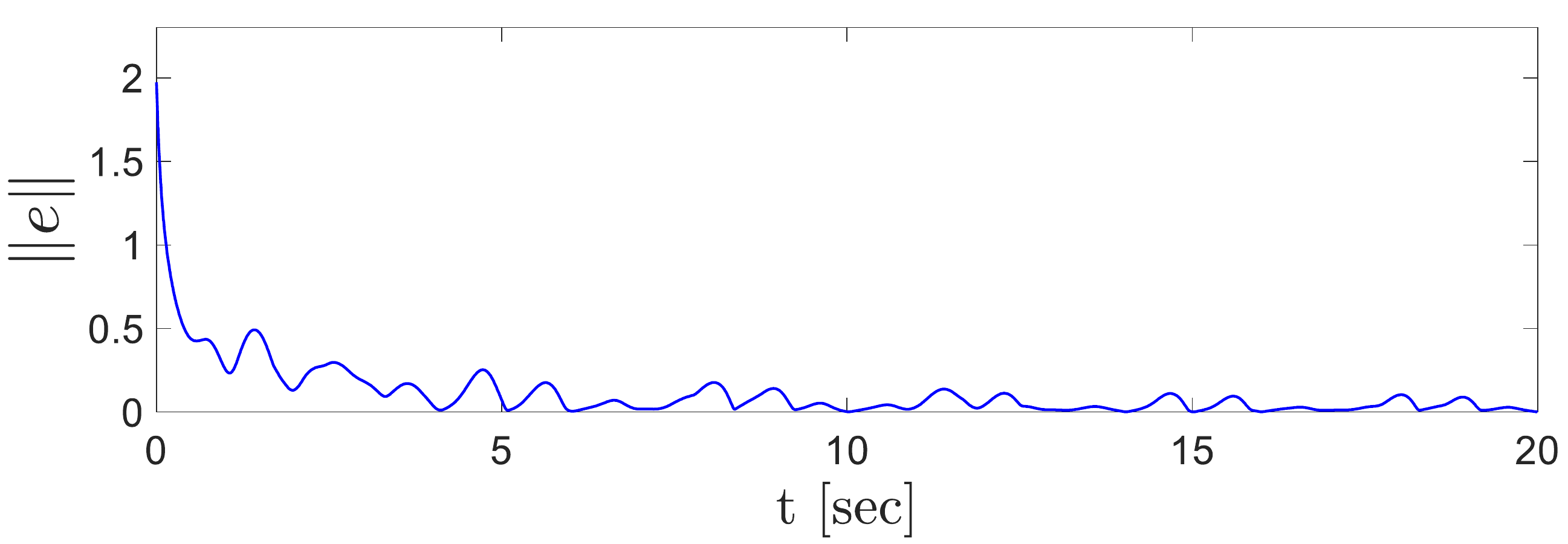}
			\caption{}
			\label{fig:robust_connv_2d}
		\end{subfigure}
		\begin{subfigure}[t]{0.236\textwidth}
			\centering
			\includegraphics[width=\textwidth]{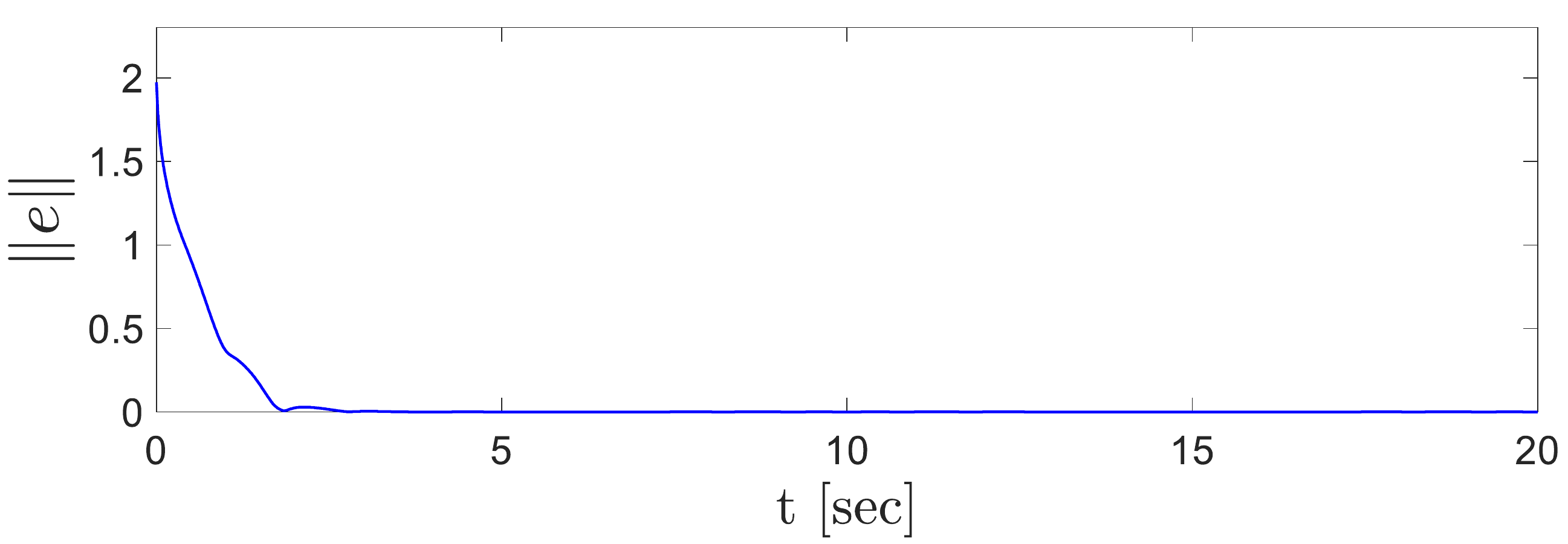}
			\caption{}
			\label{fig:PPC_2d}
		\end{subfigure}
		\caption{Comparison of results using control protocols: (a) \eqref{eq:contr_normal_robus}, and (b) \eqref{eq:u_single} when $2\delta(t)$ is considered as the external disturbance of the agents.}
		\label{fig:all_2d}
	\end{figure}

	\textbf{\textit{Case II)}} The disturbance magnitude is doubled. As Fig~\ref{fig:robust_connv_2d} illustrates, although the control law \eqref{eq:contr_normal_robus} is still able to drive the distance errors to a small neighborhood of zero (see $\|e\|$ evolution), it is not able to ensure  convergence to the desired shape. As stated in Remark \ref{rem:robus_to_Distur}, this stems from the fact that it is not able to guarantee predefined quality of the  transient response of the actual formation. Indeed, during the transient phase the actual formation distorts massively, forcing it to get closer to one of the shapes lying inside $\mathrm{Amb}(\mathcal{F}^\ast)$ before the controller dominates the effects of external disturbances. Fig~\ref{fig:PPC_2d} verifies that the proposed prescribed performance distance based formation control law still ensures convergence to the correct shape.
	
\begin{figure}[tbp]
		\flushleft
		\begin{subfigure}[t]{0.234\textwidth}
			\centering
			\includegraphics[width=\textwidth]{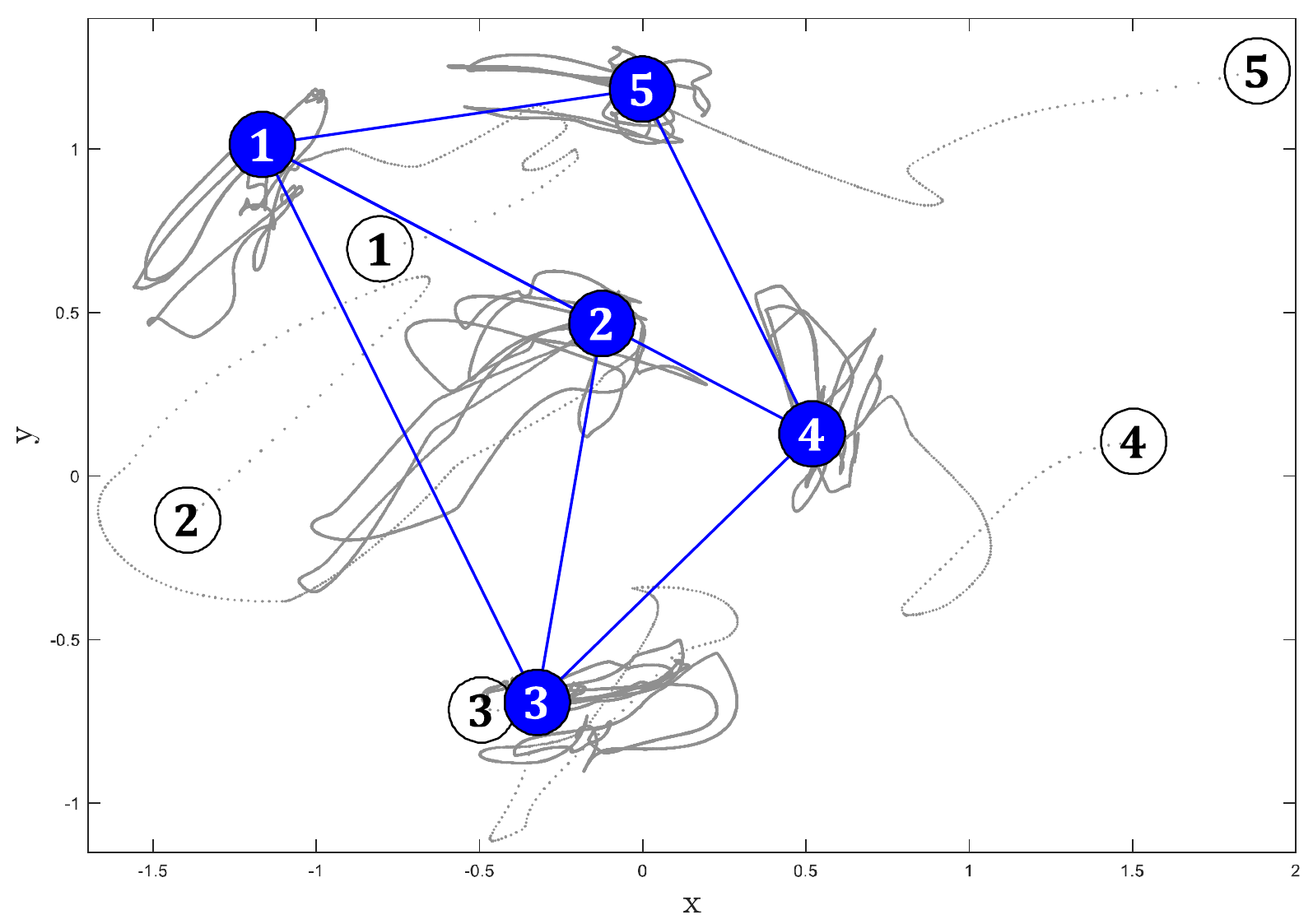}
		\end{subfigure}
		\begin{subfigure}[t]{0.236\textwidth}
			\centering
			\includegraphics[width=\textwidth]{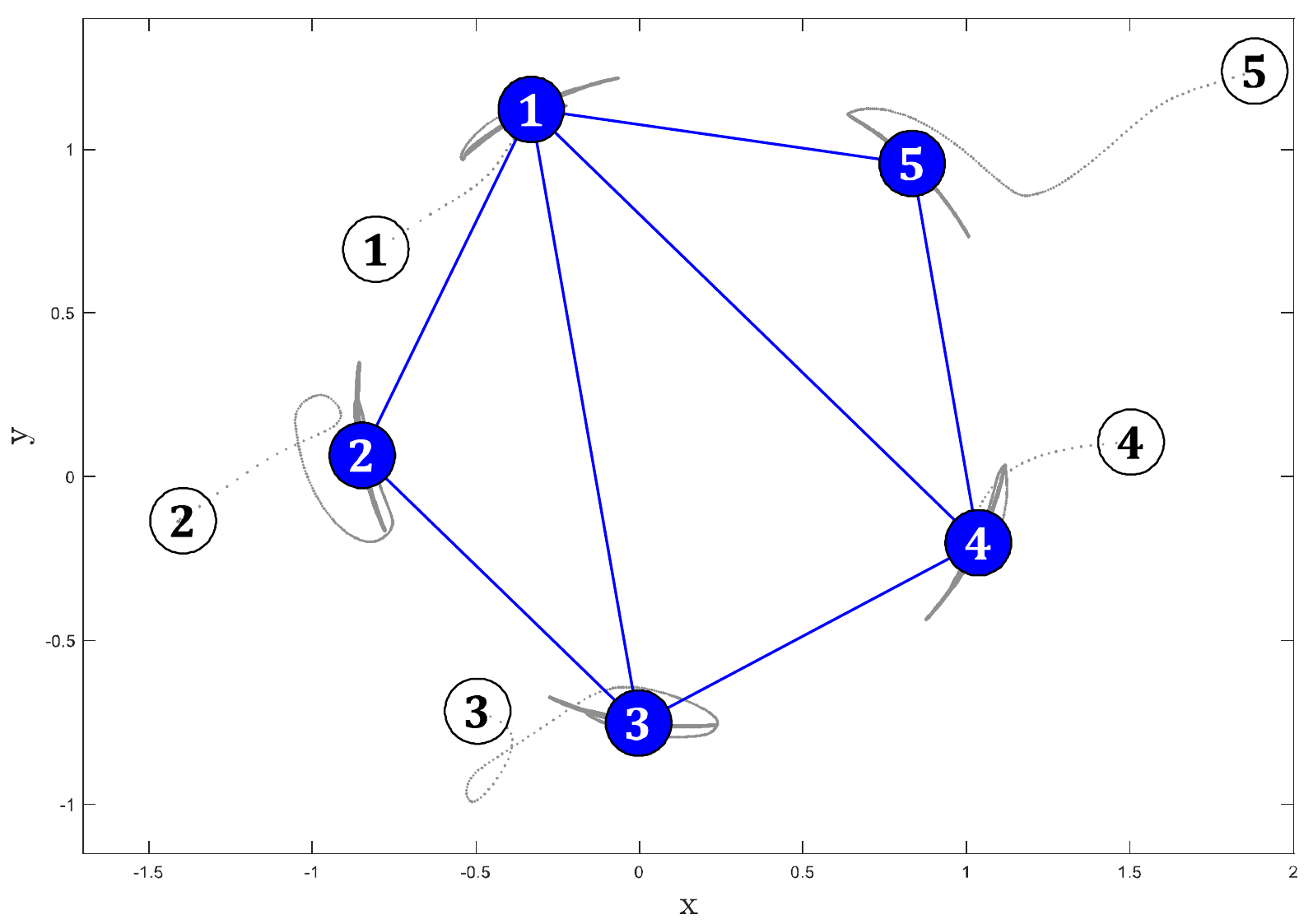}
		\end{subfigure}
		\begin{subfigure}[t]{0.236\textwidth}
			\centering
			\includegraphics[width=\textwidth]{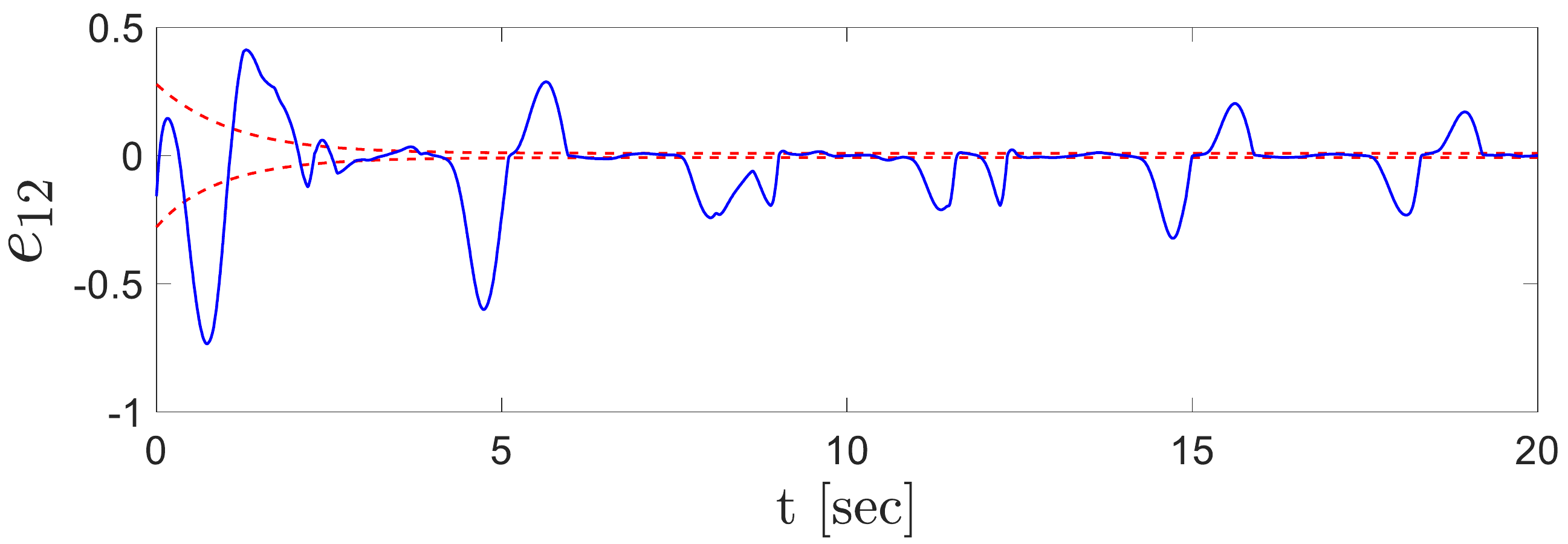}
		\end{subfigure}
		\begin{subfigure}[t]{0.236\textwidth}
			\centering
			\includegraphics[width=\textwidth]{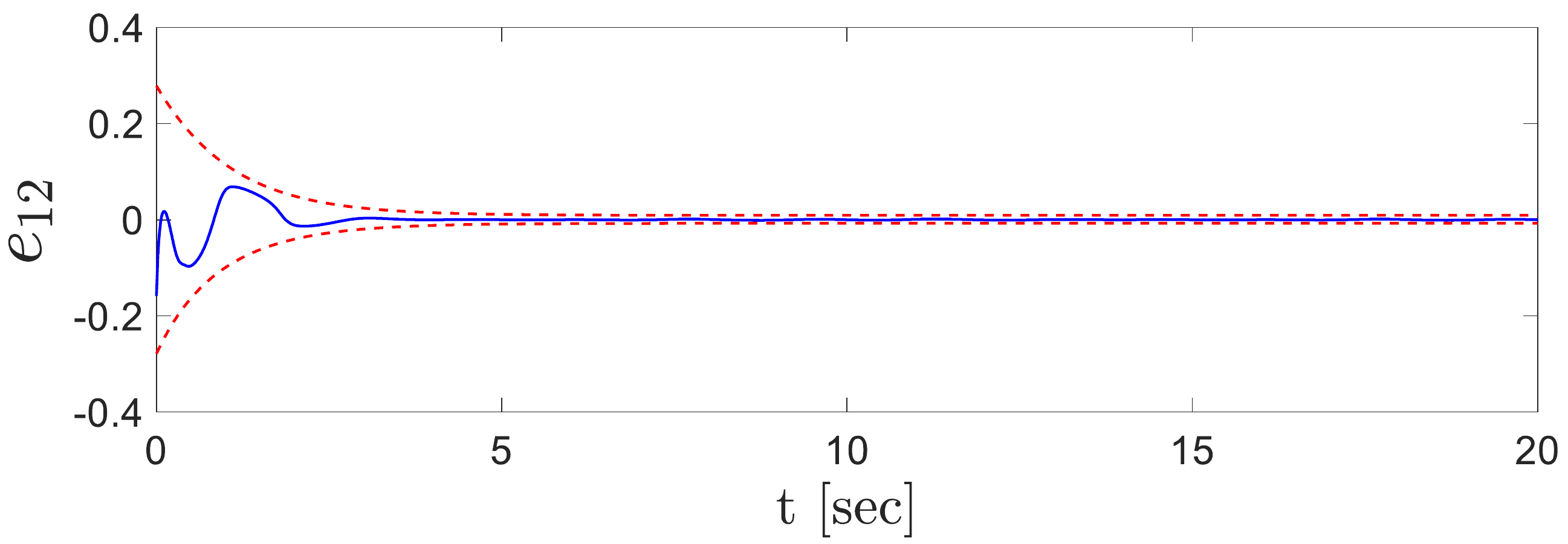}
		\end{subfigure}
		\begin{subfigure}[t]{0.236\textwidth}
			\centering
			\includegraphics[width=\textwidth]{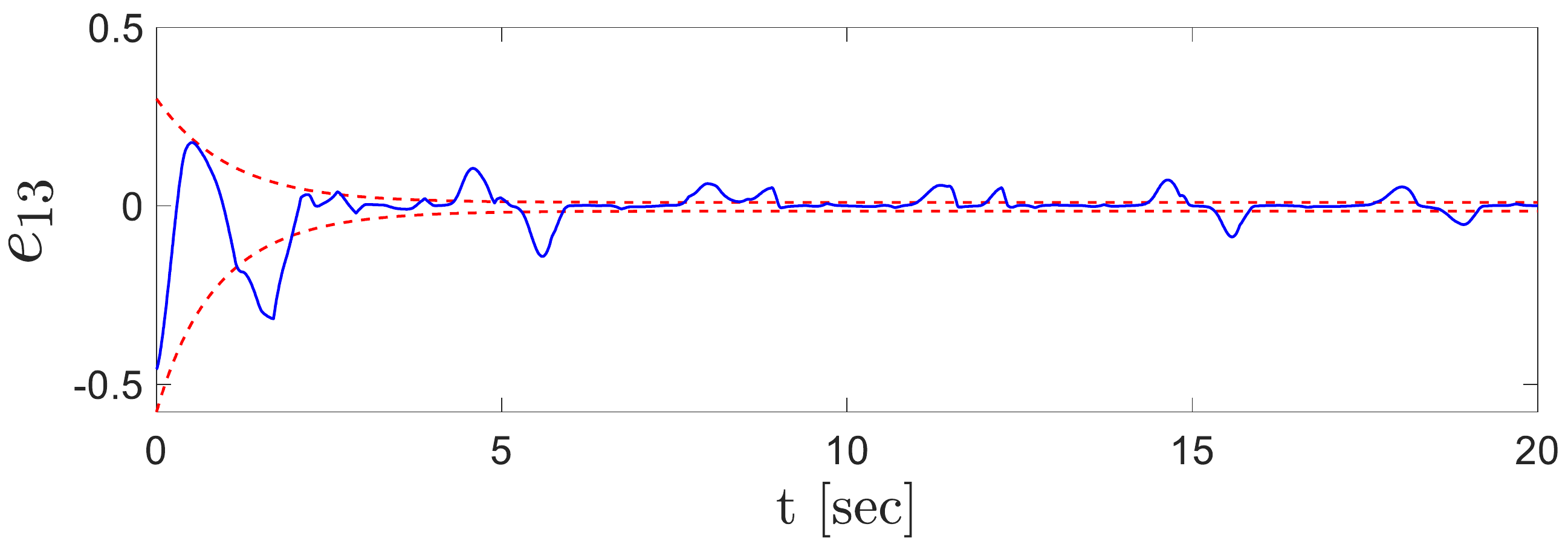}
		\end{subfigure}
		\begin{subfigure}[t]{0.236\textwidth}
			\centering
			\includegraphics[width=\textwidth]{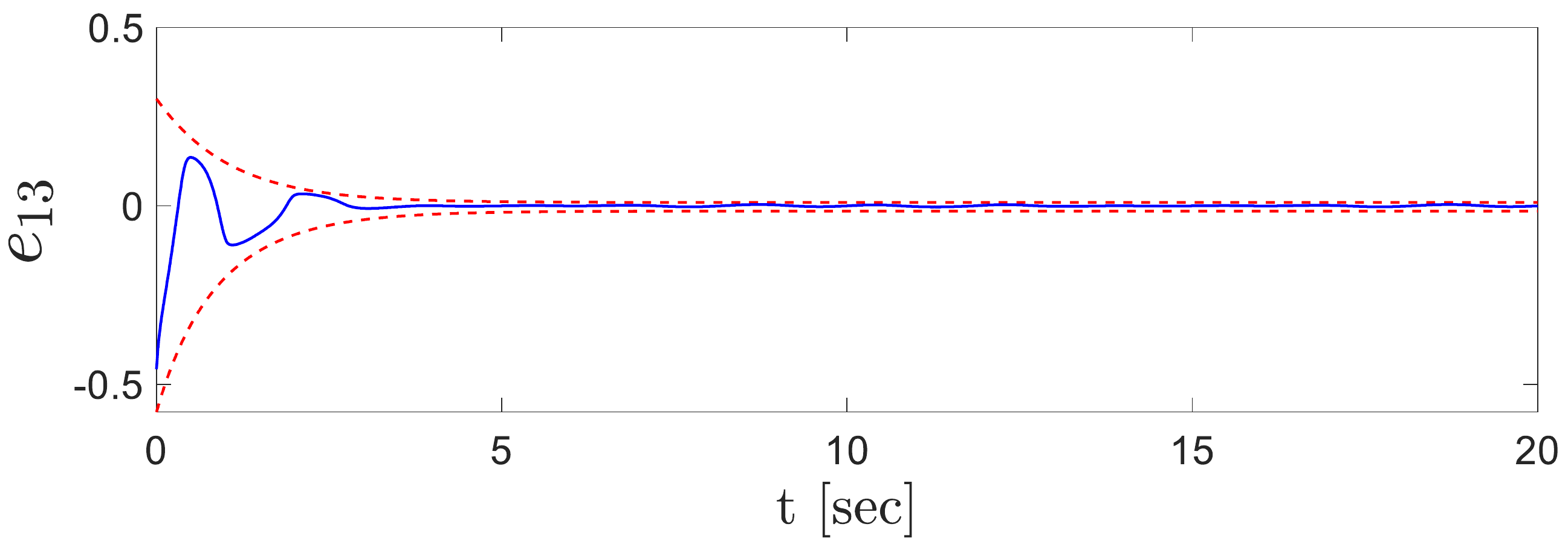}
		\end{subfigure}
		\begin{subfigure}[t]{0.236\textwidth}
			\centering
			\includegraphics[width=\textwidth]{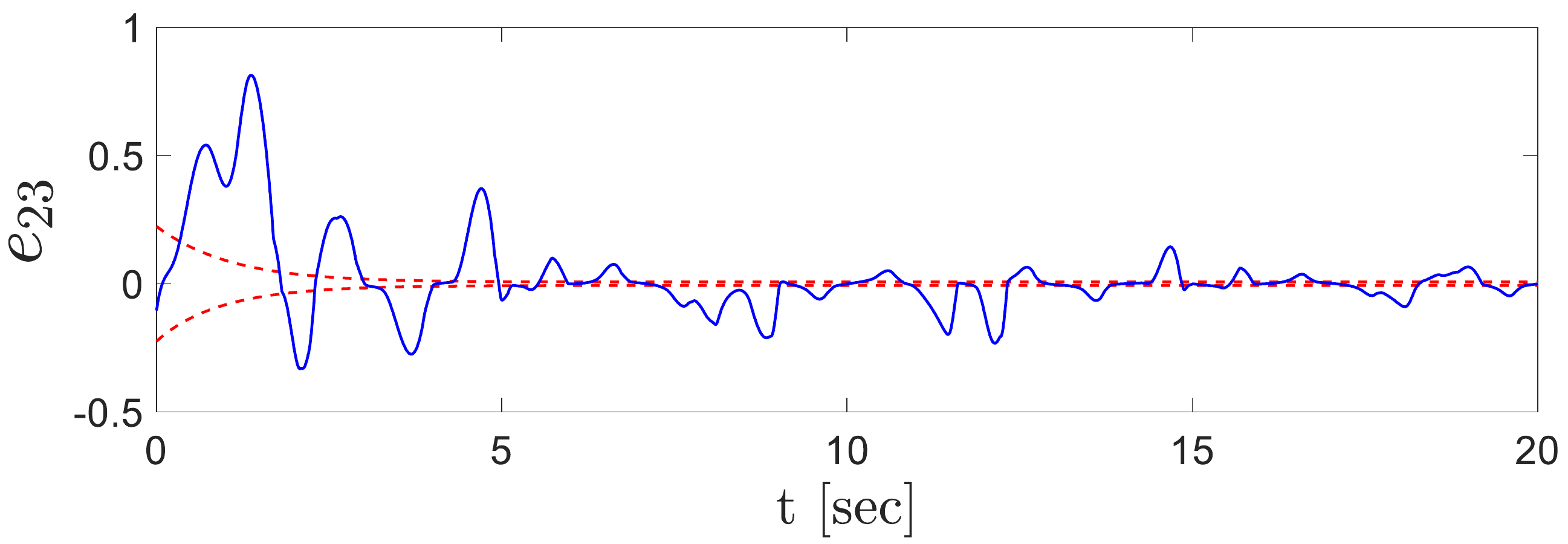}
		\end{subfigure}
		\begin{subfigure}[t]{0.236\textwidth}
			\centering
			\includegraphics[width=\textwidth]{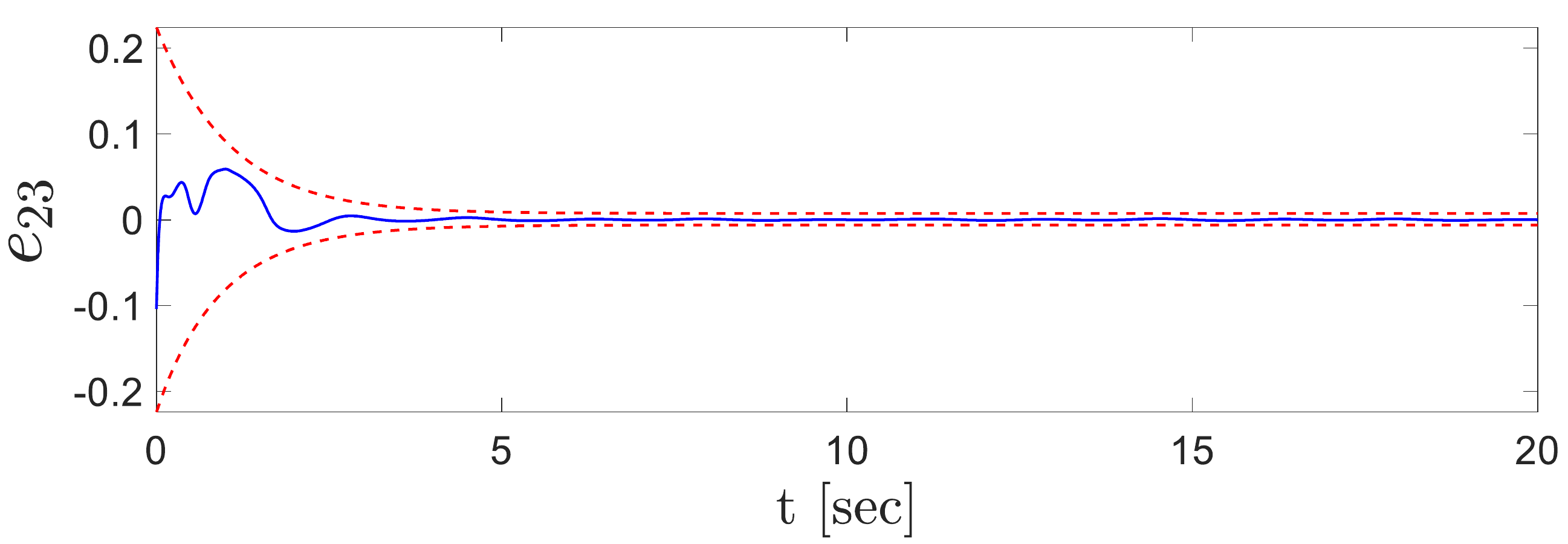}
		\end{subfigure}
		\begin{subfigure}[t]{0.236\textwidth}
			\centering
			\includegraphics[width=\textwidth]{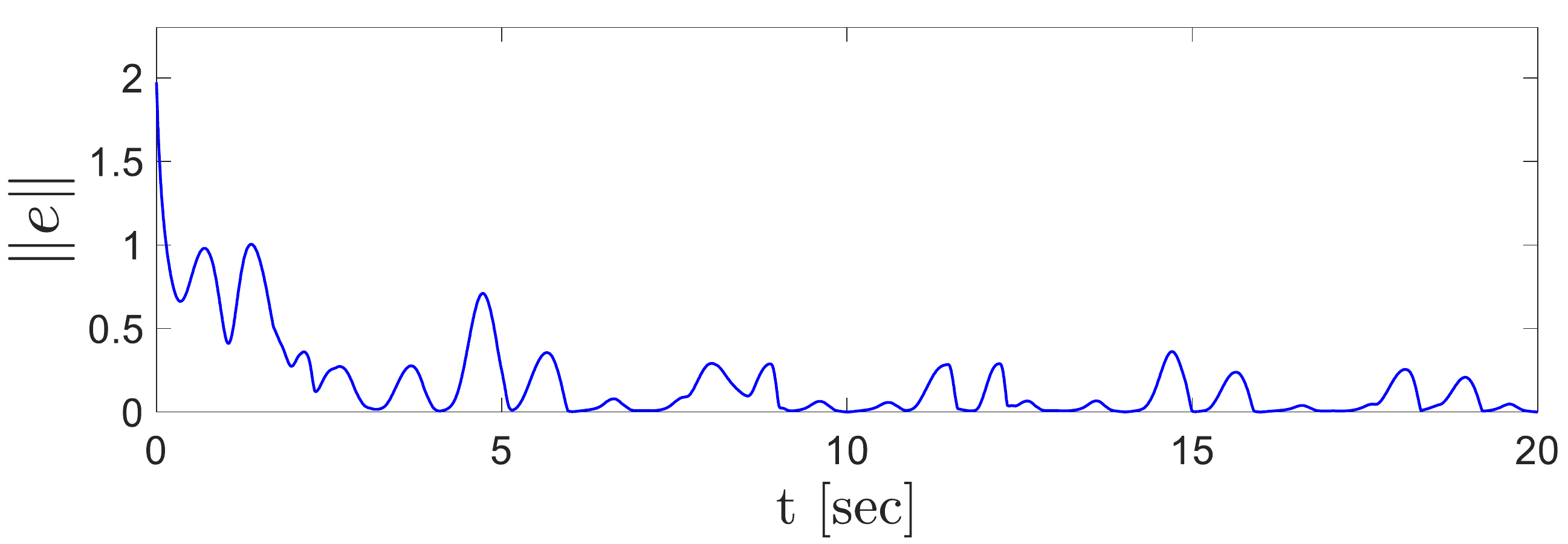}
			\caption{}
		\end{subfigure}
		\begin{subfigure}[t]{0.236\textwidth}
			\centering
			\includegraphics[width=\textwidth]{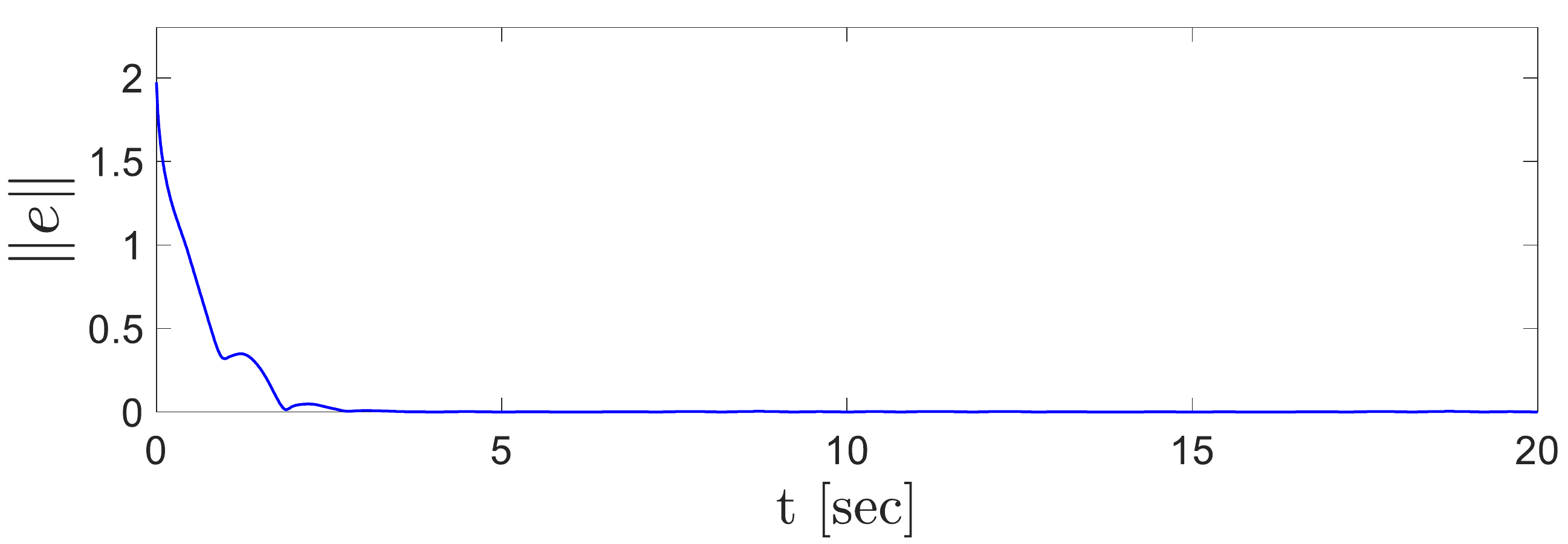}
			\caption{}
		\end{subfigure}
		\caption{Comparison of results using control protocols: (a) \eqref{eq:contr_normal_robus}, and (b) \eqref{eq:u_single} when $4\delta(t)$ is considered as the external disturbance of the agents.}
		\label{fig:all_4d}
	\end{figure}

	\textbf{\textit{Case III)}} Finally, the external disturbance was increased up to $4\delta(t)$. Similarly to the results in Fig.~\ref{fig:all_4d}, the control protocol \eqref{eq:contr_normal_robus} is not able to ensure convergence to the desired shape, whereas it is still able to stabilize inter-agent distance errors around $e=0$. On the other hand, \eqref{eq:u_single} not only ensures convergence to the correct shape but also shows an outstanding desirable performance, dealing with the high amplitude external disturbances that try to distort the formation shape all the time. Notice that, such performance is achieved without modifying the controller's gains and is purely prescribed by the performance functions in the controller design procedure.
	
	Conclusively, notice that during the transient response phase, the increased robustness against formation distortions induced by the external disturbances is a unique property of the proposed control law \eqref{eq:u_single} that ensures convergence to the desired shape thanks to the PPC method. Certainly, by increasing the control gains in \eqref{eq:contr_normal_robus}, one may be able to improve the transient and steady state behavior of the closed-loop system in order to avoid/reduce risk of $\mathcal{F}(t) \rightarrow \mathrm{Amb}(\mathcal{F}^\ast)$. However, this approach inevitably leads to a tedious trial and error procedure to achieve satisfactory performance. On the contrary, in the proposed scheme the control gains selection in the proposed scheme is significantly simplified. It is also noteworthy to mention that the control law \eqref{eq:u_single} also ensures connectivity maintenance and collision avoidance among neighboring agents even when agents are subject to high external disturbances, which cannot be guaranteed by a conventional robust control scheme.
	
	\subsection{Formation Maneuvering}
	
\begin{figure}[tbp]
		\centering
		\includegraphics[width=0.48\textwidth]{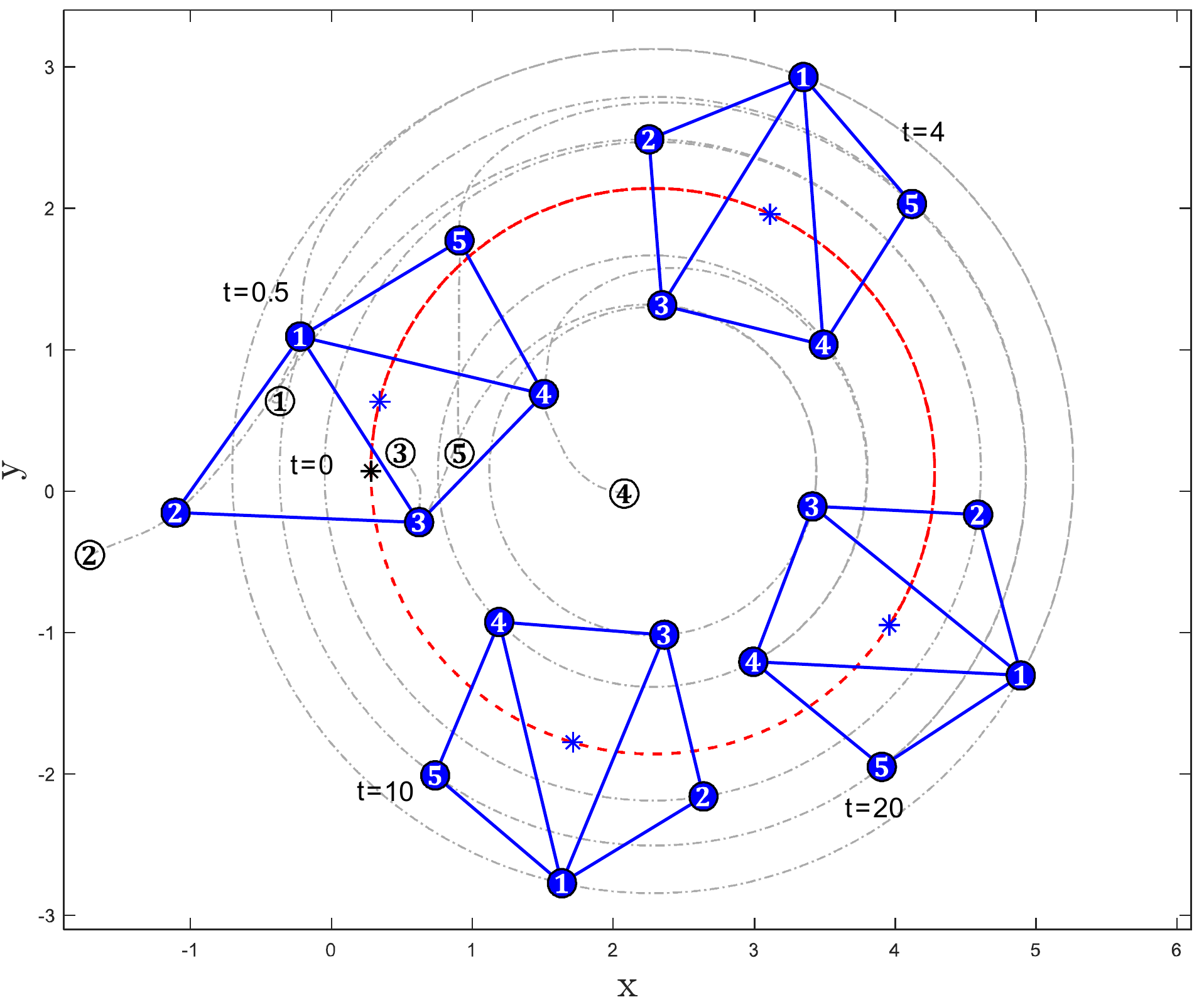}
		\caption{Agents trajectories towards the target shape as well as tracking the desired centroid velocity}
		\label{pic:maneu}
	\end{figure}

\begin{figure}[tbp]
		\flushleft
		\begin{subfigure}[t]{0.236\textwidth}
			\centering
			\includegraphics[width=\textwidth]{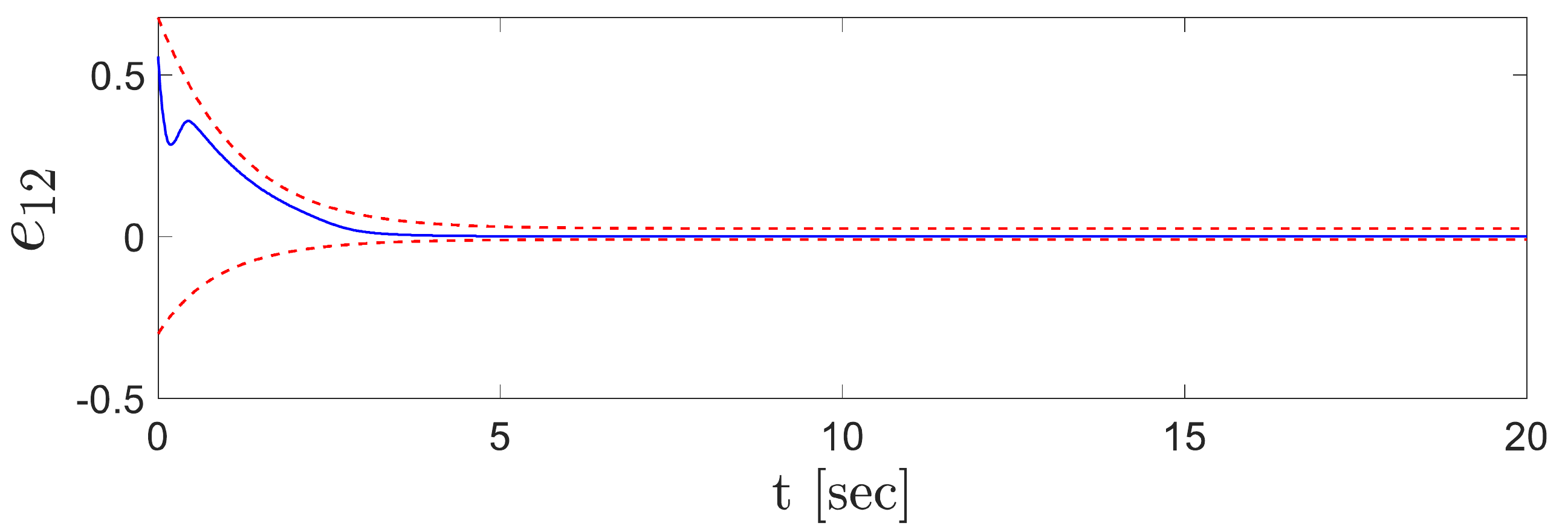}
		\end{subfigure}
		\begin{subfigure}[t]{0.236\textwidth}
			\centering
			\includegraphics[width=\textwidth]{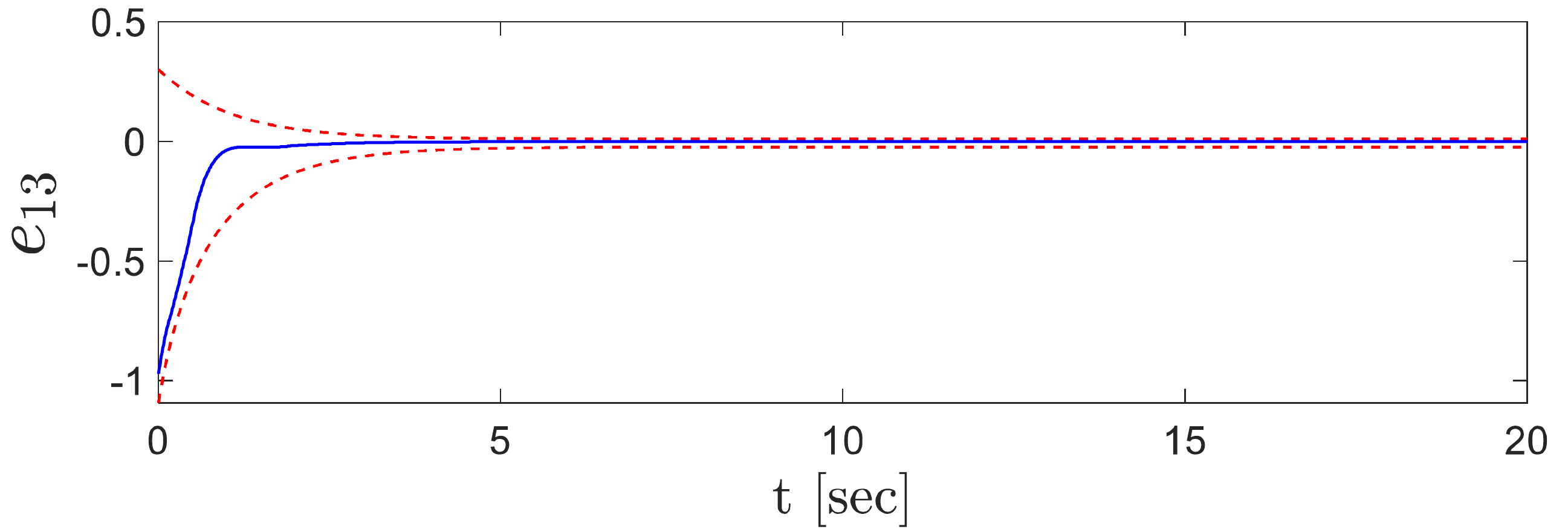}
		\end{subfigure}
		\begin{subfigure}[t]{0.236\textwidth}
			\centering
			\includegraphics[width=\textwidth]{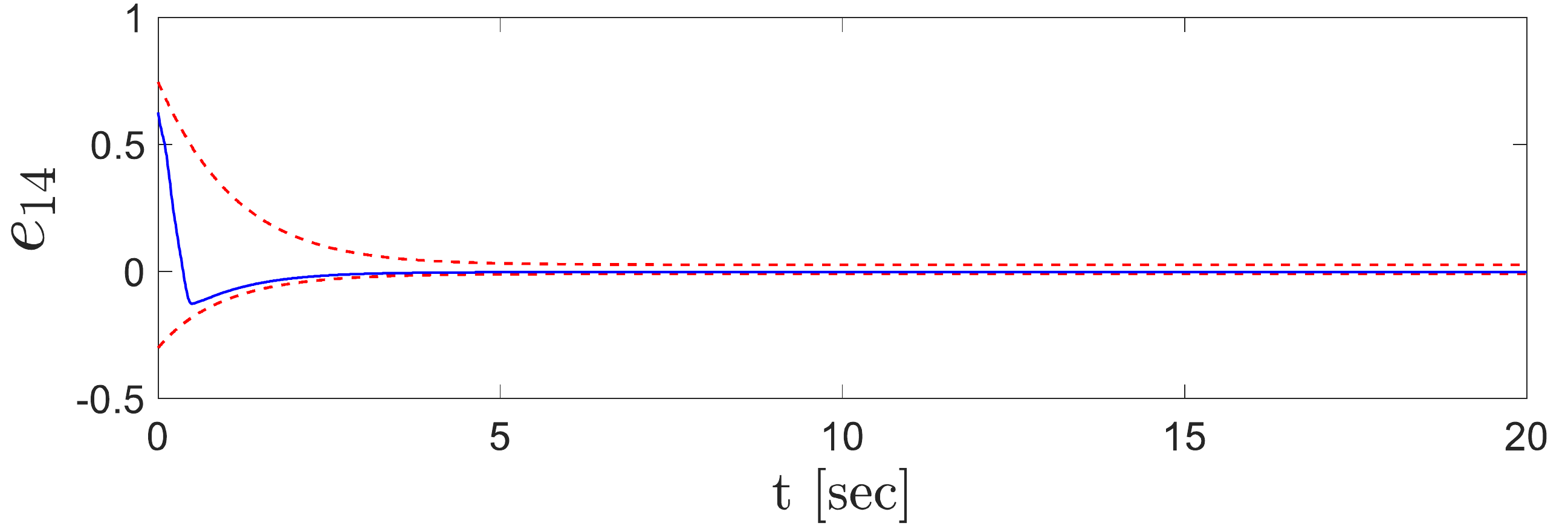}
		\end{subfigure}
		\begin{subfigure}[t]{0.236\textwidth}
			\centering
			\includegraphics[width=\textwidth]{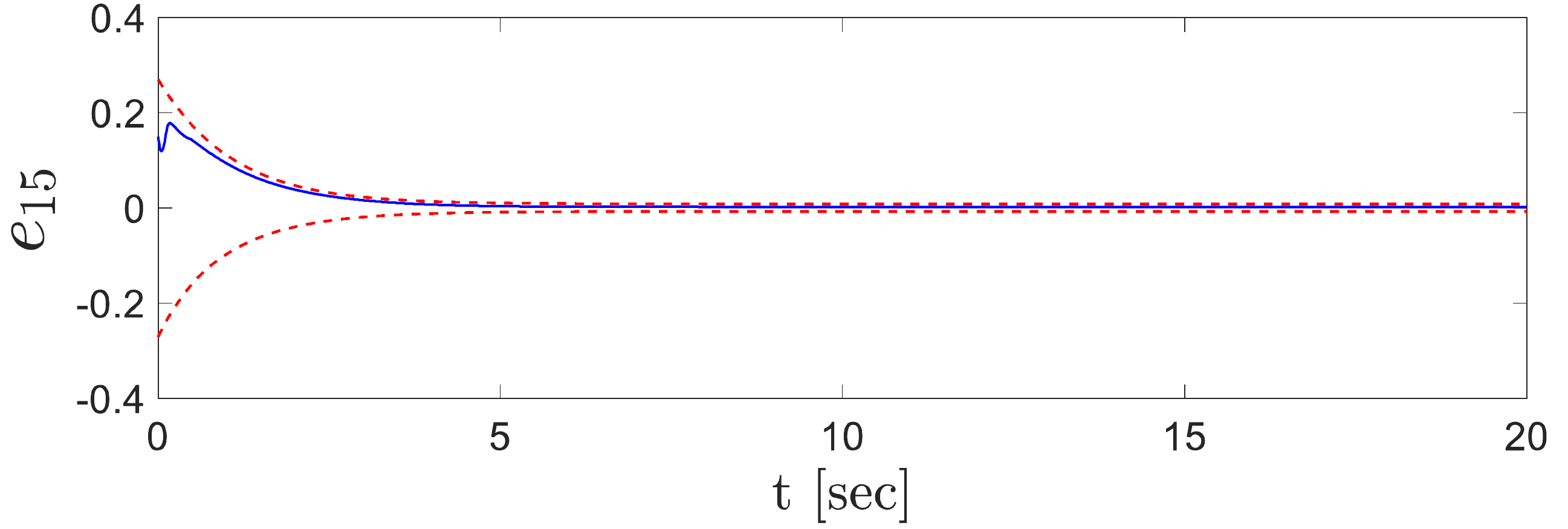}
		\end{subfigure}
		\begin{subfigure}[t]{0.236\textwidth}
			\centering
			\includegraphics[width=\textwidth]{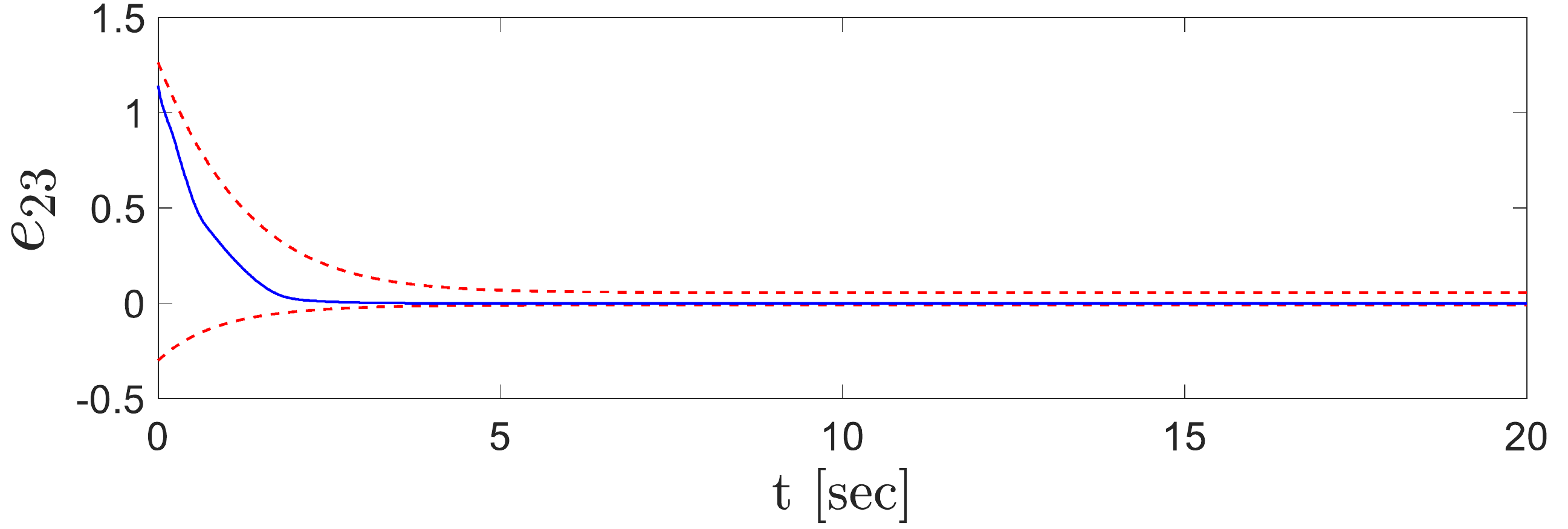}
		\end{subfigure}
		\begin{subfigure}[t]{0.236\textwidth}
			\centering
			\includegraphics[width=\textwidth]{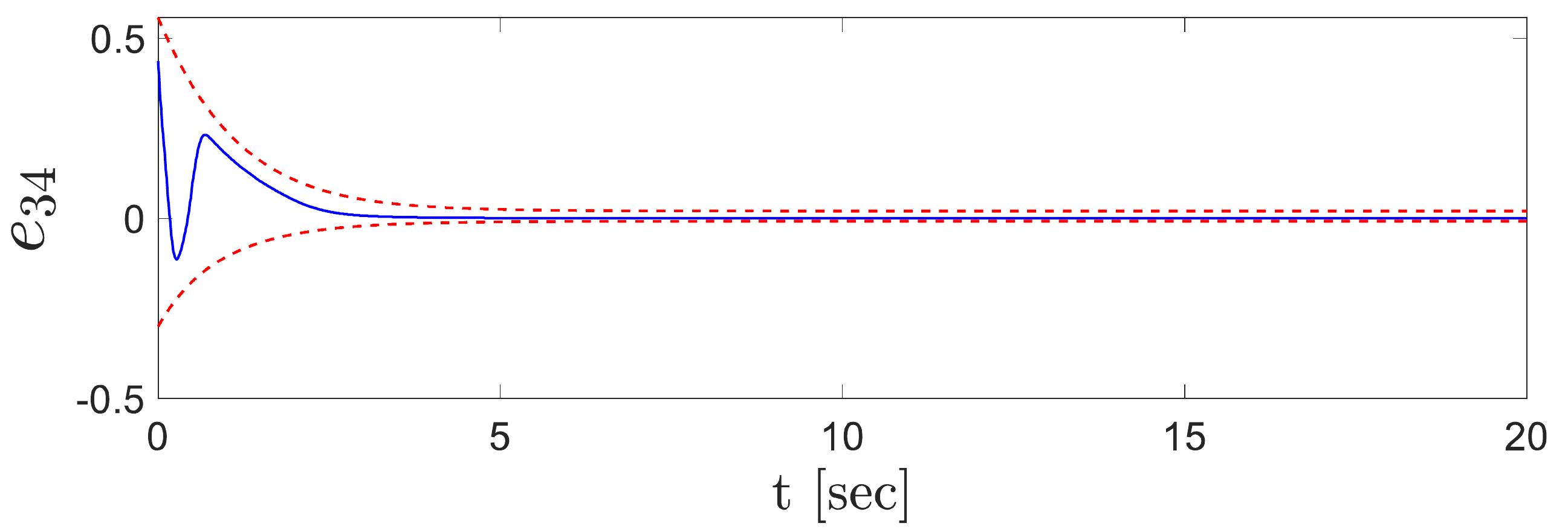}
		\end{subfigure}
		\begin{subfigure}[t]{0.236\textwidth}
			\centering
			\includegraphics[width=\textwidth]{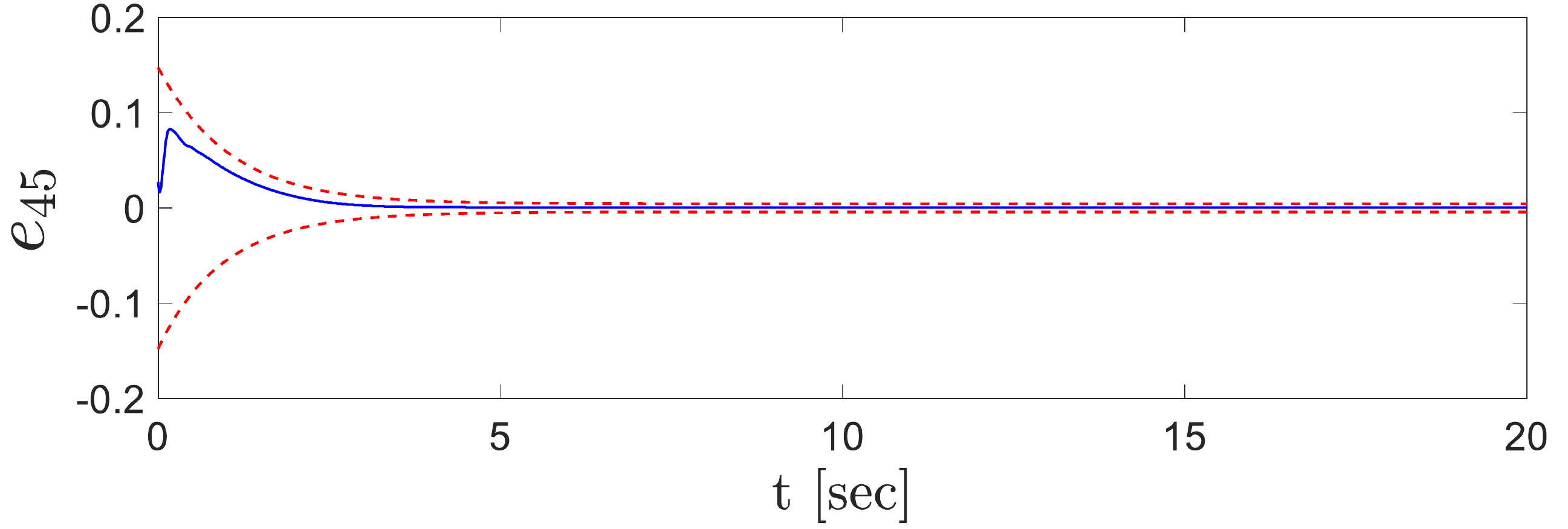}
		\end{subfigure}
		\begin{subfigure}[t]{0.236\textwidth}
			\centering
			\includegraphics[width=\textwidth]{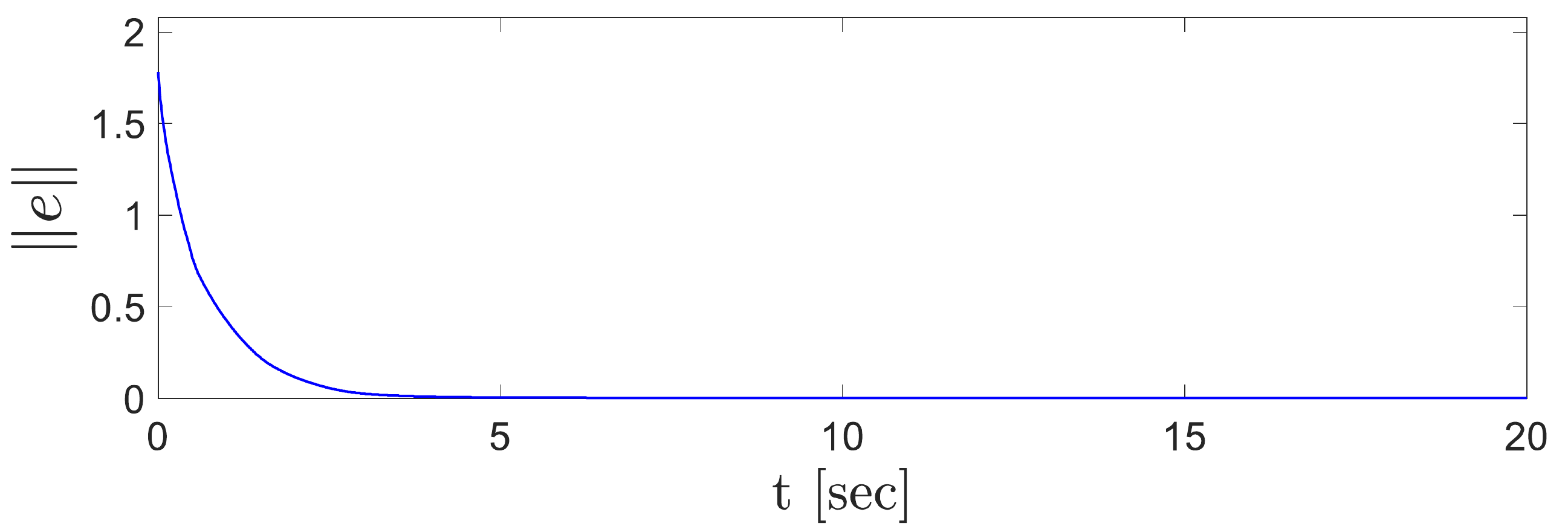}
		\end{subfigure}
		\caption{Inter-agent distance errors of the pentagon formation for $(i,j)\in \mathcal{E^\ast}$ as well as $\|e\|$ using the formation maneuvering controller \eqref{eq:u_single_maneuever}.}
		\label{fig:maneu_diagrams}
	\end{figure}

	Consider a group of five agents with single integrator dynamics \eqref{eq:singledyn_nomi} and the same desired formation as in the previous subsection. We assume that agent 5 is the leader of the group and thus has access to the desired velocity for the formation centroid and knows the total number of agents in the group. Let the desired velocity be $v_d=[\sin(0.5t), \cos(0.5t)]$ resulting in a circular motion. The initial positions of the agents are considered to be $q_1(0)=[-0.3639, \, 0.6361]$, $q_2(0)= [-1.7126,\,-0.4526]$, $q_3(0)=[0.4919,\, 0.2706]$, $q_4(0)=[2.0789, \, -0.0179]$, $q_5(0)=[0.9100,\,0.2679]$. Moreover, for the formation maneuvering control law in \eqref{eq:u_single_maneuever} it is assumed that the controller gains are set to $k_{ij}=0.2, (i,j) \in \mathcal{E^\ast}$ and $a_{ij}=1, (i,j) \in \mathcal{E^\ast}$. All other parameters (for the performance bounds and Algorithm \ref{algo}) are considered the same as in Subsection \ref{Simu:acqu}. Fig.~\ref{pic:maneu} shows the agents’ trajectories towards the desired shape while the formation centroid tracks the velocity $v_d(t)$ (the asteroid mark $\ast$ represents the centroid of the formation). Notice, based on Theorem \ref{th: single_cent_maneu}, th	e control protocol \eqref{eq:u_single_maneuever} ensures $\dot{q}_c(t)=v_d(t)$ for all time. Therefore, the formation centroid attains the desired velocity instantly. Fig.~\ref{fig:maneu_diagrams} shows the evolution of inter-agent distance errors for each edge as well as their norm. From the evolution of $\|e\|$ it is clear that the agents can maneuver cooperatively with the given time varying velocity of the formation centroid  without affecting the formation errors (formation maintenance). Moreover, since the distance errors are kept within the PPB, the proposed controller \eqref{eq:u_single_maneuever} is capable of ensuring connectivity maintenance and collision avoidance of neighboring agents.
	
	\section{Conclusion}
	\label{Conclu}
	
	In this paper, we proposed a novel decentralized robust distance-based formation control law with guaranteed performance for single integrator agents affected by unknown external disturbances. Moreover, by imposing proper predefined performance bounds in the design, we solved the problems of connectivity maintenance and collision avoidance among neighboring agents. Then, the results were extended to solve distance-based formation maneuvering for nominal single integrator agents, in which the centroid of the formation tracks a predefined time-varying velocity. It was assumed that only one agent, i.e., the leader of the group, had access to the desired centroid velocity. Moreover, we showed that the designed controllers are independent of a global coordinate system. Furthermore, we assumed that the desired formation is defined by a minimally and infinitesimally rigid graph in a 2 or 3 dimensional space. We  also highlighted that by using the proposed control scheme, which ensures  predetermined transient response, formation convergence to undesired shapes are prevented despite the presence of external disturbances. Future research efforts will be devoted towards considering directed interactions among agents and extending the results for agents with higher order and nonlinear dynamics. Another interesting topic for future research would be to check whether the proposed approach can be used directly or modified accordingly for solving the problem of distance-based formation control with neighboring agents' distance mismatch as addressed in \cite{mou2015undirected,de2014controlling}.
	\begin{ack}                               
		F. Mehdifar is an FRIA/FNRS fellow and his work is also supported by the concerted research action (ARC) ”RevealFlight”.   
	\end{ack}
	\bibliographystyle{elsarticle-num}
	\bibliography{PP_DBFC_extended_ArXiv}
\end{document}